\documentclass[11pt]{article}
\usepackage{iftex}
\ifPDFTeX
  \usepackage[T1]{fontenc}
  \usepackage[utf8]{inputenc}
\fi
\usepackage{lmodern}
\usepackage[a4paper,margin=2.2cm]{geometry}
\usepackage{amsmath,amssymb}
\usepackage{textcomp}
\usepackage{graphicx}
\usepackage{longtable,booktabs,array,calc}
\usepackage[font=small,labelformat=empty]{caption}
\usepackage[superscript]{cite}
\usepackage{placeins}
\usepackage{microtype}
\usepackage{xcolor}
\usepackage[colorlinks=true,linkcolor=blue!50!black,citecolor=blue!50!black,urlcolor=blue!50!black,breaklinks=true]{hyperref}
\usepackage{xurl}
\makeatletter
\def\fps@figure{!htbp}
\renewcommand\@biblabel[1]{#1.}
\makeatother

\title{Sham-controlled trials in cardiovascular medicine}
\author{Ibrahim Halil Tanboga, MD, PhD\\[2pt]
\small Departments of Biostatistics and Cardiology, Atlas University, Istanbul, T\"urkiye\\
\small \href{mailto:haliltanboga@gmail.com}{haliltanboga@gmail.com}}
\date{}

\begin{document}
\maketitle

\begin{abstract}
\noindent Procedural therapies can change cardiovascular physiology, symptoms and expectations at the same time, and when patients and clinicians know the allocation, randomization alone cannot separate these pathways. Sham procedures can support masking and define the incremental effect of a procedural component, but their interpretation depends on what the control contains. Improvement after sham is a placebo response that combines contextual effects, regression to the mean, natural history, measurement effects and concomitant care; it does not identify a psychological placebo effect. We integrate methodological guidance with cardiovascular trials and a detailed examination of selected protocols and statistical analysis plans. The review covers component-based control design, ethical proportionality, recruitment, treatment fidelity, masking assessment, endpoint selection, estimands, rescue treatment, missing outcomes, repeated measurements and adaptive monitoring. Cardiovascular experience shows several forms of knowledge revision, namely abandonment of unsupported procedures, reduction of exaggerated effect estimates, confirmation of symptom efficacy, restriction of benefit to selected phenotypes, and separation of physiological from patient-important effects. ORBITA and recent atrial fibrillation trials show how medication strategy, baseline symptom burden, monitoring and rescue handling can lead apparently similar trials to answer different questions. They also show why a neutral superiority test is not equivalence, why a positive supportive analysis does not replace the primary analysis, and why late allocation guesses cannot diagnose bias independently of treatment response. Credible procedural evidence requires an explicit causal contrast, proportionate control, prespecified analysis and complete reporting. Sham control serves these aims but does not by itself certify validity.


\smallskip
\noindent\emph{Evidence considered through 18 September 2026.}
\end{abstract}

\section*{Introduction}
\addcontentsline{toc}{section}{Introduction}

Cardiovascular interventions are unusually persuasive. An angiographic obstruction can disappear, an arrhythmia terminate or a pressure gradient fall immediately after an elaborate procedure. Such technical success may support biological plausibility but does not by itself show how much better patients feel or fare because of the intervention. Symptoms fluctuate, treatment often follows clinical deterioration, and an invasive encounter changes expectations, medication use and subsequent care. Improvement after an intervention is therefore not the same as improvement attributable to its proposed therapeutic component.\cite{m1,m2}

Sham-controlled randomized trials make the groups' experiences similar enough for masking while a defined component is withheld from controls. Besides identifying ineffective treatment, they can confirm benefit, calibrate its size, clarify mechanisms and show when an intervention is useful. Yet the label ``sham-controlled'' covers very different designs. Sedation without ablation, catheterization without implantation, an implanted device programmed off and vehicle injection through an otherwise identical delivery procedure each create a different contrast and risk.\cite{m1,m3}

This narrative methodological review examines these distinctions using cardiovascular evidence available through 18 September 2026. We combined a structured examination of selected trial reports, protocols, statistical analysis plans and methodological papers with targeted checks of the primary publications. It is not an exhaustive systematic review or a meta-analysis; study characteristics, documentary audits and source limitations are in the Supplementary Appendix. Our central argument is that interpretation starts from the control intervention and the estimand, not from whether a trial is called positive or negative.

\section*{What does a sham-controlled trial estimate?}
\addcontentsline{toc}{section}{What does a sham-controlled trial estimate?}

A procedural placebo resembles selected aspects of a therapeutic procedure but omits its hypothesized critical component. ``Sham'' describes how the control is built, and ``placebo-controlled'' describes its role in the experiment. Neither term guarantees biological inactivity. Diagnostic catheterization, sedation, cardioversion, medication optimization and implantation can help or harm even when the component under investigation is withheld.\cite{m3,m4}

The ORBITA investigators explicitly prefer ``placebo'' to ``sham'' in communication, arguing in a 2021 methodological article that the connotations of ``sham'' can undermine patient and clinician acceptance and hinder recruitment. This communication choice neither makes the control risk-free nor introduces a new causal design. We use both terms where they fit historically and describe the actual control whenever interpretation depends on it.\cite{m5}

Randomization between an active procedure (A) and a procedural control (S) estimates the effect of assignment to A rather than S under the trial's treatment, follow-up and rescue policies (Figure 1). Change within S compares one group over time and cannot supply the counterfactual, namely what would have happened without the trial or its control procedure. A large sham-group improvement therefore neither proves a large psychological placebo effect nor makes the active procedure unnecessary, and a small one does not exclude contextual influences.\cite{m6,m7}

\begin{figure}[!htbp]
\centering
\includegraphics[width=\textwidth,height=0.80\textheight,keepaspectratio]{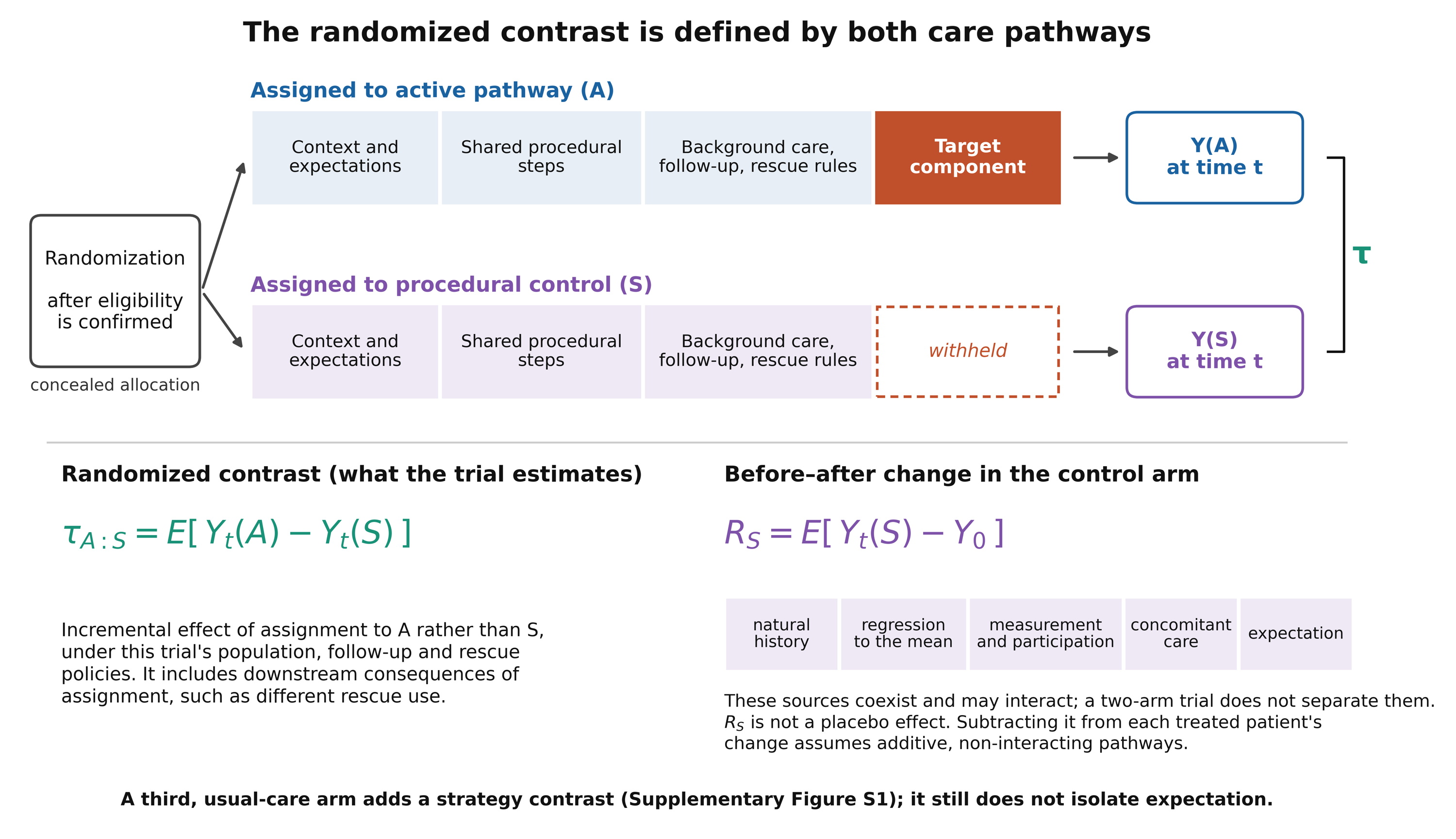}
\caption*{\textbf{Figure 1. Central illustration. The randomized contrast depends on both care pathways.} Patients are randomized, after eligibility is confirmed, to an active pathway (A) or a procedural-control pathway (S). Both pathways share selected contextual, procedural and clinical components, including background care with its follow-up and rescue rules, and the control pathway withholds the target component. Randomization estimates \ensuremath{\tau}, the difference between the assigned pathways under the trial's population, follow-up and rescue policies. This difference includes downstream consequences of assignment, such as different rescue use. It does not decompose the control group's before--after change (R\_S) into natural history, regression to the mean, measurement and research-participation effects, concomitant care and expectation, which may interact and are not separately identified in a two-arm trial. The distinction between component efficacy and the complete treatment strategy is a question of estimand and design and does not imply that contextual and biological pathways are additive. The equations illustrate a mean contrast, and other estimands use different population summaries. Segment widths are schematic. Original conceptual illustration informed by ASPIRE and DITTO.\cite{m2,m3,m7}}
\end{figure}

The distinction between these contrasts is important when much of the care is shared. In atrial fibrillation (AF) ablation trials, cardioversion, antiarrhythmic treatment and intensive surveillance may accompany both ablation and sham, so the contrast is incremental to that common care.\cite{m14,m17,m44} Device-on versus device-off trials can estimate the effect of activation after implantation and leave the total benefit and harm of implantation versus no implant unidentified.\cite{m45,m48}

A third usual-care arm can help separate the incremental component effect from the effect of the complete procedural strategy. The sham-versus-usual-care contrast, however, still mixes procedural exposure, attention, expectations, cointerventions and possible differences in outcome ascertainment, and it isolates expectation only under additional, justified design assumptions.\cite{m2,m7} Contextual and physiological pathways may also interact, so subtracting a constant ``placebo contribution'' from each patient's response is not generally valid. Figure 1 and Supplementary Figure S1 show why component definition precedes causal interpretation.

\section*{Why patients change after a control procedure}
\addcontentsline{toc}{section}{Why patients change after a control procedure}

We reserve the term placebo effect for effects attributable to mechanisms such as expectation and learning, not for every change observed in a placebo group. Table 1 separates it from the placebo response, which is the observed change after a placebo intervention. The placebo response may contain a placebo effect but also includes spontaneous variation, concurrent treatment, measurement error and secular disease change. Nocebo mechanisms can worsen symptoms or increase adverse-event reporting, so the absence of improvement does not mean the absence of contextual influence.\cite{m6}

\subsection*{\textbf{Table 1.} Sources of change after a procedural control: definitions and design consequences}
{\small\setlength{\LTpre}{4pt}
\begin{longtable}[]{@{}
  >{\raggedright\arraybackslash}p{(\linewidth - 4\tabcolsep) * \real{0.1834}}
  >{\raggedright\arraybackslash}p{(\linewidth - 4\tabcolsep) * \real{0.3305}}
  >{\raggedright\arraybackslash}p{(\linewidth - 4\tabcolsep) * \real{0.4861}}@{}}
\toprule\noalign{}
\begin{minipage}[b]{\linewidth}\raggedright
\textbf{Concept}
\end{minipage} & \begin{minipage}[b]{\linewidth}\raggedright
\textbf{Meaning}
\end{minipage} & \begin{minipage}[b]{\linewidth}\raggedright
\textbf{Consequence for a cardiovascular trial}
\end{minipage} \\
\midrule\noalign{}
\endhead
\bottomrule\noalign{}
\endlastfoot
Placebo effect & A change attributable to contextual mechanisms such as expectation or learning. & Requires a design capable of identifying those mechanisms; cannot be estimated from sham-group improvement alone. \\
Placebo response & Observed change after a placebo or sham intervention, regardless of its causes. & Report descriptively; use the randomized between-group contrast for the procedural efficacy claim. \\
Nocebo effect & Adverse change mediated by expectations or learned associations. & Standardize balanced consent and communication; collect adverse events consistently without dismissing them as psychological. \\
Regression to the mean & Less extreme repeat measurements after selection on a noisy extreme value. & Use repeated baseline assessment and concurrent randomization; avoid attributing within-arm blood-pressure or symptom change to treatment. \\
Research-participation / Hawthorne effects & Behavioral changes associated with observation, assessment or trial participation. & Describe contact intensity, adherence support and monitoring; do not assume a fixed or universally beneficial effect. \\
Natural history / progression & Recovery, fluctuation or deterioration that would occur without the tested component. & Benefit can mean less deterioration; parallel follow-up is needed even when an untreated trajectory seems predictable. \\
Ceiling / floor effects & A bounded scale restricts measurable improvement or worsening near its limits. & Examine baseline distribution and measurement range; distinguish limited headroom from biological effect modification. \\
Response shift & A change in the standards, values or interpretation used to report health status. & Use validated instruments and consistent timing; interpret patient reports together with clinically relevant complementary outcomes. \\
Expectation and allocation belief & Expected benefit and belief about which treatment was received are related but distinct. & Record wording, timing, confidence and reasons; allocation guesses alone do not quantify expectations. \\
Cointervention / rescue & Additional effective care, sometimes caused by symptoms after randomization. & Match decision rules, not necessarily realized use; specify whether rescue is included, incorporated or hypothetically excluded in the estimand. \\
Effect modification & The causal treatment contrast differs by a baseline characteristic. & Test a prespecified interaction; differences between trial averages or subgroup P values do not establish individual treatment selection. \\
\end{longtable}
}

{\footnotesize These mechanisms are conceptually distinct but may coexist or interact. A two-arm active-versus-sham trial generally does not identify their separate contributions. Sources: ASPIRE, CoPPS, blinding-method papers and methodological guidance.\cite{m2,m6,m7,m8,m9,m15,m16,m23}\par}

Regression to the mean matters most when eligibility requires severe symptoms, high blood pressure or an unusually poor questionnaire score. When measurements vary from one occasion to the next, patients selected at an extreme tend to have less extreme repeat values even without effective treatment (Figure 2A), a consequence of selection and imperfect repeatability rather than spontaneous healing or deception. Repeated baseline measurements can improve characterization and precision, yet an uncontrolled before--after comparison stays vulnerable. Concurrent randomization addresses the comparative problem because both groups pass through the same selection.\cite{m8}

\begin{figure}[!htbp]
\centering
\includegraphics[width=\textwidth,height=0.80\textheight,keepaspectratio]{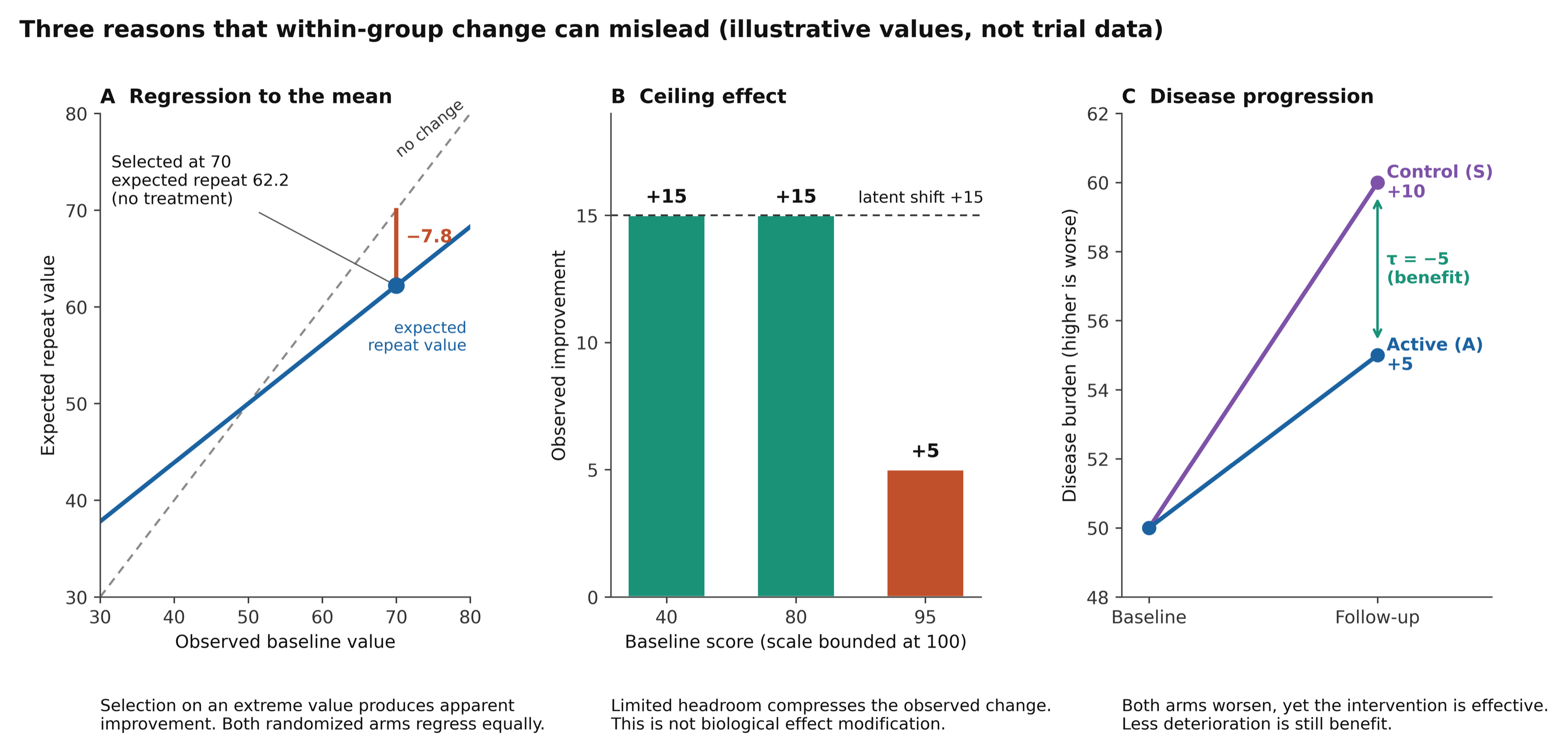}
\caption*{\textbf{Figure 2. Three reasons that within-group change can mislead.} All values are illustrative, not trial data. (A) Selection on an unusually severe baseline score can be followed by regression toward the population mean. Under a linear-normal repeated-measure model with population mean 50, between-person variance 100 and independent measurement-error variance 64, a baseline value of 70 has an expected repeat value of 50+(100/164)\ensuremath{\times}20=62.2 without treatment. (B) On a scale capped at 100, a 15-point latent improvement produces observed gains of 15, 15 and 5 when baseline scores are 40, 80 and 95. (C) An effective intervention can slow disease-related deterioration even when neither group improves from baseline. Here the control pathway worsens by 10 units and the active pathway by 5, a randomized contrast of \ensuremath{-}5. These mechanisms can coexist, and none of them can be inferred from a sham response alone.\cite{m8}}
\end{figure}

The Hawthorne effect is often cited as a universal explanation for improvement in trials. More precisely, participation, observation or awareness of assessment may change adherence, physical activity and symptom recording, and the size of these heterogeneous, hard-to-isolate research-participation effects should not be assumed from the label. Intensive follow-up can also deliver effective care, so a fall in blood pressure after sham angiography may reflect better adherence and measurement conditions rather than a purely psychological response to catheterization.\cite{m9}

Ceiling and floor effects limit how much change a bounded scale can register. A patient starting near the maximum quality-of-life score cannot improve numerically as much as a more symptomatic patient, even if both benefit (Figure 2B).\cite{m63} This is a property of the measurement, not biological effect modification. Baseline severity may also be related to prognosis, regression to the mean and treatment response, and these explanations must not be conflated. Enrichment for severe symptoms leaves more room for improvement but narrows generalizability and can amplify selection-related regression.

In progressive disease, an effective intervention may slow deterioration while both groups worsen, so the absence of within-group improvement can hide benefit (Figure 2C). Spontaneous recovery after an acute event can instead produce striking improvement in both groups. Response shift, a change in the standards patients use to interpret a questionnaire after treatment, can alter reported health without an equivalent change in the underlying construct.\cite{m64} Expectation, effort and interviewer behavior can also influence exercise tests, so ``objective'' outcomes are less protected than they seem.

These mechanisms should guide design, not serve as after-the-fact explanations for an unwelcome result. Inference rests primarily on the randomized between-group contrast. Mechanistic attribution needs additional measurements or experimental comparisons, and neither the size nor the statistical significance of the sham response can stand in for them.

Patient-reported improvement is still a legitimate therapeutic outcome, and masking is meant to make its comparison more interpretable, not to treat symptoms as less real than imaging findings. Physiological endpoints are not automatically protected from bias, because differential medication, surveillance, test completion or encouragement can affect their acquisition. Agreement between mechanistic and patient-important outcomes strengthens an interpretation. Disagreement calls for scrutiny of the proposed causal pathway, measurement sensitivity and treatment burden, not a preference for whichever endpoint is positive.

\section*{Designing the intervention, control and population}
\addcontentsline{toc}{section}{Designing the intervention, control and population}

Design begins with the clinical question. Testing a physiological component, symptom benefit under a specified medication strategy or the net value of a complete procedural pathway may each require a different control. The DITTO framework gives a sequence (deconstruct the intervention, identify its critical elements, remove the selected elements, consider risks and feasibility, optimize the control), and ASPIRE guidance addresses when an invasive placebo is necessary, acceptable and deliverable.\cite{m2,m3,m7} Supplementary Table S5 lists the resulting operational questions.

Deconstruction should cover preparation, access, imaging, anesthesia, instrumentation, the therapeutic maneuver, recovery, medication and follow-up, with an explicit reason to include, omit or simulate each component (Figure 3).\cite{m3} Sensory credibility does not require anatomical duplication. Omitting transseptal puncture from an AF sham can reduce risk and still preserve masking through sedation, venous access and a credible recovery pathway.\cite{m11,m14,m17,m44} An invasive step added only for visual symmetry is hard to justify if patients and assessors cannot perceive its absence.

\begin{figure}[!htbp]
\centering
\includegraphics[width=\textwidth,height=0.80\textheight,keepaspectratio]{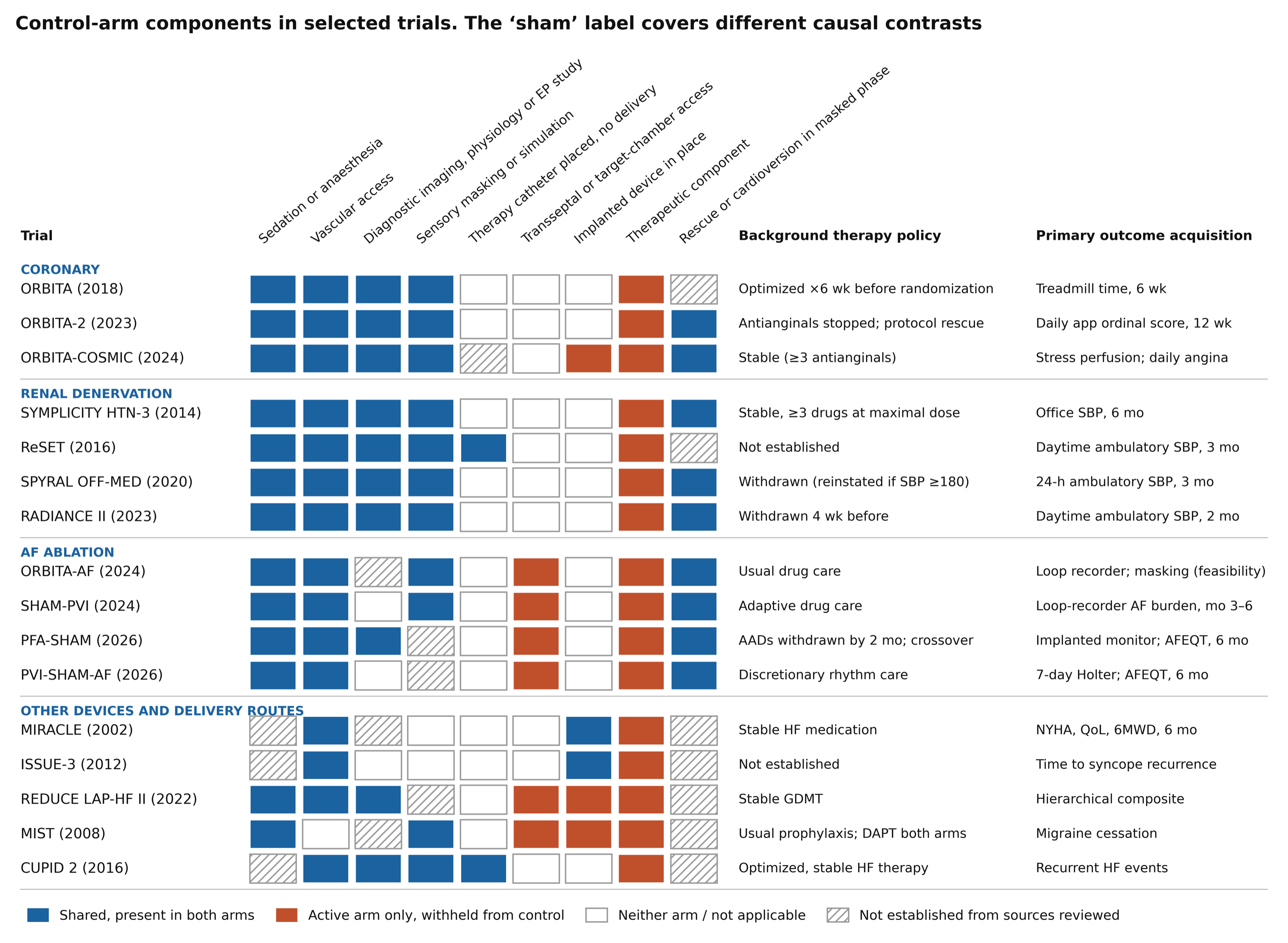}
\caption*{\textbf{Figure 3. Control-arm components in selected trials.} Each row is one randomization and each column a procedural or care component. Blue cells mark components present in both arms and red cells components present in the active arm only (withheld from control). White cells mark components absent from both arms or not applicable, and hatched cells components not established from the sources reviewed. Background therapy policy and primary outcome acquisition are summarized at right. The clinical question determines the estimand, which in turn determines the control components, conduct and analysis. A credible control minimizes avoidable risk while keeping the information needed for the contrast. Prespecified treatment delivery, rescue, masking assessment and outcome collection preserve that question during follow-up. Interpretation then returns to the population and strategy actually randomized and separates efficacy from transportability. The matrix describes the causal contrast each trial estimated. It is not a quality score, and more shared components do not make a better control. Sources were Supplementary Tables S1 and S2, primary publications and published design papers. Operational design questions are summarized in Supplementary Table S5, an author-developed synthesis that is not a validated checklist or a replacement for SPIRIT, CONSORT, ASPIRE or TIDieR-Placebo.\cite{m2,m4,m59,m60}}
\end{figure}

Proportionality is the ethical test. It asks whether the incremental research risk is justified by the importance of an otherwise unresolved question and whether a less burdensome design could answer it adequately. Proven effective care should remain available, and the Declaration of Helsinki permits placebo or withholding of an established intervention only under specified conditions, including compelling methodological reasons and protection from additional serious or irreversible harm. Consent should explain the actual control, its risks, uncertain benefit, alternatives, rescue arrangements and post-trial access. Honest checking of comprehension is more useful than reassurance that inadvertently equates ``placebo'' with harmlessness.\cite{m10}

Control design should be piloted with patients, procedural teams and outcome assessors, covering recruitment, retention and the separation of masked follow-up from unmasked procedural staff. ORBITA-AF illustrates the value of a dedicated feasibility study. A small trial can show whether a credible AF ablation placebo is deliverable without being expected to estimate efficacy precisely, and reading feasibility data as efficacy data invites overinterpretation of unstable estimates.\cite{m11}

The active procedure also needs quality assurance, because sham control cannot make up for insufficient operator experience, incomplete target treatment or an obsolete device. Eligibility, operator competence, definitions of technical success and independent assessment of delivery should be prespecified.\cite{m1} Randomizing only after procedural success (for example, successful device placement) answers a legitimate conditional question, but it excludes earlier failures and complications and cannot estimate the effect of offering the procedure to all eligible patients. Failed attempts and postrandomization crossovers ordinarily stay in the assigned-strategy analysis.

Run-in periods can establish symptom stability, optimize background care and test outcome collection, but they may also select adherent participants and lower baseline symptom burden. Optimizing medication before randomization, as in ORBITA, asks whether the intervention adds benefit to intensive treatment. Withdrawing antianginals, as in ORBITA-2, asks about procedural efficacy with limited background antianginal therapy. Neither design is inherently superior, and applicability depends on matching the trial strategy to the clinical decision.\cite{m12,m13}

Endpoints should follow the intended benefit. A procedure proposed mainly to relieve symptoms needs a patient-important symptom or functional outcome, whereas a mechanistic endpoint can establish target engagement and help interpret discordance.\cite{m19} Prespecified assessment times should allow for healing, delayed device action and plausible durability. Longer masking improves durability assessment but prolongs uncertainty and the potential need for rescue, and a longer trial is not automatically better if participants cannot ethically or practically stay on their assigned strategy.

The sample-size rationale should state the effect worth detecting, outcome variability, baseline adjustment, repeated-measure correlation, attrition and the consequences of co-primary endpoints or interim monitoring.\cite{m59} An individual patient's threshold for meaningful improvement is not automatically the between-group superiority margin. For a fixed total sample with similar variances, equal allocation often gives greater precision. Unequal allocation can help recruitment or safety characterization, but the cost of an active-arm majority should be weighed explicitly rather than assumed away.

Recruitment also affects transportability. Patients willing to accept a control procedure may differ from those seeking guaranteed intervention in symptom severity, expectations, previous treatment and willingness to tolerate rescue delays. Screening logs should distinguish clinical ineligibility, clinician non-offer and patient refusal, with limited baseline characterization where consent permits. In PVI-SHAM-AF, 262 (21.9\%) of 1,199 invited patients consented and were randomized. This limits assumptions about the wider referral population without invalidating the randomized comparison, and center concentration and specialist operator expertise similarly limit where results can reasonably be applied.\cite{m14}

Randomization should occur once eligibility can be established but before decisions that could selectively remove difficult cases. Allocation concealment, which requires an independent sequence and inaccessible future assignments, serves a different purpose from masking after allocation. Stratification by a few important factors, such as center or disease phenotype, should be reflected in the analysis. Crossover activation studies must deal with washout, carryover, period effects and changing disease state, and an irreversible intervention does not yield a valid crossover comparison simply because participants eventually receive both strategies.\cite{m1}

\section*{Conduct, masking and outcome acquisition}
\addcontentsline{toc}{section}{Conduct, masking and outcome acquisition}

Masking belongs to particular people at particular times, so reports should state whether patients, ward staff, follow-up clinicians, outcome assessors, adjudicators and analysts were masked. When the operator must know the assignment, the main safeguards are restricted access to procedural information, standardized communication and separation from later outcome assessment. Discharge letters, electronic records, device interrogation, imaging and adverse events can reveal allocation even after a convincing procedure.\cite{m4,m5}

Allocation guesses should be collected after recovery from sedation, before treatment effects could plausibly reveal allocation, and at prespecified later times when useful. Whether to offer ``do not know'' or require a choice is debated. Wording, uncertainty, confidence and reasons for guessing should be prespecified and reported, and an uncertainty question can precede a forced guess if that serves the assessment. Believing that one received active treatment differs from expecting a particular amount of improvement. Equal guesses do not prove equal expectations, and aggregate accuracy can hide different patterns in the two arms.\cite{m15,m16}

Late guesses are especially hard to interpret because clinical response can itself produce a correct guess. An association between assignment and an end-of-trial guess may reflect information leakage, recognizable adverse effects or true efficacy, and on its own it cannot diagnose bias.\cite{m15,m16} Excluding correct guessers or adjusting the treatment effect for guesses conditions on a postrandomization variable and can destroy the protection of randomization. Blinding indices are useful descriptive summaries when their formula, denominator, timing and handling of uncertainty are reported, but they are not certificates of validity or invalidity (Supplementary Figure S2).

Published indices also need computational checking. In PFA-SHAM, pre-discharge responses in the PFA arm (11 correct, 2 incorrect, 17 uncertain) give 0.69 when uncertain respondents are excluded and 0.30 with the original three-response Bang denominator (Supplementary Figure S3). The table appears to use the first convention and its footnote the second. This needs clarification before indices are compared across trials, although it does not by itself invalidate the clinical result, and raw response counts let readers separate different summaries from disagreement about masking itself.\cite{m15,m17}

Follow-up should standardize opportunities for contact, medication review, rehabilitation and repeat testing. Identical care rules need not produce identical care, since an ineffective arm may be expected to use more rescue medication. Because clinician discretion informed by accidental unmasking can contaminate this pathway, rescue criteria, decision-makers and documentation should be specified while leaving room for necessary clinical judgment.

Digital diaries can reduce long retrospective recall and describe symptom trajectories. ORBITA-2 asked about the previous day rather than capturing events in real time and achieved exceptionally complete daily reporting, but completeness is not contemporaneous accuracy, and thousands of diary entries do not create thousands of independent randomized participants. Reminder frequency, back-entry windows, help from relatives, device access and technical failures should be reported and handled consistently. Questionnaires should continue after treatment discontinuation or rescue whenever the target estimand requires them.\cite{m18,m19}

Adverse-event surveillance must cover the shared research procedures as well as the active component. A complication in a sham participant may reflect clinically meaningful exposure to the control, not merely an inconvenient feature of trial administration. Conversely, implant complications that occur in both arms of an activation trial cannot be estimated by subtracting device-off from device-on outcomes. Independent monitoring should consider cumulative benefit--risk, treatment fidelity, retention and accidental unmasking alongside formal stopping rules.

Safety comparisons also need denominators and exposure windows. Events among attempted procedures and among all randomized participants answer different questions, and both can be informative if clearly labeled. A trial powered for symptoms usually estimates uncommon stroke, tamponade or mortality risks imprecisely, and observing no events does not establish safety equivalence. Crossover after the blinded phase can give patients access to treatment and provide information about delivery. The original assignment remains randomized but may then estimate an early-versus-deferred strategy rather than sustained active treatment against an untreated control. Long-term reports should state which strategy is being compared.

\section*{Statistical analysis: preserving the question through the trial}
\addcontentsline{toc}{section}{Statistical analysis: preserving the question through the trial}

An estimand specifies the population, treatment conditions, outcome, handling of intercurrent events and population-level summary, and writing it down forces decisions about crossover, repeat intervention, rescue medication and death before a model is chosen. Under a treatment-policy strategy, outcomes after rescue remain relevant. A hypothetical strategy asks what would have happened without rescue and needs additional assumptions, whereas a composite strategy folds the event into the outcome. These are different questions, not interchangeable fixes for inconvenient data.\cite{m20}

Continuous outcomes generally benefit from prespecified adjustment for a reliable baseline measurement rather than comparison of unadjusted change scores, which improves precision and addresses chance imbalance without redefining treatment assignment. Nonlinear models need more care because conditional odds ratios, marginal probabilities, mean differences and distributional shifts summarize different aspects of treatment effect. A model-based score for a specified patient should not be reported as an observed group mean.\cite{m21}

ORBITA-2, which compared percutaneous coronary intervention (PCI) with a placebo procedure, makes these distinctions concrete. Its original protocol described an ordinal analysis of covariance, and its prespecified statistical analysis plan a Bayesian longitudinal ordinal model conditioned on the previous day's state. At the journal's request, the main publication reported a frequentist ordinal analysis of the final score (odds ratio 2.21; 95\% confidence interval {[}CI{]} 1.41--3.47), while the prespecified longitudinal analysis in the supplement gave a day-84 conditional odds ratio of 1.88 (95\% credible interval {[}CrI{]} 1.57--2.27). Both favored PCI, but their estimands and model structures differ. Because it conditions on the previous day's postrandomization state, the longitudinal odds ratio is not a marginal assigned-strategy effect. The analysis history should be reported transparently, and a journal-requested analysis does not make the prespecified one disappear.\cite{m13}

The ORBITA-2 angina symptom score also encodes clinical priorities (Supplementary Figure S4).\cite{m13} Medication units take precedence over daily episode categories, and intolerable angina, acute coronary syndrome and death rank above all ordinary grades. A day without angina on one medication unit therefore ranks worse than a day with six or more episodes and no medication. This is an explicit ordinal choice, not a cardinal equation equating one tablet with a fixed amount of suffering. Component outcomes and their clinical interpretation remain necessary, and the score is not a complete measure of net clinical benefit.

Repeated outcomes can improve precision and show when efficacy begins or wanes, provided within-patient dependence is modeled with, for example, mixed models, generalized estimating equations or ordinal transition models suited to the target. The number of participants remains central to information about treatment effects and heterogeneity. Flexible longitudinal models need enough prespecification to limit opportunistic choices about time, interactions, covariance and outcome distribution.

Covariate adjustment should rely mainly on information available before randomization.\cite{m20,m21} Achieved blood pressure, postprocedural AF burden, medication escalation and treatment guesses may lie on the causal pathway, and adding them to the primary model can remove part of the effect or introduce selection bias. Mechanistic mediation analyses are possible but need stronger assumptions than the randomized total-effect comparison. Excluding failed interventions or nonadherent controls does not make per-protocol analyses unbiased; they need a defined causal target and suitable methods, and the assigned-strategy result should be retained.

Missing outcomes need an explicit causal and clinical account. Complete-case analysis does not become intention-to-treat because patients keep their assigned labels, and multiple imputation depends on its variables and missingness assumptions without guaranteeing recovery of unobserved outcomes, so sensitivity analyses should explore plausible departures. Death differs in kind from a missed questionnaire because subsequent quality of life does not exist. Assigning a worst rank, analyzing survival jointly or asking a hypothetical survival question changes what is estimated and should be named accordingly.\cite{m22}

In PFA-SHAM, pre-rescue scores on the Atrial Fibrillation Effect on Quality-of-Life (AFEQT) questionnaire replaced the six-month scores of five sham participants who underwent rescue ablation (Supplementary Figure S5). Keeping them in their randomized groups preserves assignment, but the endpoint becomes health status at six months, or before rescue for rescued participants, rather than at six months under a strategy permitting rescue. That may answer a useful question about failure of initial management, but it is not an ordinary fixed-time treatment-policy contrast.\cite{m17}

Multiplicity and monitoring should match the claims. Co-primary outcomes that must all succeed differ from multiple outcomes of which any single success establishes efficacy, and secondary endpoints, subgroups and repeated publications from one randomization are not independent confirmation. Bayesian updating gives coherent inference under a specified model and prior but does not automatically control frequentist type I error in an adaptive trial, and a design claiming such control needs its complete stopping and decision rule evaluated, usually by simulation. PFA-SHAM's documented switch to Bayesian analysis before its first reported interim analysis (Supplementary Figure S6) differs from an originally Bayesian design, but the amendment should not be assumed to have been outcome-driven.\cite{m17}

A nonsignificant superiority test does not establish equivalence; interpretation should consider the estimate, its interval and clinically relevant alternatives. ORBITA's exercise-time difference of 16.6 seconds (95\% CI \ensuremath{-}8.9 to 42.0 seconds) encompassed its planned 30-second effect. In PVI-SHAM-AF, the primary result was similarly inconclusive for superiority, although a prespecified adjusted mixed-model sensitivity analysis favored ablation. Such supportive analyses can assess robustness or answer related questions but cannot silently replace the confirmatory endpoint. Heterogeneity of treatment effect needs an interaction and credibility assessment, not significance in one subgroup and nonsignificance in another.\cite{m12,m14,m23}

The ORBITA-2 secondary analyses of physiological severity, stress-induced ischemia and symptom typicality did test treatment interactions and so offer more than prognostic associations. They nevertheless reuse the same randomized cohort, involve multiple analytic opportunities and have not independently validated a treatment-selection rule. A continuum of greater average benefit with greater ischemic or symptom burden does not define a sharp threshold below which no patient benefits, and enrichment rules need prospective evaluation rather than acceptance merely because an interaction was detected.\cite{m24,m25,m26}

An analysis amendment should be judged by its rationale, timing, access to accumulating outcomes and effect on interpretation. A change before database lock is not necessarily free of outcome knowledge, and a change late in recruitment is not necessarily improper. Versioned protocols, signed analysis plans, access records and a clear split between prespecified and added analyses make that judgment possible (Supplementary Figure S6).\cite{m14,m59,m60} A sensitivity analysis tests assumptions for the same estimand, whereas a supplementary analysis may target a different outcome summary or rescue strategy, and calling both ``sensitivity'' can hide a substantive change in the question.

\section*{What cardiovascular trials have changed}
\addcontentsline{toc}{section}{What cardiovascular trials have changed}

The history is more varied than a sequence of procedural reversals (Table 2; Figure 4). Blinded studies of internal mammary artery ligation challenged that operation rather than modern coronary bypass surgery. The percutaneous laser evidence includes the small positive BELIEF trial alongside larger neutral comparisons, so sham evidence should be read across all relevant trials and not only the best-known negative study.\cite{m27,m28,m29,m30,m31}

\subsection*{\textbf{Table 2.} How cardiovascular sham and masked-device trials revised prior knowledge}
{\small\setlength{\LTpre}{4pt}
\begin{longtable}[]{@{}
  >{\raggedright\arraybackslash}p{(\linewidth - 6\tabcolsep) * \real{0.1598}}
  >{\raggedright\arraybackslash}p{(\linewidth - 6\tabcolsep) * \real{0.2268}}
  >{\raggedright\arraybackslash}p{(\linewidth - 6\tabcolsep) * \real{0.3144}}
  >{\raggedright\arraybackslash}p{(\linewidth - 6\tabcolsep) * \real{0.2990}}@{}}
\toprule\noalign{}
\begin{minipage}[b]{\linewidth}\raggedright
\textbf{Intervention / pivotal trials}
\end{minipage} & \begin{minipage}[b]{\linewidth}\raggedright
\textbf{Evidence before the pivotal comparison}
\end{minipage} & \begin{minipage}[b]{\linewidth}\raggedright
\textbf{What the controlled trial showed}
\end{minipage} & \begin{minipage}[b]{\linewidth}\raggedright
\textbf{What changed and what remained unresolved}
\end{minipage} \\
\midrule\noalign{}
\endhead
\bottomrule\noalign{}
\endlastfoot
Internal mammary artery ligation

Cobb; Dimond (1959--1960) & Apparent angina relief after surgical ligation in uncontrolled series. & Two small blinded surgical comparisons found improvement without a persuasive incremental ligation effect.\cite{m27,m28} & Undermined this operation's proposed efficacy. These were not trials of modern bypass grafting. \\
Percutaneous myocardial laser

Stone; BELIEF; DIRECT & Uncontrolled experience and unblinded randomized studies suggested symptomatic benefit. & Larger blinded comparisons were neutral; the smaller BELIEF study had a positive angina endpoint.\cite{m29,m30,m61} & Weakened claims of a reliable laser-specific benefit. Do not describe every sham laser trial as negative. \\
Renal denervation

SYMPLICITY HTN-3; SPYRAL; RADIANCE & Large office-pressure reductions in early uncontrolled and unblinded randomized studies. & HTN-3: office systolic difference \ensuremath{-}2.39 mmHg (95\% CI \ensuremath{-}6.89 to 2.12). Later trials often found modest ambulatory reductions; some remained neutral.\cite{m33,m34,m35,m36} & Recalibrated efficacy and stimulated better devices, procedural delivery and medication assessment. Hard-outcome benefit remains unproven by these trials. \\
Stable coronary PCI

ORBITA; ORBITA-2 & Symptom benefit supported by clinical experience and unblinded RCTs. & ORBITA: exercise-time difference 16.6 s (95\% CI \ensuremath{-}8.9 to 42.0) after intensive medication. ORBITA-2: final symptom-score OR 2.21 (1.41--3.47) with little background antianginal therapy.\cite{m12,m13} & Clarified the size and conditions of symptom efficacy. Cross-trial differences do not randomize medication strategy or establish a treatment sequence. \\
Chronic total occlusion PCI

ORBITA-CTO (2026) & Registries and heterogeneous unblinded RCTs; symptom effects susceptible to expectations, crossover and treatment of other lesions. & N=50: angina symptom-score OR 4.38 (95\% CrI 1.57--12.69), favoring PCI over six months.\cite{m40} & Confirmed symptom efficacy in selected single-vessel CTO without bystander disease; no reliable mortality inference. \\
Coronary sinus reducer

COSIRA; ORBITA-COSMIC & Encouraging uncontrolled refractory-angina experience. & COSIRA supported angina improvement. ORBITA-COSMIC: primary symptom outcome improved despite neutral primary perfusion outcome.\cite{m41,m42} & Strengthened symptom evidence while questioning a simple perfusion-mediated explanation. No established prognostic effect. \\
AF ablation

SHAM-PVI; PFA-SHAM; PVI-SHAM-AF & Substantial rhythm-control evidence and symptom/QoL benefit from unblinded comparisons; the masked incremental symptom effect remained uncertain. & SHAM-PVI and PFA-SHAM showed rhythm and QoL benefit. PVI-SHAM-AF primary AFEQT contrast 2.6 (95\% CI \ensuremath{-}2.7 to 8.0); prespecified adjusted mixed-model sensitivity contrast 5.4 (1.0--9.8).\cite{m14,m17,m44} & Confirmed efficacy in some settings and exposed heterogeneity in symptom benefit and estimands. The trials do not compare energy sources with one another. \\
Pacing for reflex syncope

VPS II; ISSUE-3; BIOSync CLS & Large effects in earlier open-label pacing trials. & Broad early masked comparison inconclusive; selected asystolic/cardioinhibitory populations benefited in later trials.\cite{m45,m46,m47} & Restricted the indication to a better-defined physiological phenotype. On/off comparisons do not estimate total implantation risk versus no device. \\
Resynchronization / rate-adaptive pacing

MIRACLE; EchoCRT; RAPID-HF & Mechanistic plausibility and early functional signals encouraged wider pacing indications. & MIRACLE supported functional benefit; EchoCRT narrow-QRS mortality HR 1.81 (95\% CI 1.11--2.93). RAPID-HF raised exercise heart rate without improving exercise capacity.\cite{m48,m49,m62} & Confirmed selected CRT benefit and prevented physiological reasoning from justifying harmful or unproven extensions. \\
HF shunting / autonomic intervention

REDUCE LAP-HF II; RELIEVE-HF; REBALANCE-HF & Early hemodynamic, feasibility or uncontrolled clinical signals. & Larger sham comparisons did not establish overall clinical or exercise-hemodynamic benefit.\cite{m50,m51,m52} & Limited routine extrapolation. Phenotype signals require prospective confirmation; a positive mechanistic pilot is insufficient. \\
Indirect mitral annuloplasty

REDUCE-FMR & Uncontrolled functional-MR studies suggested reverse remodeling and symptomatic benefit. & N=120: reduction in regurgitant volume; paired primary imaging available in 68. Clinical functional outcomes were inconclusive.\cite{m43} & Established a physiological signal with important missing-imaging limitations, not definitive patient-important efficacy. \\
PFO closure for migraine

MIST; PREMIUM & Observational associations and symptom improvement after closure. & Neither trial met its primary migraine efficacy endpoint.\cite{m56,m57} & Weakened the migraine indication; cannot be extrapolated to secondary stroke prevention. \\
Pulmonary artery denervation

PADN-CFDA & Promising small or incompletely masked studies with varying drug strategies. & N=128: six-minute walk difference 33.8 m (95\% CI 16.7--50.9) favoring denervation.\cite{m58} & Supported functional efficacy under the study conditions. Mortality and broader treatment-strategy applicability remain unresolved. \\
Conditioning / regenerative therapy

ERICCA; CUPID 2; DREAM-HF & Biological plausibility and encouraging early studies, some already placebo-controlled. & Larger controlled trials were neutral on their primary outcomes.\cite{m53,m54,m55} & Challenged selected mechanisms or delivery strategies. Changed conclusions cannot be attributed solely to adding sham, nor generalized to every therapy in the field. \\
\end{longtable}
}

{\footnotesize Selected evidence transitions, not a count of all trials or a formal assessment of certainty. N denotes randomized participants unless otherwise stated. CI, confidence interval; CrI, credible interval; AF, atrial fibrillation; AFEQT, Atrial Fibrillation Effect on Quality-of-Life; CRT, cardiac resynchronization therapy; CTO, chronic total occlusion; HF, heart failure; MR, mitral regurgitation; OR, odds ratio; PCI, percutaneous coronary intervention; PFO, patent foramen ovale; QoL, quality of life; RCT, randomized controlled trial. The Supplementary Appendix includes an expanded inventory, actual control components and documentary limitations.\par}

\begin{figure}[!htbp]
\centering
\includegraphics[width=\textwidth,height=0.80\textheight,keepaspectratio]{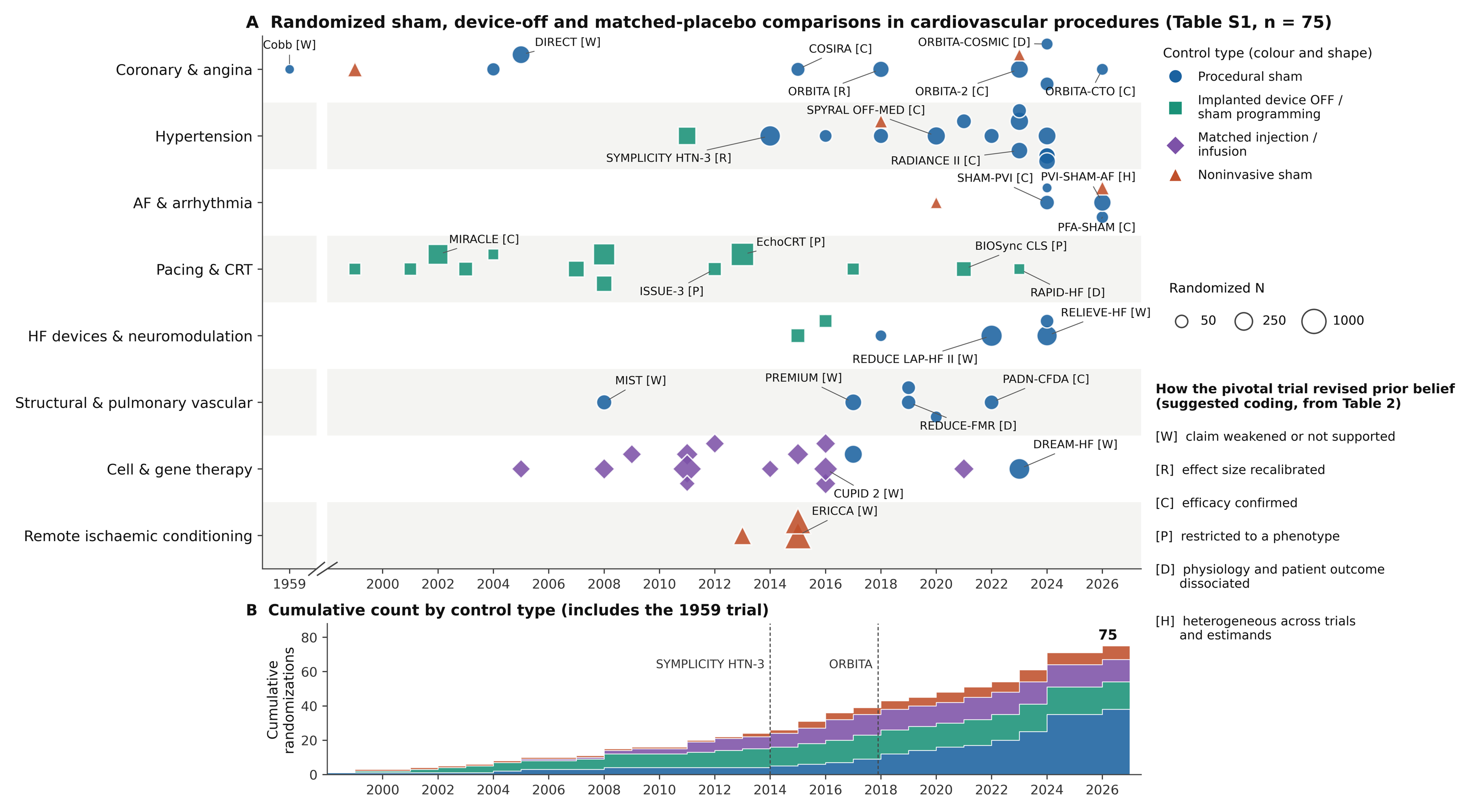}
\caption*{\textbf{Figure 4. Cardiovascular sham-controlled, device-off and matched-placebo randomizations, 1959--2026.} (A) Each marker is one randomization from Supplementary Table S1 (n = 75, boundary studies excluded), placed by year of the principal publication and by clinical domain. Marker area is proportional to randomized N, and color and shape denote the control type. Labeled trials are the principal comparisons discussed in Table 2. Bracketed letters give a suggested classification of how each trial revised prior belief (W, claim weakened or not supported; R, effect size recalibrated; C, efficacy confirmed; P, restricted to a phenotype; D, physiological and patient-important outcomes dissociated; H, heterogeneous across trials and estimands). The 1959 trial is shown on a broken axis, and vertical offsets within a domain only prevent overlap. (B) Cumulative number of randomizations by control type. The inventory is targeted rather than exhaustive, so the counts describe the evidence assembled for this review and not the complete output of the field.}
\end{figure}

Renal denervation shows recalibration and subsequent redevelopment (Table 2; Supplementary Figure S7). Large office-pressure reductions in early unblinded studies were followed by a neutral SYMPLICITY HTN-3 primary comparison, despite reductions within both arms. Device generation, procedural completeness, changes in antihypertensive medication and measurement strategy complicated interpretation. Later SPYRAL and RADIANCE trials showed more modest ambulatory effects under defined conditions, although not every later sham comparison was positive. The evidence supports blood-pressure lowering in selected settings, but neither the original dramatic effect nor a reduction in cardiovascular mortality.\cite{m32,m33,m34,m35,m36}

In stable coronary disease, clinical experience and unblinded randomized evidence, including ACME, RITA-2 and COURAGE, supported symptom benefit before ORBITA.\cite{m37,m38,m39} ORBITA challenged the assumption that opening a severe stenosis necessarily adds a large exercise benefit after intensive medication optimization, and ORBITA-2 then demonstrated symptom efficacy with little background antianginal treatment. The two findings refine the conditions of benefit rather than contradict each other. ORBITA-2 did not randomize PCI against optimized medication, however, so comparing the two trials cannot establish a medication-by-PCI interaction or an optimal treatment sequence (Supplementary Figure S8). ORBITA-CTO extended placebo-controlled symptom evidence to selected patients with a single symptomatic chronic total occlusion, without establishing prognostic benefit or applicability to all occlusion anatomy (Table 2).\cite{m12,m13,m40}

COSIRA supported symptomatic benefit from coronary sinus reduction, and ORBITA-COSMIC found improvement in its primary symptom outcome despite no clear benefit in its primary perfusion outcome. This dissociation questions a simple mechanistic explanation without nullifying the patient-reported effect. REDUCE-FMR showed a signal of reduced regurgitant volume after indirect mitral annuloplasty, but incomplete paired imaging and inconclusive clinical outcomes limited what could be claimed about functional benefit.\cite{m41,m42,m43}

AF ablation offers a particularly instructive comparison (Figure 5). SHAM-PVI (126 randomized patients) reduced AF burden, with an adjusted AFEQT advantage of 18.39 points, and PFA-SHAM (60 severely symptomatic patients) found substantial rhythm and quality-of-life effects, with a primary AFEQT change contrast of 32.6 points. PVI-SHAM-AF (262 patients) improved rhythm measurements, yet its primary AFEQT contrast was neutral (2.6 points; 95\% CI \ensuremath{-}2.7 to 8.0) and a prespecified adjusted mixed-model sensitivity estimate was positive (5.4 points; 95\% CI 1.0 to 9.8). Because these two analyses also handled death differently, model sensitivity has to be distinguished from a change in outcome summary.\cite{m14,m17,m44}

\begin{figure}[!htbp]
\centering
\includegraphics[width=\textwidth,height=0.80\textheight,keepaspectratio]{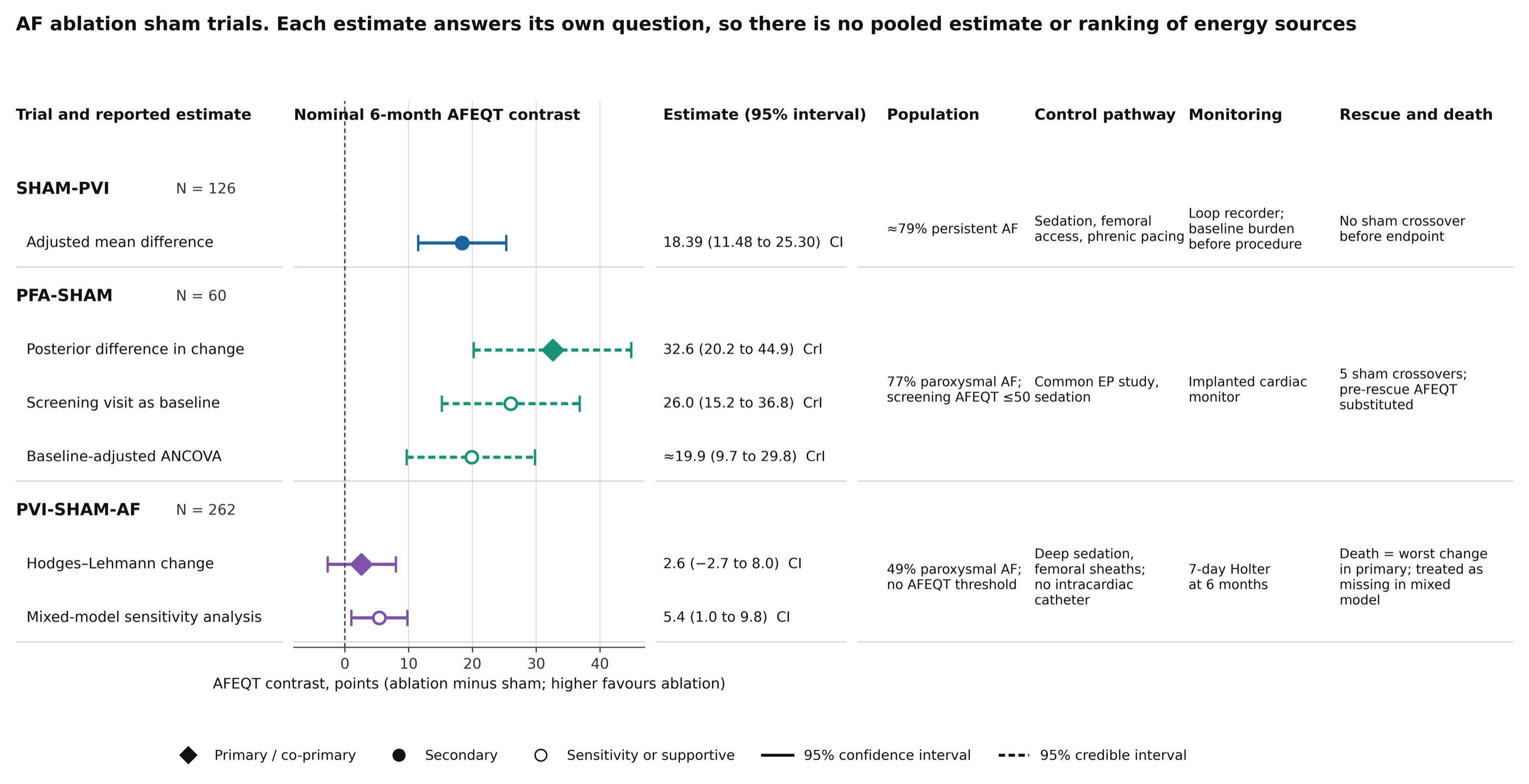}
\caption*{\textbf{Figure 5. Similar instruments do not guarantee the same estimand in AF ablation trials.} Reported AFEQT contrasts (ablation minus sham) at a nominal six-month assessment are shown separately for each trial, and there is no pooled estimate. Diamonds mark primary or co-primary analyses, filled circles a secondary outcome and open circles sensitivity or supportive analyses. Solid lines are 95\% CIs and dashed lines 95\% CrIs. In SHAM-PVI, the adjusted mean difference in a secondary outcome was 18.39 points (95\% CI 11.48--25.30). In PFA-SHAM, the posterior mean difference in change for a co-primary outcome was 32.6 points (95\% CrI 20.2--44.9). The corresponding estimate was 26.0 points (15.2--36.8) with the screening visit as baseline and approximately 19.9 points (9.7--29.8) with baseline-adjusted ANCOVA. Pre-rescue questionnaires replaced six-month values in five sham participants. In PVI-SHAM-AF, the primary Hodges--Lehmann location-shift estimate was 2.6 points (95\% CI \ensuremath{-}2.7 to 8.0) and the prespecified adjusted mixed-model sensitivity estimate 5.4 points (1.0--9.8). Deaths were assigned the worst possible change in the primary analysis and handled as missing in the mixed model. The trials differ in population, background care, monitoring, rescue, baseline symptom burden and statistical summary (right-hand columns), so their ordering is not a randomized comparison of ablation energy. AFEQT, Atrial Fibrillation Effect on Quality-of-Life; CI, confidence interval; CrI, credible interval.\cite{m14,m17,m44}}
\end{figure}

These estimates are not randomized comparisons of cryoballoon, radiofrequency and pulsed-field energy. Baseline symptom severity, AF phenotype, medication policy, cardioversion, rhythm monitoring, rescue handling and statistical summary all differed (Figures 3 and 5), changing both the clinical question and its measurement. PFA-SHAM required a screening AFEQT score of 50 or less, and PVI-SHAM-AF did not. The widely discussed 27\% versus 48\% recurrence comparison in PVI-SHAM-AF was a post-hoc combination of clinical recurrence and Holter findings that counted missing Holters as no detected AF, and prespecified clinical recurrence alone was inconclusive.\cite{m14,m17} Baseline severity plausibly contributes to the discordance, but aggregate comparisons cannot prove that it explains the discordance.

Pacing trials show how masking can refine indications (Table 2; Supplementary Figure S7). Early open-label vasovagal syncope trials suggested large effects.\cite{m65,m66} The masked VPS II and SYNPACE trials were inconclusive,\cite{m45,m67} whereas later trials that selected patients with documented asystole or specified cardioinhibitory responses demonstrated benefit. In heart failure, masked resynchronization trials supported functional efficacy in appropriate patients, while EchoCRT identified harm when resynchronization was extended to a narrow-QRS population. Sham or device-off control can therefore confirm a useful therapy while preventing inappropriate extrapolation.\cite{m45,m46,m47,m48,m49}

Other programs separate mechanism from clinical value (Table 2). Early physiological or uncontrolled findings with interatrial shunting and autonomic modulation in heart failure were not consistently confirmed in larger controlled studies, and remote ischemic conditioning and gene or cell therapies produced neutral larger trials, sometimes after encouraging studies that were themselves placebo-controlled. Because sample size, population, treatment delivery and endpoints changed at the same time, these revisions cannot be attributed to sham alone. Nor do they justify erasing positive smaller studies by claiming that a whole biological field has failed.\cite{m50,m51,m52,m53,m54,m55}

Evidence also remains specific to the indication. MIST and PREMIUM did not meet their primary migraine endpoints after patent foramen ovale closure, and these findings cannot be extrapolated to closure for secondary stroke prevention. PADN-CFDA supported an improvement in six-minute walk distance after pulmonary artery denervation under its trial conditions, without establishing a mortality effect. A procedure should not receive a universal verdict from a study of one indication, population or outcome.\cite{m56,m57,m58}

Synthesis across sham trials therefore requires more than grouping by device name, and populations, control components, background therapy, outcome definitions, rescue policies and follow-up should be aligned before pooling. A confidence interval for an adjusted mean difference and a credible interval for a posterior change contrast may both be valid yet unsuitable for naive graphical ranking. Aggregate relationships between baseline symptoms and trial effect cannot identify individual treatment-effect modification.\cite{m23} Individual-participant analyses with prespecified interactions would be more informative but must keep trial-specific treatment strategies and distinguish baseline modifiers from postrandomization mediators.

Across these programs, the most consequential errors have been both interpretative and operational. Within-arm change has been equated with causation, physiological success extrapolated to symptoms or survival, selected subgroups treated as confirmation and trials compared without aligned estimands, while unstandardized medication, inadequate procedural delivery, selective postrandomization exclusions and incomplete outcome ascertainment can further weaken causal interpretation. None of these problems is remedied merely by describing a trial as sham-controlled.

\section*{Reporting and future priorities}
\addcontentsline{toc}{section}{Reporting and future priorities}

Protocols and reports should make the comparison reproducible by listing each procedural component, stating who was masked and when, preserving the history of endpoint and analysis amendments, reporting rescue and missingness by assignment, and giving component outcomes alongside composite scores. SPIRIT 2025 and CONSORT 2025 supply the general structure, and ASPIRE and TIDieR-Placebo add procedural and control detail. CoPPS offers useful conceptual principles, but its stated scope excludes surgical and device interventions, so it should not be presented as a directly applicable standard for cardiovascular trials.\cite{m2,m4,m6,m59,m60}

Future cardiovascular procedural trials should combine credible masking with explicit questions about durability, patient selection and delivery in routine care. Placebo-controlled efficacy studies and pragmatic comparative-effectiveness trials complement each other, together showing whether a defined component helps, under which conditions, and whether offering the complete strategy justifies its burdens. Sham control can bring greater causal clarity, but only through disciplined design, conduct and interpretation throughout the trial.

\section*{Declarations}
\addcontentsline{toc}{section}{Declarations}
\textbf{Funding.} This work received no specific funding.

\textbf{Competing interests.} The author declares no competing interests.

\textbf{Use of generative AI.} Generative AI tools were used for language editing and to assist in preparing draft text. The author reviewed and revised all content and takes full responsibility for the manuscript.

\FloatBarrier
\renewcommand{\refname}{References}

\clearpage
\appendix
\section*{Supplementary Appendix}
\addcontentsline{toc}{section}{Supplementary Appendix}
\emph{4 methodological notes \textbar{} 5 supplementary tables \textbar{} 8 supplementary figures. Supplementary references are numbered S1, S2, \ldots}

\subsection*{Scope, evidence identification and source appraisal}
\addcontentsline{toc}{subsection}{Scope, evidence identification and source appraisal}
This appendix supports a narrative methodological review, with a clinical evidence inventory and a focused documentary appraisal of selected cardiovascular trials. The evidence cutoff was 18 September 2026. The purpose was to explain what different procedural controls identify, how design and analysis influence interpretation, and how controlled evidence changed earlier beliefs. The inventory is deliberately broad but is not a registered systematic review, a quantitative meta-analysis or a complete census of every small cardiovascular device experiment. No pooled effect or formal GRADE rating was constructed.

The corpus was supplemented by targeted searches and primary-publication verification using PubMed/Europe PMC and publisher records. Searches combined procedural-control terms (``sham'', ``placebo procedure'', ``placebo-controlled'', ``device on/off'' and ``randomized'') with coronary intervention, renal denervation, atrial fibrillation ablation, pacing, heart failure, interatrial shunt, pulmonary artery denervation, conditioning, cell therapy and gene therapy. Trial names and DOIs were used to resolve reports and updates. Reference lists were used to locate earlier evidence and later definitive comparisons. Current methodological documents were checked at the issuing organizations. This was targeted verification and reference chaining, not a reproducible database search claiming exhaustive retrieval.

For each selected trial, extraction distinguished the randomized population from implanted, treated and analyzed subsets; the control components; the primary outcome and its timing; the direction and uncertainty of the between-group effect; rescue and masking; and the earlier evidence being revised. Protocols, statistical plans and final reports were compared where supplied. Documentary observations were classified as legitimate design tradeoffs, interpretative limitations, unresolved reporting questions or demonstrable discrepancies. They were not assumed to imply misconduct. Citation metadata were checked separately from the clinical interpretation. A primary publication, subgroup report and extended follow-up from one cohort remained one randomization.

\subsection*{Supplementary Note 1. Components, estimands and assumptions}
\addcontentsline{toc}{subsection}{Supplementary Note 1. Components, estimands and assumptions}
The estimand should be written in clinical language before a statistical model is selected. For example: among symptomatic patients meeting the trial's eligibility criteria, what is the difference in six-month health status between assignment to an ablation pathway and assignment to a specified control pathway, with cardioversion and medication rescue available under the same decision rules? This statement identifies a strategy comparison. It differs from the effect of technically successful ablation, an effect with all antiarrhythmic drugs withheld, or a hypothetical effect in the absence of rescue. The ICH E9(R1) framework provides the structure for separating these questions.\cite{s1}

In potential-outcome notation, a randomized contrast can be represented as E{[}Y(A,t){]}\ensuremath{-}E{[}Y(S,t){]}, where A and S denote complete assigned pathways. The control response, E{[}Y(S,t)\ensuremath{-}Y(0){]}, uses a baseline measurement instead of the counterfactual outcome without that pathway. The latter is not a causal placebo-effect estimate. Nor does the randomized contrast necessarily isolate a purely biological direct effect: differential downstream rescue may be part of the intervention's consequences, and contextual effects may interact with target engagement. These are identification issues, not merely adjustments that can be repaired by choosing a more elaborate regression model.\cite{s2,s3}

A three-arm design adds information when both component efficacy and the complete-strategy effect are clinically important. Active versus procedural control estimates one contrast; active versus usual care estimates another. Procedural control versus usual care can quantify the effect of offering the control bundle but still combines attention, expectations, procedural exposure and other care. Lack of masking in the usual-care arm is part of that strategy contrast and can also affect outcome acquisition. Additional masking of assessors and equal surveillance remain useful. Supplementary Figure S1 illustrates how identical incremental effects can coexist with different complete-strategy benefits.

Intercurrent-event strategies should be outcome-specific. Rescue medication can be included in a treatment-policy outcome, incorporated into an ordinal symptom-and-treatment score, or excluded hypothetically. Those choices should reflect the clinical question rather than which analysis yields the largest effect. Treatment discontinuation, nonadherence and failed procedures likewise require specification. Keeping participants in their randomized groups is necessary for an assigned-strategy analysis, but the ``intention-to-treat'' label alone does not identify the endpoint or recover missing measurements. If rescue changes observation time, the reported outcome must make that change explicit.

For a bounded questionnaire, a simple average-change contrast can be affected by baseline headroom, while a rank-based analysis asks about the relative ordering of outcomes. A Hodges--Lehmann shift estimate is most straightforward under a common-shape location-shift interpretation; distributions with differing shape or mass at boundaries warrant additional description. An adjusted mean difference, proportional-odds ratio and responder probability are not numerical translations of the same statistic. Presenting distributions, clinically interpretable probabilities and component outcomes can reveal features hidden by one summary. Individual responder thresholds should be justified separately from a population-average superiority margin.

Continuous baseline adjustment should preserve information where appropriate rather than divide severity into arbitrary categories. Prespecified flexible functions can capture nonlinearity, subject to sample size and model stability. A baseline-by-treatment interaction can test effect modification, but selection on a baseline extreme also produces regression to the mean in both arms. An association between a trial's mean baseline score and its treatment effect is ecological: it cannot establish the benefit expected for an individual patient with that score. Analyses of postrandomization AF burden, achieved pressure or allocation belief require a distinct causal interpretation because those variables can be affected by treatment.\cite{s4,s5}

For repeated measurements, the analysis should identify whether its target is an endpoint state, average burden over a period, cumulative symptom-free time, time to sustained improvement or a conditional transition. Adequate modeling of dependence prevents a dense diary from being mistaken for a much larger randomized sample. The number and timing of observed records, delayed entry, floor and ceiling masses, and influential participants should be described. If a model produces marginal probabilities from conditional transitions, the integration or prediction procedure is part of the analysis specification and should be reproducible.

Missing-data planning begins with preventing unnecessary missingness and continuing the outcome measurements needed for the estimand. Imputation models should be compatible with assignment, baseline prognosis, outcome distribution and the analysis model. Multiple imputation under a missing-at-random assumption should be supplemented, where materially relevant, by analyses allowing worse or better unobserved outcomes than predicted. Such exercises should be anchored in plausible clinical scenarios rather than an arbitrary statistical threshold. A death-related missing questionnaire cannot be treated as merely an administrative omission without changing the scientific question.\cite{s6}

Sensitivity and supplementary analyses serve different purposes. Repeating the primary analysis under alternative plausible assumptions about missing outcomes can probe robustness of the same estimand. Replacing a rank outcome with a conditional mean, changing the handling of death or using a different rescue strategy may instead estimate a related but different quantity. Reporting both is useful; choosing the significant result as the definitive one is not. The PVI-SHAM-AF comparison in main-text Figure 5 illustrates why the distinction should be made before interpreting P values.

Adaptive designs require an explicit schedule, information definition, prior where relevant, stopping rule, decision threshold and handling of enrollment already in progress. A trial with six-month endpoints can meet a stopping boundary after nearly all planned patients have already enrolled. Operating-characteristic simulations can evaluate false-positive decisions, power, expected sample size and estimation bias under plausible null and alternative scenarios. Bayesian posterior probabilities remain conditional on the model and prior; optional monitoring does not remove the need to justify the trial's decision rule. Early stopping also limits precision for safety, durability and subgroup effects.

\subsection*{Supplementary Note 2. Recruitment, masking and operational safeguards}
\addcontentsline{toc}{subsection}{Supplementary Note 2. Recruitment, masking and operational safeguards}
Patient participation should begin with the question that matters to patients, the burden of the comparator and the acceptability of rescue and follow-up. Consent should describe the actual steps in each arm and the uncertainty being resolved. Exaggerated emphasis on the active procedure or dismissive descriptions of control can affect both recruitment and expectations. ASPIRE's ethical framework supports transparent, proportionate consent; its later guidance clarifies that rigorous consent is owed in all trials rather than suggesting that ordinary randomized research may use a lower ethical standard.\cite{s7,s8}

A control-component matrix should be agreed before recruitment. For each step, the team should record the therapeutic rationale, potential cue to allocation, incremental risk, choice to retain or simulate it, and method of fidelity assessment. The aim is sufficient credibility for the estimand, not maximum invasiveness. CoPPS usefully emphasizes component and contextual reporting, but explicitly excludes surgery and device trials from its formal scope. Its principles are used here by analogy, alongside the directly relevant ASPIRE and TIDieR-Placebo guidance.\cite{s2,s3,s9}

Role-based information access should extend beyond the catheterization laboratory. An apparently masked patient can receive an unmasked procedural letter; a follow-up clinician can infer assignment from imaging or medication instructions; an assessor can see a device interrogation. The protocol should specify which information can be withheld safely, who retains emergency access and how accidental disclosures are logged. Necessary unmasking for clinical care should be documented without automatically discarding subsequent patient outcomes. Treatment guesses, adverse effects and procedural clues should be described separately from formally authorized unmasking.

Masking questionnaires should state the exact question, response options and timing relative to treatment, symptom assessment and discussion with staff. A belief that active treatment was received is not the same construct as confidence in improvement. A ``do not know'' response can capture uncertainty, while a subsequent forced guess can support methods that require binary answers. The preferred approach is disputed and should be specified rather than retrospectively selected. With unequal allocation, naive 50\% accuracy is not an adequate universal benchmark. Raw counts by arm permit alternative indices and reveal the effect of uncertain responses.\cite{s10,s11}

The original three-response Bang index for an arm can be written as (n\_correct\ensuremath{-}n\_incorrect)/N, including uncertain respondents in N. Calculating the same numerator over only definite guessers yields a different conditional summary. For example, 11 correct, 2 incorrect and 17 uncertain responses give 0.30 with N=30, but 0.69 among the 13 definite guessers. Both arithmetic results can be described, but they are not interchangeable. PFA-SHAM's table and explanatory footnote appear to use different denominators; clarification is needed before comparing its printed indices with other AF trials.\cite{s10,s12}

Early masking assessment is informative about procedural credibility, whereas late assessment increasingly reflects the clinical course. An efficacious intervention may lead participants to guess correctly because they improve. Conversely, early uncertainty cannot rule out later disclosure. Neither observation alone identifies the amount or direction of bias. Analyses conditioned on late guesses can introduce postrandomization selection; the assigned-group comparison should remain primary. Supplementary Figure S2 shows these time-ordered relationships without assuming that every pathway occurs in every trial.

Quality assurance should cover delivery of the active component, delivery of the control, follow-up fidelity and outcome acquisition. A monitoring dashboard may track technical success, protocol deviations, medication changes, completion and timing of patient reports, unintended unmasking and serious events by randomized assignment. The independent monitoring committee requires adequate access to safety information while the investigators maintain masking. Prespecified corrective actions for recruitment or delivery problems are preferable to waiting until an unexpected efficacy result triggers retrospective explanations.

\subsection*{Supplementary Note 3. Interpreting and auditing the ORBITA trials}
\addcontentsline{toc}{subsection}{Supplementary Note 3. Interpreting and auditing the ORBITA trials}
ORBITA and ORBITA-2 answer related but different questions. ORBITA tested PCI added to intensive antianginal treatment; ORBITA-2 tested PCI with regular antianginals withdrawn or minimized, while allowing blinded rescue. In ORBITA, the primary exercise-time contrast was 16.6 seconds (95\% CI \ensuremath{-}8.9 to 42.0; P=0.20). This superiority result did not establish equivalence: its interval includes no benefit and appreciable benefit. It also does not demonstrate that PCI has no symptomatic effect under other medication conditions. Conversely, ORBITA-2 cannot show that withdrawing medication improves PCI efficacy, because background medication was not randomized between the two trials. Their comparison changes population, treatment context, outcome and follow-up (Supplementary Figure S8). \cite{s13,s14}

ORBITA-2 randomized 301 participants with stable symptoms, obstructive coronary disease and demonstrable ischemia. The causal contrast was assignment to PCI versus an invasive placebo procedure within a shared treatment pathway. Secondary prevention continued, and antianginal drugs required for other indications could continue. Patient-triggered contact prompted rescue prescribing by blinded staff according to an individualized protocol incorporating symptoms and preference. Consequently, ``off medication'' identifies the intended treatment context, not universal absence of medication throughout follow-up. This is neither PCI versus optimized medical therapy nor an estimate assuming rescue never occurs. \cite{s14}

The primary outcome integrates symptoms, prescribed antianginal units and specified clinical events (Supplementary Figure S4). Ordinary grades rise first with medication category and then with episodes: no symptoms on one unit scores 7, whereas six or more attacks without medication scores 6. This ordering prevents successful pharmacological rescue from hiding continued treatment requirements, but gives medication burden priority over observed symptoms. It is an ordinal preference structure, not a cardinal utility scale. One medication unit is not numerically equivalent to seven attacks, and death is not 79 times one attack. The consultation supporting the ordering does not establish equal intervals or a universal minimal important difference. \cite{s14}

Intolerable-angina unblinding, acute coronary syndrome and death occupy grades 77--79. Rescue medication is therefore incorporated through a composite strategy, rather than simply censored. Hospitalization alone, stroke and bleeding are not overriding primary-score categories, so the score is not a complete net-clinical-benefit measure. Periprocedural infarctions did contribute, despite excluding routine diary days 0--1 from the longitudinal analysis. Four ACS events occurred with PCI and six with placebo, with no deaths; these small counts cannot establish prognostic superiority or safety equivalence. PRN glyceryl trinitrate was excluded from the primary score. \cite{s14}

The main publication used a baseline-adjusted frequentist ordinal analysis at the Journal's request, reporting OR 2.21 (95\% CI 1.41--3.47). The SAP, completed 2.2 years before database lock, specified Bayesian longitudinal modeling, reported in the appendix. Its day-84 OR was 1.88 (95\% credible interval 1.57--2.27). The former summarizes the final outcome distribution; the latter concerns conditional daily transitions. Different estimands and conditioning sets explain why their numerical values need not coincide. The publication transparently describes the change; characterizing it as concealed outcome switching would be inaccurate. Model-predicted scores of 2.9 and 5.6 are neither raw group means nor counts of attacks. \cite{s14}

The longitudinal analysis uses a first-order Markov ordinal model, conditioning each day on the preceding state, baseline mean score, treatment and time. Nonlinear terms and treatment--time interaction permit changing effects. Dependence is recognised when transition probabilities become state-occupancy estimates. Patient-days are therefore not simply treated as independently randomized participants. Nevertheless, first-order conditional dependence, link and constraint assumptions require assessment. The final appendix relaxes proportional odds for treatment as well as time; the literal SAP mentions time alone. This documented refinement merits an explicit rationale and timing statement, without assuming that it was selected to manufacture benefit. \cite{s14}

A null-centred treatment prior assigned 5\% probability below OR 0.25 and 5\% above OR 4. A near-flat-prior sensitivity analysis produced a similar day-84 estimate, OR 1.86 (95\% credible interval 1.55--2.25). This supports robustness to that prior choice within the model. It does not test every dependence structure, missingness mechanism or outcome preference. A posterior probability exceeding 99.9\% means high probability of benefit conditional on the model, prior and data; it is not the probability of important benefit for every patient. The specified interim reviews considered safety, not efficacy. \cite{s14}

Daily reporting was 99.7\% complete, supported by reminders and readily available staff. Two placebo participants required last-observation carry-forward for the main final-score analysis. The SAP separately specified coarsened states for missing diary components and continued ACS/death ascertainment after unblinding. These rules should not be conflated. The extracted model output does not fully reconstruct every event-state or coarsening implementation. Near-complete ordinary diaries limit, but do not abolish, missing-data assumptions. COVID-related missing exercise and echocardiography assessments were examined with multiple imputation, with concordant results. Diary completion also does not verify ingestion of prescribed drugs or equal physical activity. \cite{s14,s15}

Masking incorporated deep sedation, auditory isolation, standardized documentation and separation of unblinded procedural personnel from subsequent care. Discharge allocation guesses supported concealment. At twelve weeks, patients guessed more accurately, while staff indices remained near zero. Late guessing may result from genuine symptom improvement as well as disclosure; it cannot by itself identify the mechanism or quantify bias. Conversely, successful discharge masking cannot prove that expectations remained identical throughout follow-up. Reporting the timing of guesses, who supplied them and whether they preceded outcome discussions is more informative than declaring blinding simply successful or failed. \cite{s14}

The original protocol described 90\% power and 342 randomized participants; the final protocol and SAP described 80\% power and 284 (Supplementary Figure S6C). Actual randomization of 301 exceeded the revised target. The planned approximately 400 participants referred to enrollment, not the required randomized cohort. Planning used a mean-difference approximation rather than simulation of the final ordinal Markov model. Future reporting should explain this revision, distinguish planned and implemented rescue/washout details, and publish executable data-preparation rules. These are transparency improvements, not demonstrated reasons to disregard the positive result. \cite{s14}

Secondary outcomes and later interaction analyses require restraint about multiplicity and independence. Physiology, stress echocardiography and symptom substudies reuse the same randomization, with different available-data subsets. An interaction OR compares treatment effects across predictor values; it is not the overall PCI effect or a validated treatment-selection threshold. A stronger effect with typical angina does not prove zero benefit without it. Collectively, these analyses refine biological interpretation but do not replicate the trial independently or test the proposed diagnostic pathways. ORBITA-2 establishes twelve-week symptomatic efficacy in its treatment context, while comparative sequencing, long-term prognosis and broader applicability remain separate questions. \cite{s14,s16,s17,s18}

\subsection*{Supplementary Note 4. Interpreting the atrial fibrillation sham trials}
\addcontentsline{toc}{subsection}{Supplementary Note 4. Interpreting the atrial fibrillation sham trials}
The four randomized AF ablation studies address related but non-identical questions. ORBITA-AF evaluated feasibility; SHAM-PVI tested continuously measured AF burden; PFA-SHAM tested arrhythmia recurrence and disease-specific quality of life as co-primary outcomes; PVI-SHAM-AF prioritized the AFEQT summary score. Their differences cannot be reduced to positive versus negative verdicts on ablation. Each estimates the consequences of adding a specified ablation strategy to its chosen comparator, in a selected population, over a defined observation period. None directly measures the psychological placebo effect or determines what proportion of improvement after routine ablation is attributable to expectation.\cite{s12,s19,s20,s21}

The control pathways included active clinical care. Cardioversion was performed when required, medications could change, and structured follow-up accompanied procedural simulation. PFA-SHAM randomized participants after a common electrophysiological study; its comparator therefore included diagnostic catheterization, not sedation alone. PVI-SHAM-AF used deep sedation and femoral sheaths without intracardiac catheters, whereas ORBITA-AF and SHAM-PVI included phrenic pacing. All avoided sham transseptal puncture. These differences specify the contrast rather than automatically compromise it. Reproducing a hazardous procedural step requires evidence that its contribution to credibility or identification justifies the additional risk. Successful initial masking with limited instrumentation weighs against treating anatomical duplication as an unconditional requirement.

Population and measurement also constrain comparison. SHAM-PVI predominantly enrolled persistent AF, whereas 77\% of PFA-SHAM participants had paroxysmal AF and screening required AFEQT \ensuremath{\le}50. PVI-SHAM-AF imposed no symptom-score threshold and included approximately equal proportions of paroxysmal and persistent AF. Continuous implantable monitoring in ORBITA-AF, SHAM-PVI and PFA-SHAM cannot be equated with a seven-day Holter at six months. Conversely, detecting a brief recurrence is not equivalent to substantial time in AF or impaired quality of life. These are complementary outcomes; their differing sensitivity and clinical meaning should inform interpretation rather than an automatic hierarchy favoring whichever endpoint is positive.

PVI-SHAM-AF illustrates the importance of preserving the primary statistical question. Its confirmatory rank-based analysis estimated a Hodges--Lehmann AFEQT change contrast of 2.6 points (95\% CI \ensuremath{-}2.7 to 8.0; P=.36). A prespecified mixed-model sensitivity analysis, adjusted for sex and trial site, estimated 5.4 points (1.0--9.8). The latter does not replace the primary result, but neither should it disappear from an account of robustness. The analyses summarize treatment effects differently and use different modeling assumptions. The defensible conclusion is failure to demonstrate superiority on the prespecified primary analysis, with uncertainty encompassing modest benefit; equivalence was not established.\cite{s21}

Death further changes the interpretation of those analyses. The primary analysis assigned the worst possible AFEQT change, \ensuremath{-}100, to patients who died, whereas the mixed model treated death like other missing observations, implicitly addressing quality of life had the patient survived. One patient died in each group, but balanced counts do not make these estimands equivalent. Paired questionnaires were available for 241 of 262 randomized participants; multiple imputation retained the full randomized population. Reporting the imputation model and sensitivity to informative missingness matters more than describing an analysis as intention-to-treat without qualification.

The rhythm findings require a similar hierarchy. Clinically detected recurrence in PVI-SHAM-AF was inconclusive (HR 0.66; 95\% CI 0.35--1.24), while Holter-detected AF and burden favored ablation. The often cited 27\% versus 48\% recurrence proportions came from a post-hoc composite of clinical recurrence or Holter detection; patients lacking analyzable Holters were counted as having no detected AF. These proportions should not be called continuously verified AF freedom. Index cardioversion occurred in 90 of 168 treated ablation patients and 32 of 89 sham patients, distinguishing the comparator from untreated natural history.\cite{s21}

PVI-SHAM-AF also documents late amendments without evidence that its clinical primary endpoint changed (Supplementary Figure S6A). The final protocol amendment, with additional secondary outcomes and subgroups, was submitted to the ethics committee on April 1, 2026. The SAP was signed May 15, database lock occurred June 1, the amendment was approved June 2 and the writing team was unmasked June 8. These pre-lock specifications should be distinguished from explicitly post-hoc analyses. Approval timing and imperfect reconciliation of subgroup definitions warrant transparent documentation, but do not independently demonstrate outcome-driven revision. The primary endpoint and rank-based method were already described in the design publication.\cite{s22}

PFA-SHAM provides a different example of adaptation (Supplementary Figure S6B). Its initial frequentist design was changed after enrollment began, following SHAM-PVI\textquotesingle s publication. The supplement dates SAP finalization to October 21, 2024, before the first documented interim analysis. Both Bayesian stopping criteria were required: evidence for arrhythmia freedom and for AFEQT improvement. They were met at 50 evaluable patients, when 60 had already been randomized. This shows how delayed endpoints limit the recruitment savings from early stopping. Bayesian adaptation can support proportionate enrollment, but posterior updating alone does not establish a desired frequentist false-positive rate; the full decision rule requires operating-characteristic evaluation.\cite{s12,s23}

Rescue is particularly important to PFA-SHAM\textquotesingle s patient-reported endpoint (Supplementary Figure S5). Five sham participants crossed over, one at day 11 because symptomatic AF required hospitalization and four between days 131 and 157. For early exits, pre-rescue AFEQT and HADS assessments replaced six-month measurements. This prespecified approach limits contamination by subsequent ablation and protects symptomatic participants, but estimates change to six months or earlier rescue, rather than six-month quality of life under an assigned strategy including rescue. Complete follow-up does not eliminate this distinction. Excluding the day-11 crossover in a per-protocol analysis preserved the direction of benefit without making the estimands interchangeable.\cite{s12,s23}

PFA-SHAM\textquotesingle s efficacy conclusion was robust, while its numerical magnitude was analysis-dependent. The primary AFEQT contrast was 32.6 points (95\% credible interval 20.2--44.9); using screening as baseline produced 26.0 points (15.2--36.8), and baseline-adjusted ANCOVA produced approximately 19.9 points (9.7--29.8). Improvement before randomization, especially in controls, helps explain why baseline definition matters. All estimates favored ablation. However, their shared direction should not obscure their clinically material numerical differences, or justify comparing the largest estimate directly with another trial\textquotesingle s differently defined primary contrast.\cite{s12}

Masking results require comparable denominators and timing (Supplementary Figure S3). PFA-SHAM\textquotesingle s tabulated Bang indices correspond to the difference between correct and incorrect guesses among definite guessers, excluding uncertain respondents. The original three-response Bang index includes uncertain respondents in the denominator; PFA-SHAM\textquotesingle s explanatory footnote then reports values consistent with that latter calculation. Raw response counts allow this inconsistency to be recognized without assuming defective conduct.\cite{s10,s12} SHAM-PVI assessed masking at discharge, three months and six months: patients\textquotesingle{} guesses became more informative while staff generally remained uncertain. Clinical response can cause correct guesses, but absent reported reasons and confidence data, that explanation cannot exclude information leakage. Early uncertainty alone does not certify sustained masking.

SHAM-PVI\textquotesingle s principal result also needs correct scale interpretation. Its estimate of 0.25 (95\% CI 0.15--0.42) is an adjusted geometric-mean burden ratio, despite ``difference'' terminology. The 60.31 versus 35.0 percentage-point reductions describe within-group mean changes. AFEQT\textquotesingle s 18.39-point advantage was concordant secondary evidence, with multiplicity limitations. More controls resumed antiarrhythmics, consistent with responsive clinical care and a treatment-policy comparison; adjusting naively for this postrandomization exposure would not isolate PVI\textquotesingle s direct effect. The evidence supports short-term incremental efficacy, not the stronger claim that no clinically relevant placebo effect exists.\cite{s20,s24}

ORBITA-AF\textquotesingle s contribution is narrower. Its 20-patient feasibility objective and imprecise masking estimates should not be recast as confirmatory efficacy evidence. Reported late quality-of-life values among participants maintaining sinus rhythm condition on a postrandomization outcome and cannot preserve an intention-to-treat comparison. Unreconciled recurrence totals and differences between report and participant-information timings warrant clarification, with no inference of misconduct.\cite{s19} Across these studies, the useful synthesis is therefore conditional: ablation can improve physiology and patient experience beyond an active control pathway, while the size and durability of that benefit depend on population, background treatment and the outcome actually estimated. Source omissions, reporting inconsistencies, legitimate design compromises and demonstrated analytical limitations deserve distinct judgments.

\subsection*{Reading the expanded trial inventory}
\addcontentsline{toc}{subsection}{Reading the expanded trial inventory}
Table S1 groups related reports when they share a randomization or a sequential cohort. It distinguishes surgical or catheter-based sham procedures, inactive device programming, simulated noninvasive treatment and matched injection/vehicle controls. Placebo drug trials without a procedural or device component are outside this review. Some boundary studies are listed separately because masking device-derived information or comparing different active drug pathways is not the same as a sham procedure. N refers to randomized participants wherever this can be determined; implantation, treatment and analysis subsets are identified when materially different.

``Before'' describes the evidence preceding the highlighted comparison, not a formal causal explanation for why its results differed. A subsequent neutral trial may change the population, technology, expertise, endpoint and sample size at the same time as masking. Earlier evidence was sometimes already randomized and masked. It is therefore inaccurate to attribute every revision of knowledge to sham alone. Likewise, post-hoc phenotypes, extended follow-up and secondary papers do not increase the number of independent randomized experiments.

Tables S2--S4 provide focused AF and ORBITA comparisons and documentary audits. Table S5 is an author-developed operational aid, not a validated risk-of-bias instrument or a consensus endorsement. Findings categorized as questions requiring clarification should not be treated as proven protocol violations. Original protocols, amendments, statistical plans, final publications and accessible executable analyses remain the appropriate basis for a definitive assessment.

\subsection*{\textbf{Table S1.} Coronary interventions and angina}
CTO PCI: Before ORBITA-CTO, six independent open-label RCTs addressed differing symptom, imaging and clinical-event questions: DECISION-CTO, EuroCTO, EXPLORE, COMET-CTO, REVASC and IMPACTOR-CTO. Symptom evidence was mixed and prognosis unestablished; the prior evidence was not merely observational. Coronary sinus reducer: Anatomical rationale and uncontrolled refractory-angina series suggested benefit. COSIRA supplied early positive sham evidence; subsequent mechanistic testing therefore followed an already sham-supported symptom signal. External counterpulsation: Early physiological and uncontrolled clinical observations suggested ischemia and angina improvement; the specific contribution beyond a credible low-pressure control was uncertain. Internal mammary intervention: Historical uncontrolled series claimed angina relief after internal mammary artery ligation. Modern catheter occlusion studies subsequently tested collateral physiology with a different intervention; neither should be confused with coronary bypass grafting. Myocardial laser revascularization: Unblinded procedural studies reported symptom improvement in refractory angina. Masked trials were needed because symptom reporting and effort could be influenced by treatment knowledge. PCI for stable angina: Clinical experience and unblinded randomized trials, including COURAGE, supported symptom relief in selected stable coronary disease. The incremental effect beyond a procedural control, under specified antianginal treatment, remained uncertain. Therapeutic ultrasound for angina: Angiogenic and perfusion mechanisms and early clinical observations suggested that external acoustic therapies might relieve refractory angina; device, dose and outcome differences preclude treating all ultrasound/shock-wave trials as one intervention.

{\footnotesize\setlength{\LTpre}{4pt}
\begin{longtable}[]{@{}
  >{\raggedright\arraybackslash}p{(\linewidth - 6\tabcolsep) * \real{0.1186}}
  >{\raggedright\arraybackslash}p{(\linewidth - 6\tabcolsep) * \real{0.3093}}
  >{\raggedright\arraybackslash}p{(\linewidth - 6\tabcolsep) * \real{0.2268}}
  >{\raggedright\arraybackslash}p{(\linewidth - 6\tabcolsep) * \real{0.3454}}@{}}
\toprule\noalign{}
\begin{minipage}[b]{\linewidth}\raggedright
\textbf{Trial / randomized N}
\end{minipage} & \begin{minipage}[b]{\linewidth}\raggedright
\textbf{Control and primary endpoint}
\end{minipage} & \begin{minipage}[b]{\linewidth}\raggedright
\textbf{Result}
\end{minipage} & \begin{minipage}[b]{\linewidth}\raggedright
\textbf{Knowledge contribution and limitations}
\end{minipage} \\
\midrule\noalign{}
\endhead
\bottomrule\noalign{}
\endlastfoot
ORBITA-CTO (2026)

N=50 & Procedural sham. CTO PCI versus angiographic placebo procedure in symptomatic single-vessel CTO without bystander disease.

Primary: Daily symptom/medication/event ordinal angina score over six months. & OR for a better score 4.38 (95\% CrI 1.57--12.69), Pr(benefit)=.996; +30.6 angina-free days (11.1--50.7).\cite{s25} & Added direct masked evidence of symptom benefit to previously mixed open-label CTO trials. Pilot-sized selected population and expert operators; cannot establish survival or uncommon-harm effects.

Denominator: 50 randomized, 25/25; all 50 in the primary analysis despite one procedural-complication withdrawal. \\
COSIRA (2015)

N=104 & Procedural sham. Coronary sinus reducer implantation versus sham catheter procedure.

Primary: Improvement by at least two CCS classes at six months. & 35\% versus 15\%; P=.02.\cite{s26} & Strengthened the claim of symptom efficacy beyond a matched procedure. Exercise/perfusion and prognosis were not established by the positive symptom endpoint.

~ \\
ORBITA-COSMIC (2024)

N=51 & Procedural sham. Reducer implantation versus sham catheter procedure.

Primary: Stress myocardial blood flow; designated primary symptom outcome: daily angina. & Stress-flow difference +0.06 mL/min/g (95\% CrI \ensuremath{-}0.09 to 0.20); less-angina OR 1.40 (1.08--1.83).\cite{s27} & Supported symptom efficacy while failing to confirm the proposed transmural-perfusion mechanism. Small trial with discordant perfusion and symptom outcomes; symptoms do not validate the hypothesized mechanism.

Denominator: 51 randomized (25 reducer, 26 control); 50 included in the reported primary analysis (24/26). \\
MUST-EECP (1999)

N=139 & Noninvasive sham procedure. 35 hours of active versus low-pressure sham counterpulsation.

Primary: Exercise-related efficacy, including exercise duration and time to ST depression. & Time to ST depression improved (P=.01); total exercise duration was not superior.\cite{s28} & Provided a partial ischemia signal rather than universal confirmation of functional benefit. Multiple efficacy measures, limited duration; no survival inference.

~ \\
Cobb internal mammary artery ligation (1959)

N=17 & Procedural sham. Artery ligation versus skin-incision sham operation.

Primary: Angina response and exercise evaluation in a historical trial. & No convincing symptom benefit beyond sham was demonstrated.\cite{s29} & Challenged uncontrolled surgical-success claims. Very small historical study; not evidence about modern CABG or catheter occlusion.

~ \\
CLIMACCS (2024)

N=101 & Procedural sham. Permanent internal mammary artery device occlusion versus sham.

Primary: Change in exercise duration at six weeks. & +23 versus +4 s; P=.44; collateral-flow physiology improved.\cite{s30} & A physiological improvement did not establish functional benefit. Different intervention from historical ligation; limited precision and duration.

~ \\
BELIEF (2004)

N=82 & Procedural sham. Percutaneous myocardial laser plus medical therapy versus sham plus medical therapy.

Primary: Improvement by at least two CCS angina classes at 12 months. & 35\% versus 14\% improved; P=.04. Exercise time and medication use were similar.\cite{s31} & A small positive masked symptom trial; the laser evidence was not uniformly neutral. Small sample; symptoms and objective exercise findings were not concordant.

~ \\
DIRECT (2005)

N=298 & Procedural sham. Low-dose or high-dose laser versus catheter sham without myocardial channels.

Primary: Change in exercise duration at six months. & Sham +28.0 s, low dose +33.2 s, high dose +28.0 s; P=.94.\cite{s32} & Failed to confirm the benefit suggested by open laser studies. Dose groups and procedural harms matter; a 30-day MI signal accompanied laser treatment.

~ \\
ORBITA (2018)

N=200 & Procedural sham. PCI versus placebo procedure after intensive antianginal optimization.

Primary: Change in treadmill exercise duration at six weeks. & Between-group difference +16.6 s (95\% CI \ensuremath{-}8.9 to 42.0); P=.200.\cite{s13} & Did not demonstrate incremental exercise benefit under optimized medication; narrowed an overbroad efficacy claim. Short duration, selected single-vessel disease, uncertain smaller effects; not an equivalence trial or proof that all PCI is ineffective.

~ \\
ORBITA-2 (2023)

N=301 & Procedural sham. PCI versus placebo procedure with antianginals withdrawn or minimal at baseline.

Primary: Daily ordinal angina symptom score over 12 weeks, incorporating medication and clinical events. & Mean scores 2.9 versus 5.6; odds ratio for a better score 2.21 (95\% CI 1.41--3.47), P\textless.001.\cite{s14} & Confirmed symptomatic efficacy in a different medication context from ORBITA. Selected symptomatic population; the contrast does not establish prognostic benefit or a pure placebo fraction.

~ \\
Shindo LIPUS pilot (2023)

N=56 & Noninvasive sham procedure. Low-intensity pulsed ultrasound versus matched application without therapeutic energy.

Primary: Change in weekly nitroglycerin use at three months. & \ensuremath{-}3.06 versus \ensuremath{-}3.10 uses/week; no superiority.\cite{s33} & The angiogenic rationale did not yield a positive primary clinical result. Small pilot; do not combine indiscriminately with shock-wave platforms.

~ \\
\end{longtable}
}

\subsection*{\textbf{Table S1.} Hypertension procedures and devices}
Baroreflex activation for resistant hypertension: Uncontrolled baroreflex-activation studies reported large blood-pressure falls and motivated a pivotal blinded test. Renal denervation: Initial uncontrolled series and unblinded SYMPLICITY HTN-2 suggested large office-BP reductions. After HTN-3, the prior evidence included both neutral and positive sham trials; device generation, renal-artery coverage, background drugs and adherence assessment changed across programs.

{\footnotesize\setlength{\LTpre}{4pt}
\begin{longtable}[]{@{}
  >{\raggedright\arraybackslash}p{(\linewidth - 6\tabcolsep) * \real{0.1186}}
  >{\raggedright\arraybackslash}p{(\linewidth - 6\tabcolsep) * \real{0.3093}}
  >{\raggedright\arraybackslash}p{(\linewidth - 6\tabcolsep) * \real{0.2268}}
  >{\raggedright\arraybackslash}p{(\linewidth - 6\tabcolsep) * \real{0.3454}}@{}}
\toprule\noalign{}
\begin{minipage}[b]{\linewidth}\raggedright
\textbf{Trial / randomized N}
\end{minipage} & \begin{minipage}[b]{\linewidth}\raggedright
\textbf{Control and primary endpoint}
\end{minipage} & \begin{minipage}[b]{\linewidth}\raggedright
\textbf{Result}
\end{minipage} & \begin{minipage}[b]{\linewidth}\raggedright
\textbf{Knowledge contribution and limitations}
\end{minipage} \\
\midrule\noalign{}
\endhead
\bottomrule\noalign{}
\endlastfoot
Rheos Pivotal (2011)

N=265 & Implanted device delayed activation. Immediate baroreflex activation versus 6 months OFF followed by activation; 2:1 randomization, double blind.

Primary: Five co-primary objectives: acute systolic-pressure response, sustained response, procedural safety, stimulation safety and device safety. & Sustained response, stimulation safety and device safety criteria were met; acute-response and procedural-safety criteria were not. A prespecified ancillary analysis found SBP \ensuremath{\le}140 mm Hg in 42\% versus 24\% (P=.005).\cite{s34} & Rheos converted large uncontrolled efficacy expectations into mixed pivotal efficacy and safety findings. The ancillary blood-pressure result must not replace failed co-primary criteria. Shared implantation and postimplant randomization limit the surgical-risk contrast.

Denominator: 265 implanted participants randomized 1 month after implantation; delayed activation was the control during the initial 6 months. \\
SYMPLICITY HTN-3 (2014)

N=535 & Procedural sham. Renal angiography sham without denervation versus first-generation single-electrode RF denervation.

Primary: Office systolic BP at six months. & Difference \ensuremath{-}2.39 mmHg (95\% CI \ensuremath{-}6.89 to 2.12); ambulatory difference \ensuremath{-}1.96 (\ensuremath{-}4.97 to 1.06); superiority not demonstrated.\cite{s35} & Did not reproduce the large office-BP effect from earlier unblinded evidence. Specific device/protocol; medication changes, denervation completeness and operator issues do not erase the neutral result or prove all future RDN ineffective.

~ \\
ReSET (2016)

N=69 & Procedural sham. Sedation, angiography and dummy imaging with catheter in place but no ablation.

Primary: Daytime ambulatory systolic BP at three months. & \ensuremath{-}6.2 versus \ensuremath{-}6.0 mmHg; no superiority.\cite{s36} & Experienced-operator neutrality weakened an explanation based only on HTN-3 operator inexperience. Single-center small study; cannot exclude a modest effect.

~ \\
RADIANCE-HTN SOLO (2018)

N=146 & Procedural sham. Endovascular ultrasound denervation versus renal angiography after drug withdrawal.

Primary: Daytime ambulatory systolic BP at two months. & Adjusted difference \ensuremath{-}6.3 mmHg (95\% CI \ensuremath{-}9.4 to \ensuremath{-}3.1), P=.0001.\cite{s37} & Confirmed short-term BP efficacy for ultrasound RDN off medication. Selected off-medication efficacy setting; not a hard-outcome trial.

~ \\
WAVE IV (2018)

N=81 & Noninvasive sham procedure. External therapeutic ultrasound versus diagnostic-level bilateral ultrasound with coded device settings.

Primary: Office systolic BP at 24 weeks; ambulatory BP was a key secondary measure. & Office change \ensuremath{-}12.8 versus \ensuremath{-}23.0 mmHg, P=.133; principal BP comparisons were not positive.\cite{s38} & Failed to confirm uncontrolled WAVE I--III optimism. Stopped early for futility; later efficacy measurements used a depleted per-protocol sample.

Denominator: 81 randomized; 2017 online/2018 print publication. \\
SPYRAL HTN-OFF MED program (2020)

N=331 & Procedural sham. Multielectrode renal RF ablation versus renal-angiography sham after medication washout.

Primary: 24-hour ambulatory systolic BP at three months. & Pilot difference \ensuremath{-}5.0 mmHg (95\% CI \ensuremath{-}9.9 to \ensuremath{-}0.2); pivotal Bayesian difference \ensuremath{-}3.9 (95\% CrI \ensuremath{-}6.2 to \ensuremath{-}1.6), Pr(superiority)\textgreater.999.\cite{s39,s40} & Established a more modest BP effect with improved technique and medication verification. Pilot/pivotal information is dependent; no clinical-event efficacy claim.

Denominator: 331 total unique randomized participants in the reported program: 80 pilot plus 251 newly randomized pivotal participants. Bayesian primary analysis borrows pilot information; do not add 80 to 331. \\
RADIANCE-HTN TRIO (2021)

N=136 & Procedural sham. Ultrasound RDN versus renal-angiography sham on a standardized triple-drug single pill.

Primary: Daytime ambulatory systolic BP at two months. & Median between-group difference \ensuremath{-}4.5 mmHg (95\% CI \ensuremath{-}8.5 to \ensuremath{-}0.3), adjusted P=.022.\cite{s41} & Demonstrated an additional effect under standardized resistant-hypertension treatment. Adherence verification was available only for sampled participants; medication policy and follow-up affect transportability.

~ \\
REQUIRE (2022)

N=143 & Procedural sham. Endovascular ultrasound RDN versus sham under antihypertensive treatment in Japan/South Korea.

Primary: 24-hour ambulatory systolic BP at three months. & Difference \ensuremath{-}0.1 mmHg (95\% CI \ensuremath{-}5.5 to 5.3), P=.971.\cite{s42} & Showed that modern ultrasound RDN was not positive in every drug-treatment context. Drug standardization/adherence and masking design may matter; primary result remains neutral.

~ \\
RADIANCE II (2023)

N=224 & Procedural sham. Ultrasound RDN versus renal-angiography sham after medication withdrawal.

Primary: Daytime ambulatory systolic BP at two months. & Adjusted difference \ensuremath{-}6.3 mmHg (95\% CI \ensuremath{-}9.3 to \ensuremath{-}3.2), P\textless.001.\cite{s43} & Replicated off-medication efficacy in a larger ultrasound trial. Short-term surrogate endpoint; cannot infer mortality benefit.

~ \\
SPYRAL HTN-ON MED program (2023)

N=337 & Procedural sham. RF denervation versus renal-angiography sham while taking one to three antihypertensives.

Primary: 24-hour ambulatory systolic BP at six months; Bayesian primary analysis in the expanded report. & Pilot adjusted difference \ensuremath{-}7.0 mmHg (95\% CI \ensuremath{-}12.0 to \ensuremath{-}2.1). Expanded primary difference \ensuremath{-}0.03 (95\% interval \ensuremath{-}2.82 to 2.77), Pr(superiority)=.51; conventional difference \ensuremath{-}1.9 (\ensuremath{-}4.4 to 0.5), P=.12.\cite{s44,s45} & An early positive masked signal was not confirmed by the larger primary analysis; secondary office-BP improvement did not rescue it. Medication intensity rose more in sham; pilot and expanded cohort overlap.

Denominator: 337 in expanded report including the 80-patient pilot; do not add 80+337. \\
TARGET BP OFF-MED (2023)

N=106 & Procedural sham. Alcohol-mediated renal denervation versus sham after medication withdrawal.

Primary: 24-hour ambulatory systolic BP at eight weeks. & \ensuremath{-}2.9 versus \ensuremath{-}1.4 mmHg; P=.27.\cite{s46} & Did not establish primary off-medication efficacy for this alcohol platform. Later medication-burden findings are secondary and do not change the neutral primary test.

~ \\
Iberis-HTN (2024)

N=217 & Procedural sham. Multielectrode RF RDN versus sham on standardized triple therapy.

Primary: 24-hour ambulatory systolic BP at six months. & Adjusted difference \ensuremath{-}9.4 mmHg (95\% CI \ensuremath{-}12.8 to \ensuremath{-}5.9), P\textless.001.\cite{s47} & Showed that some selected modern trials had larger effects than a universal 3--6 mmHg shorthand. Younger selected Chinese population, specific device and standardized treatment limit cross-trial attribution.

~ \\
Netrod-HTN (2024)

N=205 & Procedural sham. Six-electrode RF RDN versus sham on standardized nifedipine/hydrochlorothiazide.

Primary: Office systolic BP at six months. & Difference \ensuremath{-}19.0 mmHg (95\% CI \ensuremath{-}23.0 to \ensuremath{-}15.0); secondary ambulatory difference \ensuremath{-}8.7 (\ensuremath{-}12.4 to \ensuremath{-}5.0); both P\textless.001.\cite{s48} & Demonstrated a large office effect with a smaller ambulatory effect in a selected population. Primary measure was office BP; younger Chinese population and reduced analysis denominators require qualification.

Denominator: 205 randomized; outcome-specific analysis denominators were lower. \\
SMART (2024)

N=220 & Procedural sham. Mapped selective renal RF ablation versus sham, with BP-targeted drug titration in both groups.

Primary: Coprimary noninferiority of office-BP control and superiority in Drug Index change at six months. & BP control 95.4\% versus 92.8\% met noninferiority; Drug Index difference \ensuremath{-}3.25 (95\% CI \ensuremath{-}5.56 to \ensuremath{-}0.94), P=.003.\cite{s49} & Supported a strategy of similar achieved control with less additional medication. Not proof of a larger achieved BP reduction; 24-hour systolic reductions were similar.

~ \\
TARGET BP I (2024)

N=301 & Procedural sham. Alcohol-mediated RDN versus sham on two to five antihypertensives.

Primary: 24-hour ambulatory systolic BP at three months. & Difference \ensuremath{-}3.2 mmHg (95\% CI \ensuremath{-}6.3 to 0.0), P=.0487; office-BP difference was not significant.\cite{s50} & Established a modest, borderline primary ambulatory-BP effect. Do not generalize one positive domain to all BP outcomes or clinical events.

~ \\
\end{longtable}
}

\subsection*{\textbf{Table S1.} AF ablation and auricular neuromodulation}
AF catheter ablation: Substantial unblinded randomized evidence already supported rhythm control and patient-reported improvement. A sham comparison addressed incremental efficacy beyond sedation, cardioversion, monitoring and follow-up; it did not create the first randomized evidence for ablation. Auricular neuromodulation: Animal and acute human electrophysiological experiments supported autonomic modulation. Later ambulatory sham trials tested arrhythmia burden, with no presumption that an AF result transfers to ventricular ectopy.

{\footnotesize\setlength{\LTpre}{4pt}
\begin{longtable}[]{@{}
  >{\raggedright\arraybackslash}p{(\linewidth - 6\tabcolsep) * \real{0.1186}}
  >{\raggedright\arraybackslash}p{(\linewidth - 6\tabcolsep) * \real{0.3093}}
  >{\raggedright\arraybackslash}p{(\linewidth - 6\tabcolsep) * \real{0.2268}}
  >{\raggedright\arraybackslash}p{(\linewidth - 6\tabcolsep) * \real{0.3454}}@{}}
\toprule\noalign{}
\begin{minipage}[b]{\linewidth}\raggedright
\textbf{Trial / randomized N}
\end{minipage} & \begin{minipage}[b]{\linewidth}\raggedright
\textbf{Control and primary endpoint}
\end{minipage} & \begin{minipage}[b]{\linewidth}\raggedright
\textbf{Result}
\end{minipage} & \begin{minipage}[b]{\linewidth}\raggedright
\textbf{Knowledge contribution and limitations}
\end{minipage} \\
\midrule\noalign{}
\endhead
\bottomrule\noalign{}
\endlastfoot
ORBITA-AF feasibility (2024)

N=20 & Procedural sham. PVI with cardioversion if needed versus placebo procedure with real cardioversion.

Primary: Maintenance of participant masking, measured by the blinding index. & Masking proved feasible in a 20-patient pilot; not powered for efficacy.\cite{s19} & Made a credible AF-ablation sham design concrete. Feasibility is not definitive efficacy; selected sinus-rhythm survivor QoL analyses do not preserve ITT.

~ \\
SHAM-PVI (2024)

N=126 & Procedural sham. Cryoballoon PVI versus sedation, femoral venous access and phrenic pacing without transseptal puncture; cardioversion when needed.

Primary: Continuously monitored AF burden during months 3--6, after a three-month blanking period. & Adjusted geometric-mean ratio 0.25 (95\% CI 0.15--0.42), P\textless.001; secondary AFEQT contrast +18.39 points (11.48--25.30).\cite{s20} & Confirmed rhythm and secondary QoL benefit beyond a matched procedural-care package. Predominantly persistent AF; rescue treatment differed under common rules; positive contrast does not measure absence of placebo effects.

~ \\
PFA-SHAM (2026)

N=60 & Procedural sham. Pulsed-field ablation versus sham after EP study; antiarrhythmics/cardioversion permitted during blanking.

Primary: Coprimary arrhythmia recurrence and AFEQT quality of life at six months. & Recurrence 2/30 versus 25/30; AFEQT posterior contrast +32.6 points (95\% CrI 20.2--44.9); both superiority probabilities \textgreater.99.\cite{s12} & Confirmed rhythm and QoL efficacy in a highly symptomatic selected population. Small cohort with screening AFEQT \ensuremath{\le}50; five sham crossovers used pre-rescue PROM values for the six-month comparison.

~ \\
PVI-SHAM-AF (2026)

N=262 & Procedural sham. PVI versus deep sedation and femoral sheaths, without intracardiac/transseptal catheterization; real cardioversion if needed.

Primary: Change in AFEQT at six months; primary Hodges--Lehmann contrast. & Primary +2.6 points (95\% CI \ensuremath{-}2.7 to 8.0), P=.36; prespecified longitudinal sensitivity +5.4 (1.0--9.8). Holter AF positivity was 21\% versus 41\%.\cite{s21} & Did not establish primary incremental QoL superiority although rhythm measures favored PVI; clarified endpoint and estimand dependence. Not equivalence. The 27\% versus 48\% recurrence figure is a post hoc clinical-or-Holter composite; prespecified clinical-recurrence HR 0.66 (0.35--1.24) was inconclusive.

~ \\
TREAT-AF (2020)

N=53 & Noninvasive sham procedure. One hour/day tragus stimulation versus earlobe stimulation for six months.

Primary: AF burden at six months. & Median burden ratio 0.15 (95\% CI 0.03--0.65), P=.011.\cite{s51} & Extended acute mechanistic signals to ambulatory AF burden beyond sham. Small trial; no stroke or mortality efficacy inference.

~ \\
TREAT-PVC (2026)

N=100 & Noninvasive sham procedure. Tragus versus earlobe stimulation for six months.

Primary: Change in PVC burden. & Median burden fell from 18.7\% to 11.1\% with active treatment and 16.7\% to 12.6\% with sham; no between-group superiority.\cite{s52} & An AF neuromodulation signal did not generalize to ventricular ectopy. Both-arm improvement does not identify a pure placebo effect.

Denominator: 100 randomized; 96 analyzed. \\
\end{longtable}
}

\subsection*{\textbf{Table S1.} Pacing and cardiac resynchronization}
CRT in narrow-QRS heart failure: Mechanical dyssynchrony on echocardiography suggested that benefits seen with wide QRS might extend to narrower electrical complexes. Cardiac contractility modulation: Acute hemodynamic and uncontrolled findings suggested benefit, while an earlier small blinded pilot had not established efficacy. Cardiac resynchronization therapy: Acute hemodynamic and uncontrolled functional observations suggested that correcting ventricular activation delay could improve systolic heart failure. Pacing for obstructive hypertrophic cardiomyopathy: Uncontrolled studies suggested that dual-chamber pacing might offer a broad alternative to surgery in drug-refractory obstructive HCM. Pacing for reflex syncope: Observational series and early unblinded trials suggested large reductions in recurrent vasovagal syncope after pacemaker implantation. Rate-adaptive pacing for HFpEF: Reduced exercise heart rate correlated with impaired aerobic capacity, suggesting that pacing might correct a causal limitation.

{\footnotesize\setlength{\LTpre}{4pt}
\begin{longtable}[]{@{}
  >{\raggedright\arraybackslash}p{(\linewidth - 6\tabcolsep) * \real{0.1186}}
  >{\raggedright\arraybackslash}p{(\linewidth - 6\tabcolsep) * \real{0.3093}}
  >{\raggedright\arraybackslash}p{(\linewidth - 6\tabcolsep) * \real{0.2268}}
  >{\raggedright\arraybackslash}p{(\linewidth - 6\tabcolsep) * \real{0.3454}}@{}}
\toprule\noalign{}
\begin{minipage}[b]{\linewidth}\raggedright
\textbf{Trial / randomized N}
\end{minipage} & \begin{minipage}[b]{\linewidth}\raggedright
\textbf{Control and primary endpoint}
\end{minipage} & \begin{minipage}[b]{\linewidth}\raggedright
\textbf{Result}
\end{minipage} & \begin{minipage}[b]{\linewidth}\raggedright
\textbf{Knowledge contribution and limitations}
\end{minipage} \\
\midrule\noalign{}
\endhead
\bottomrule\noalign{}
\endlastfoot
RethinQ (2007)

N=172 & Implanted device ON/OFF. CRT ON versus OFF in patients with an ICD indication, narrow QRS and echocardiographic dyssynchrony; 6 months.

Primary: Proportion with an increase in peak oxygen consumption of at least 1.0 mL/kg/min at 6 months. & The primary response occurred in 46\% with CRT versus 41\% with control; no significant difference.\cite{s53} & RethinQ did not establish exercise benefit from extending CRT on the basis of echocardiographic dyssynchrony. A subgroup signal does not establish efficacy for the overall narrow-QRS population; all participants had implanted devices.

Denominator: 172 randomized after CRT-device implantation. \\
EchoCRT (2013)

N=809 & Implanted device ON/OFF. CRT ON versus OFF after device implantation in systolic heart failure with QRS \textless130 ms and echocardiographic dyssynchrony.

Primary: All-cause death or first hospitalization for worsening heart failure. & Primary events occurred in 28.7\% versus 25.2\% (HR 1.20, 95\% CI 0.92-1.57; P=.15). Mortality was 11.1\% versus 6.4\% (HR 1.81, 95\% CI 1.11-2.93; P=.02).\cite{s54} & EchoCRT restricted extrapolation of CRT into narrow-QRS disease and identified a clinically important mortality signal. Efficacy in a wide-QRS phenotype does not imply efficacy or safety in another; the trial does not estimate the net effect of implanting a device versus no implantation.

Denominator: 809 randomized: 404 CRT ON and 405 OFF; the trial stopped early for futility. \\
FIX-HF-4 (2008)

N=164 & Implanted device ON/OFF. CCM ON versus OFF/sham in two 3-month crossover periods; double blind.

Primary: Co-primary changes in peak oxygen consumption and Minnesota Living with Heart Failure Questionnaire score. & Active treatment periods improved peak oxygen consumption by 0.52 mL/kg/min (95\% CI 0.04-0.99; P=.032) and questionnaire score by 2.93 points (95\% CI 0.29-5.56; P=.030) relative to sham.\cite{s55} & FIX-HF-4 supported small functional and quality-of-life effects beyond sham stimulation. Crossover and period effects require consideration; the trial does not establish mortality or hospitalization benefit, and early improvement in both arms is not a pure placebo-effect estimate.

Denominator: 164 randomized to crossover sequence: 80 active-first and 84 sham-first. \\
MUSTIC-SR (2001)

N=58 & Implanted device ON/OFF. Atriobiventricular pacing versus inactive backup VVI pacing at 40 beats/min; single-blind crossover in sinus rhythm with wide QRS.

Primary: Distance walked in 6 minutes. & Among phase completers, mean 6-minute walking distance was 399 versus 326 m, a 22\% difference (P\textless.001); quality of life and peak oxygen uptake also improved.\cite{s56} & MUSTIC-SR confirmed functional benefit beyond inactive programming in selected patients with systolic heart failure and major conduction delay. Attrition and crossover analysis affect interpretation. The separate MUSTIC atrial-fibrillation cohort has an active pacing comparator and is not included here.

Denominator: 67 enrolled and implanted; 9 withdrew before randomization, leaving 58 randomized; 48 completed both crossover phases. \\
MIRACLE (2002)

N=453 & Implanted device ON/OFF. CRT ON versus OFF with conventional heart-failure therapy maintained in both arms; double blind.

Primary: Co-primary outcomes: NYHA functional class, quality of life and 6-minute walking distance at 6 months. & Walking-distance change was +39 versus +10 m (P=.005); quality-of-life change -18 versus -9 points (P=.001), and functional class improved. LVEF and heart-failure hospitalization findings were supportive.\cite{s57} & MIRACLE strengthened the evidence for genuine symptomatic and functional CRT benefit in wide-QRS systolic heart failure. This sham-controlled trial should not be credited with all later CRT mortality evidence; its primary outcomes and selected population must remain explicit.

Denominator: 453 randomized: 228 CRT and 225 control. Implantation failure and procedural complications are separate from this randomized denominator. \\
REVERSE (2008)

N=610 & Implanted device ON/OFF. CRT ON versus OFF, with or without defibrillator capability, for 12 months.

Primary: Heart-failure clinical composite response, analyzed as the proportion worsening. & Clinical worsening was 16\% ON versus 21\% OFF (P=.10). Prespecified secondary LV end-systolic volume index and time to first heart-failure hospitalization favored CRT.\cite{s58} & REVERSE extended evidence for reverse remodeling and supported selected secondary clinical benefits, while the primary comparison was neutral. Positive secondary outcomes do not convert the primary endpoint into a positive result; benefits concern mild disease with prolonged QRS.

Denominator: 610 randomized: 419 CRT ON and 191 OFF; one cohort despite later follow-up reports. \\
M-PATHY (1999)

N=48 & Implanted device sham programming. DDD pacing versus AAI backup at 30 beats/min; double-blind 3-month crossover periods followed by an open extension.

Primary: Symptoms and exercise capacity during the randomized blinded periods, including NYHA class, quality of life, exercise time and peak oxygen consumption. & No significant blinded-period differences in symptoms or exercise measures. During open follow-up subjective measures improved without improved peak oxygen consumption; the outflow gradient fell from 82 to 48 mm Hg over the reported follow-up.\cite{s59} & M-PATHY separated gradient reduction and open subjective improvement from a convincing randomized functional benefit. The gradient before-after change is not itself the randomized treatment contrast. Small exploratory responder subsets do not establish a general pacing indication.

Denominator: 48 randomized in the blinded crossover phase; the subsequent open extension is the same cohort, not a second trial. \\
VPS II (2003)

N=100 & Implanted device ON/OFF. Dual-chamber rate-drop-response pacing versus ODO sensing without pacing; all received an actual pacemaker; double blind.

Primary: Time to first recurrence of syncope over 6 months. & Estimated 6-month syncope risk was 31\% with pacing versus 40\% with sensing only; relative risk reduction 30\% (95\% CI -33\% to 63\%), one-sided P=.14.\cite{s60} & VPS II did not confirm the large benefit previously attributed to pacing in broadly selected vasovagal syncope. The confidence interval still permits clinically important benefit. Shared implantation means the randomized contrast does not estimate implant surgery versus no implant.

Denominator: 100 randomized after pacemaker implantation: 48 active and 52 sensing only. \\
SYNPACE (2004)

N=29 & Implanted device ON/OFF. Pacemaker ON versus OFF in severe recurrent tilt-induced vasovagal syncope; double blind.

Primary: Recurrence of syncope and time to first relapse. & During median follow-up of 715 days, syncope recurred in 8/16 (50\%) ON versus 5/13 (38\%) OFF; no significant difference.\cite{s61} & SYNPACE further limited claims of routine pacing benefit based on tilt positivity alone. Very small trial; a neutral result is not equivalence or proof that every cardioinhibitory phenotype is unresponsive.

Denominator: 29 randomized after pacemaker implantation: 16 ON and 13 OFF. \\
ISSUE-3 (2012)

N=77 & Implanted device ON/OFF. DDD rate-drop-response pacing versus sensing only in patients aged at least 40 years with severe asystolic neurally mediated syncope; double blind.

Primary: Recurrent syncope, assessed by time-to-event analysis. & Estimated 2-year recurrence was 25\% ON versus 57\% OFF; relative risk reduction 57\% (95\% CI 4\% to 81\%); log-rank P=.039 against the specified .04 threshold.\cite{s62} & ISSUE-3 supported efficacy in a selected spontaneous-asystole phenotype rather than reinstating a broad vasovagal indication. The loop-recorder screening population must not be counted as randomized. Selection limits generalizability.

Denominator: 511 initially received loop recorders; 89 met criteria for implantation; 77 were randomized. \\
SPAIN (2017)

N=54 & Implanted device sham programming. DDD closed-loop stimulation versus sham low-rate, subthreshold DDI programming; double-blind crossover in patients aged at least 40 years with recurrent cardioinhibitory vasovagal syncope.

Primary: Proportion with at least 50\% reduction in syncopal episodes. & At least 50\% reduction occurred in 72\% during DDD-CLS versus 28\% during sham (P=.017); events occurred in 4/46 versus 21/46 treatment periods.\cite{s63} & SPAIN supported a benefit of closed-loop pacing beyond sham programming in a selected population. Small crossover study with incomplete protocol completion and permitted early crossover; sequence and period effects matter.

Denominator: 54 randomized; 46 completed the crossover protocol. Reported treatment-period event percentages use the 46 completers. \\
BIOSync CLS (2021)

N=127 & Implanted device ON/OFF. DDD closed-loop stimulation ON versus ODO OFF; patients and independent outcome assessors masked.

Primary: Time to first recurrence of syncope. & After median follow-up of 11.2 months, syncope occurred in 16\% ON versus 53\% OFF; HR 0.23, P=.00005. Estimated 2-year recurrence was 22\% versus 68\%.\cite{s64} & BIOSync CLS supported efficacy in patients aged at least 40 years with severe recurrent syncope and tilt-induced asystole. Patient selection and pacing algorithm differ from earlier negative trials; the historical change cannot be attributed to masking alone.

Denominator: 127 randomized: 63 pacing ON and 64 OFF. \\
RAPID-HF (2023)

N=29 & Implanted device ON/OFF. Rate-adaptive atrial pacing versus no pacing; double-blind crossover with washout.

Primary: Oxygen consumption at the anaerobic threshold. & Pacing increased peak exercise heart rate by 14 beats/min but did not improve the primary endpoint: difference +0.3 mL/kg/min (95\% CI -0.5 to 1.0; P=.46). Stroke volume fell by 24 mL; device-related events occurred in 6/29 participants.\cite{s65} & RAPID-HF showed that correcting a correlated physiological variable did not improve exercise capacity in this selected HFpEF population. Small trial; the 6/29 device-event rate is descriptive for the implanted cohort, not a randomized implant-versus-no-implant harm difference.

Denominator: 32 recruited; 29 underwent implantation and randomization. All 29 experienced the implantation exposure. \\
\end{longtable}
}

\subsection*{\textbf{Table S1.} Heart-failure devices and neuromodulation}
Interatrial shunting for heart failure: Open studies suggested that left-atrial decompression could reduce exercise filling pressures and improve symptoms; clinical efficacy required a blinded comparison. Spinal-cord neuromodulation for HFrEF: Preclinical and small clinical studies suggested that spinal neuromodulation might alter adverse neurohormonal pathways in systolic heart failure. Splanchnic neuromodulation for HFpEF: Small open studies of splanchnic blockade or ablation suggested reduced filling pressures and improved functional status. Vagal neuromodulation for HFrEF: Experimental and open clinical studies suggested that restoring vagal activity could improve remodeling, exercise capacity and symptoms.

{\footnotesize\setlength{\LTpre}{4pt}
\begin{longtable}[]{@{}
  >{\raggedright\arraybackslash}p{(\linewidth - 6\tabcolsep) * \real{0.1186}}
  >{\raggedright\arraybackslash}p{(\linewidth - 6\tabcolsep) * \real{0.3093}}
  >{\raggedright\arraybackslash}p{(\linewidth - 6\tabcolsep) * \real{0.2268}}
  >{\raggedright\arraybackslash}p{(\linewidth - 6\tabcolsep) * \real{0.3454}}@{}}
\toprule\noalign{}
\begin{minipage}[b]{\linewidth}\raggedright
\textbf{Trial / randomized N}
\end{minipage} & \begin{minipage}[b]{\linewidth}\raggedright
\textbf{Control and primary endpoint}
\end{minipage} & \begin{minipage}[b]{\linewidth}\raggedright
\textbf{Result}
\end{minipage} & \begin{minipage}[b]{\linewidth}\raggedright
\textbf{Knowledge contribution and limitations}
\end{minipage} \\
\midrule\noalign{}
\endhead
\bottomrule\noalign{}
\endlastfoot
REDUCE LAP-HF I (2018)

N=44 & Procedural sham. Femoral venous access and intracardiac echocardiography without shunt placement; participants and follow-up assessors masked.

Primary: Exercise PCWP at 1 month, modeled across available exercise stages; a separate 1-month major-event safety endpoint. & The all-stage PCWP analysis favored shunting (P=.028); the peak-only PCWP change was -3.5 versus -0.5 mm Hg (P=.14).\cite{s66} & REDUCE LAP-HF I supported the physiological mechanism beyond sham, without establishing clinical outcome benefit. Small mechanistic trial; exercise-stage modeling and peak-only results answer different questions, and hemodynamic improvement is not clinical efficacy.

Denominator: 94 enrolled; 44 randomized: 22 shunt and 22 sham. Later follow-up is not a new randomized cohort. \\
REDUCE LAP-HF II (2022)

N=626 & Procedural sham. Atrial shunt implantation versus a masked sham procedure without shunt; EF at least 40\%.

Primary: Hierarchical composite of cardiovascular death/nonfatal ischemic stroke, total heart-failure events and change in KCCQ overall summary score. & Win ratio 1.0 (95\% CI 0.8-1.2; P=.85); no benefit in the individual primary components.\cite{s67} & REDUCE LAP-HF II showed that prior physiological benefit did not translate into overall clinical benefit in the tested population. Subgroup and subsequent responder hypotheses require validation; follow-up reports must not be counted as independent randomized trials.

Denominator: 1072 enrolled; 626 randomized: 314 shunt and 312 sham. The report used a modified intention-to-treat definition excluding postrandomization ineligibility. \\
RELIEVE-HF (2024)

N=508 & Procedural sham. Interatrial shunt implantation versus placebo procedure, with randomization stratified by EF \ensuremath{\le}40\% versus \textgreater40\%.

Primary: Hierarchical ranking of death, transplant/LVAD, heart-failure events and KCCQ change through maximum 2-year follow-up; separate short-term safety performance goal. & Overall effectiveness was neutral: win ratio 0.86 (95\% CI 0.61-1.22; P=.20). KCCQ did not differ. Prespecified exploratory EF strata showed opposing cardiovascular-event effects.\cite{s68} & RELIEVE-HF did not support generalized clinical shunt benefit and generated an important phenotype-dependent benefit/harm hypothesis. Exploratory stratum findings do not establish confirmed HFrEF efficacy. Meeting a safety performance goal is not equivalent to proving overall net safety in a randomized comparison.

Denominator: 508 randomized: 250 shunt and 258 placebo procedure; 206 had EF \ensuremath{\le}40\% and 302 EF \textgreater40\%. \\
DEFEAT-HF (2016)

N=66 & Implanted device ON/OFF. Spinal cord stimulation ON versus OFF; 3:2 randomization, patient masked.

Primary: Change in left-ventricular end-systolic volume index at 6 months. & LVESVi changed by +2.1 mL/m\textsuperscript{2} ON versus -2.2 mL/m\textsuperscript{2} OFF (P=.30); secondary efficacy endpoints were also neutral.\cite{s69} & DEFEAT-HF did not confirm a meaningful remodeling or functional benefit from this SCS strategy. A small device-specific trial does not establish universal futility of neuromodulation; implantation exposure was shared.

Denominator: 81 enrolled; 66 successfully randomized and implanted. Do not report 81 as randomized. \\
REBALANCE-HF (2024)

N=90 & Procedural sham. Femoral venous access and a sheath advanced into the azygos vein, without ablation catheter placement or energy delivery.

Primary: Change in legs-up and exercise PCWP at 1 month; separate serious device/procedure-event safety endpoint. & Between-group PCWP difference -0.03 mm Hg (95\% CI -2.5 to 2.5; P=.95); secondary efficacy outcomes were also neutral.\cite{s70} & REBALANCE-HF did not confirm the earlier hemodynamic promise in the broad population studied. Post hoc responder analyses remain hypotheses; a small nonsignificant safety comparison cannot establish absence of important harm.

Denominator: 90 randomized: 44 splanchnic ablation and 46 sham. \\
NECTAR-HF (2015)

N=95 & Implanted device ON/OFF. Vagal nerve stimulation ON versus OFF after implantation; 2:1 randomization over 6 months.

Primary: Change in left-ventricular end-systolic diameter at 6 months. & LVESD changed by -0.04 versus -0.08 cm (P=.60); other remodeling, peak oxygen uptake and biomarker endpoints were neutral. Some quality-of-life measures favored stimulation.\cite{s71} & NECTAR-HF failed to confirm objective remodeling or functional benefit with this stimulation regimen despite subjective signals. Active stimulation was perceptible to many participants, limiting interpretation of subjective signals. The primary analysis required paired data; 96 is the implanted rather than randomized denominator.

Denominator: 118 enrolled; 96 implanted, but one died before randomization. The randomized arms comprised 63 ON and 32 OFF; 87 had paired primary-endpoint data. \\
\end{longtable}
}

\subsection*{\textbf{Table S1.} Structural and pulmonary vascular procedures}
Functional mitral regurgitation: Single-arm Carillon studies suggested reduced functional mitral regurgitation and clinical improvement. A catheter sham comparison was needed to assess the incremental device effect with common heart-failure treatment. PFO closure for migraine: Observational improvement in migraine after PFO closure for other indications encouraged a causal treatment hypothesis. Migraine remains a neurological endpoint despite the intracardiac procedure. Pulmonary artery denervation: Mechanistic rationale and early uncontrolled denervation studies suggested pulmonary hemodynamic improvement. Subsequent sham trials differed importantly in pulmonary-hypertension phenotype and whether the comparator also received an alternative pulmonary vasodilator.

{\footnotesize\setlength{\LTpre}{4pt}
\begin{longtable}[]{@{}
  >{\raggedright\arraybackslash}p{(\linewidth - 6\tabcolsep) * \real{0.1186}}
  >{\raggedright\arraybackslash}p{(\linewidth - 6\tabcolsep) * \real{0.3093}}
  >{\raggedright\arraybackslash}p{(\linewidth - 6\tabcolsep) * \real{0.2268}}
  >{\raggedright\arraybackslash}p{(\linewidth - 6\tabcolsep) * \real{0.3454}}@{}}
\toprule\noalign{}
\begin{minipage}[b]{\linewidth}\raggedright
\textbf{Trial / randomized N}
\end{minipage} & \begin{minipage}[b]{\linewidth}\raggedright
\textbf{Control and primary endpoint}
\end{minipage} & \begin{minipage}[b]{\linewidth}\raggedright
\textbf{Result}
\end{minipage} & \begin{minipage}[b]{\linewidth}\raggedright
\textbf{Knowledge contribution and limitations}
\end{minipage} \\
\midrule\noalign{}
\endhead
\bottomrule\noalign{}
\endlastfoot
REDUCE-FMR (2019)

N=120 & Procedural sham. Carillon indirect mitral annuloplasty versus matching jugular venous/coronary-sinus catheter procedure with withdrawal before device deployment.

Primary: Core-laboratory quantitative mitral-regurgitant volume change at 12 months. & \ensuremath{-}7.1 versus +3.3 mL/beat; between-group reduction 10.4 mL/beat (95\% CI 0.1--20.7), P=.049.\cite{s72} & Strengthened a mechanistic imaging-efficacy claim after single-arm evidence. Only 68 paired primary measurements; six-minute walk and KCCQ comparisons were not significant; clinical events underpowered.

Denominator: 87 device/33 sham randomized; 73/87 implanted; paired primary MR-volume measurements in 55 device/13 sham participants. \\
MIST (2008)

N=147 & Procedural sham. PFO closure versus sham groin incision under general anaesthesia.

Primary: Complete migraine cessation. & 3/74 versus 3/73; primary outcome not positive.\cite{s73} & Did not confirm observational claims that closure eliminates refractory migraine. Cardiovascular procedure with neurological endpoint; not evidence against PFO closure for a different indication.

~ \\
PREMIUM (2017)

N=230 & Procedural sham. PFO closure versus right-heart-catheterization sham plus medical therapy.

Primary: At least 50\% reduction in migraine attacks. & 38.5\% versus 32.0\%, P=.32; secondary headache-day change \ensuremath{-}3.4 versus \ensuremath{-}2.0/month.\cite{s74} & A second major sham program again failed its primary migraine endpoint. Secondary and frequent-aura subgroup signals require confirmation; do not convert them into primary success.

~ \\
PADN-5 (2019)

N=98 & Procedural sham; unequal drug strategy. PADN plus heart-failure therapy versus sham PADN plus sildenafil and heart-failure therapy.

Primary: Change in six-minute walk distance at six months. & Increase 83 versus 15 m; reported adjusted difference 66 m (95\% CI 38.2--98.8), P\textless.001.\cite{s75} & Supported a functional treatment-strategy benefit in combined pre/postcapillary PH. Unequal sildenafil exposure prevents interpreting this as a clean PADN add-on effect; no established mortality benefit.

Denominator: 98 randomized; first online presentation/publication in 2018, journal issue 2019. \\
Romanov residual-CTEPH PADN (2020)

N=50 & Procedural sham; alternative drug strategy. PADN strategy versus mapping without ablation plus riociguat after pulmonary endarterectomy.

Primary: Change in pulmonary vascular resistance at 12 months. & PVR reduction 258 versus 149 dyn\ensuremath{\cdot}s\ensuremath{\cdot}cm\ensuremath{-}5; difference 109 (95\% CI 45--171), P=.001.\cite{s76} & Extended masked hemodynamic testing to residual CTEPH. Only 50 selected participants; alternative treatment strategies rather than identical background-drug exposure.

~ \\
PADN-CFDA (2022)

N=128 & Procedural sham. Catheter positioned without ablation versus PADN; PDE5 inhibition in both groups.

Primary: Change in six-minute walk distance at six months. & Adjusted difference +33.8 m (95\% CI 16.7--50.9), P\textless.001.\cite{s77} & Provided a cleaner matched-drug sham contrast supporting functional efficacy in PAH. A functional endpoint in selected PAH does not establish mortality efficacy or transport to all PH phenotypes.

~ \\
\end{longtable}
}

\subsection*{\textbf{Table S1.} Cell and gene delivery}
CD34 cell therapy for angina: Small blinded placebo-injection studies, including ACT34-CMI, generated an antianginal signal. The larger RENEW program tested confirmation and included a separate open standard-care arm. Cardiopoietic cell therapy: The earlier C-CURE study suggested functional improvement against a non-sham comparison. CHART-1 tested a more rigorous procedural control and a hierarchical clinical/functional endpoint. Limb-ischemia regenerative therapy: Angiogenic and cell-therapy rationale plus small studies suggested limb benefit. TALISMAN was already placebo controlled and its positive amputation finding was secondary despite a neutral ulcer-healing primary endpoint. Myocardial VEGF gene therapy: Angiogenic mechanisms and early studies suggested improved myocardial perfusion. Catheter-delivered placebo injection controls tested the biological product on top of common mapping and injection procedures. Myocardial cell therapy for heart failure: Preclinical rationale and small, heterogeneous clinical studies suggested ventricular or functional improvement. Earlier evidence included controlled and placebo-controlled trials, not only uncontrolled case series; cell products and delivery methods differ. SERCA2a gene therapy: Preclinical and early-phase work motivated intracoronary gene transfer. The encouraging randomized CUPID phase II evidence was itself blinded and placebo controlled; CUPID 2 tested replication at a larger scale.

{\footnotesize\setlength{\LTpre}{4pt}
\begin{longtable}[]{@{}
  >{\raggedright\arraybackslash}p{(\linewidth - 6\tabcolsep) * \real{0.1186}}
  >{\raggedright\arraybackslash}p{(\linewidth - 6\tabcolsep) * \real{0.3093}}
  >{\raggedright\arraybackslash}p{(\linewidth - 6\tabcolsep) * \real{0.2268}}
  >{\raggedright\arraybackslash}p{(\linewidth - 6\tabcolsep) * \real{0.3454}}@{}}
\toprule\noalign{}
\begin{minipage}[b]{\linewidth}\raggedright
\textbf{Trial / randomized N}
\end{minipage} & \begin{minipage}[b]{\linewidth}\raggedright
\textbf{Control and primary endpoint}
\end{minipage} & \begin{minipage}[b]{\linewidth}\raggedright
\textbf{Result}
\end{minipage} & \begin{minipage}[b]{\linewidth}\raggedright
\textbf{Knowledge contribution and limitations}
\end{minipage} \\
\midrule\noalign{}
\endhead
\bottomrule\noalign{}
\endlastfoot
ACT34-CMI (2011)

N=167 & Matched placebo injection/infusion. Two CD34-cell doses versus equal-volume diluent injected intramyocardially.

Primary: Weekly angina frequency at six months. & Low-dose 6.8 versus placebo 10.9 episodes/week, P=.020; exercise capacity improved. High-dose comparisons were not similarly significant.\cite{s78} & Positive blinded phase-II signal supported further confirmation. Dose inconsistency, multiple comparisons and modest size limit certainty.

~ \\
RENEW (2016)

N=112 & Placebo injection plus open usual-care arm. CD34 cells, intramyocardial placebo injection, and an additional open standard-care arm.

Primary: Change in total exercise time at 12 months. & Cell versus placebo difference +36.6 s (95\% CI \ensuremath{-}56.1 to 129.2), P=.43.\cite{s79} & An incompletely executed confirmatory study left the earlier signal unresolved rather than definitively refuted. Sponsor stopped recruitment for strategic reasons; open standard-care arm answers an additional strategy question.

Denominator: 112 randomized of a planned 444; the target sample size is not the final N. \\
CHART-1 (2017)

N=315 & Procedural sham. Endomyocardial cardiopoietic cells versus sham catheter procedure without myocardial injection.

Primary: Hierarchical clinical/functional endpoint at 39 weeks. & Mann--Whitney estimator 0.54 (95\% CI 0.47--0.61), P=.27.\cite{s80} & Did not confirm the early C-CURE functional signal in the overall population. LV-volume subgroup findings are exploratory; sham without injection is distinct from matched placebo injection.

Denominator: 315 randomized; 271 received the assigned procedure. \\
TALISMAN 201 (2008)

N=125 & Matched placebo injection/infusion. Intramuscular NV1FGF versus placebo on the same injection schedule.

Primary: Ulcer healing in critical limb ischemia. & 19.6\% versus 14.3\%, P=.514; secondary amputation signal favored gene therapy.\cite{s81} & Provided a secondary clinical signal, not a positive primary endpoint, for later phase-III testing. Primary failure and secondary multiplicity must be retained in the historical narrative.

Denominator: 125 randomized; 107 evaluable. \\
TAMARIS (2011)

N=525 & Matched placebo injection/infusion. Intramuscular NV1FGF versus visually matched placebo injections over four sessions.

Primary: Major amputation or death at one year. & 36\% versus 33\%; HR 1.11 (95\% CI 0.83--1.49), P=.48.\cite{s82} & Did not confirm TALISMAN's secondary amputation signal in a larger blinded clinical-outcome trial. Earlier evidence was already placebo controlled; this is failure of confirmation rather than first removal of procedural expectation bias.

~ \\
JUVENTAS (2015)

N=160 & Matched placebo injection/infusion. Three intra-arterial bone-marrow mononuclear-cell infusions versus placebo.

Primary: Major amputation at six months. & 19\% versus 13\%; RR 1.46 (95\% CI 0.62--3.42); no benefit demonstrated.\cite{s83} & Did not validate an amputation benefit inferred from earlier improvement observations. Secondary symptom/perfusion measures improved in both arms; that change does not isolate a placebo mechanism.

~ \\
Euroinject One (2005)

N=80 & Matched placebo injection/infusion. NOGA-guided myocardial VEGF plasmid versus placebo-plasmid injections.

Primary: Stress myocardial perfusion at three months. & No primary perfusion superiority; angina improved in both groups without a between-group difference.\cite{s84} & Did not confirm a perfusion or symptom benefit beyond matched injection. Secondary regional-wall-motion findings cannot replace the primary result.

~ \\
NORTHERN (2009)

N=93 & Matched placebo injection/infusion. NOGA-guided endocardial VEGF165 plasmid versus buffered-saline injection.

Primary: SPECT perfusion at three and six months. & No superiority in perfusion, exercise duration or angina.\cite{s85} & Both-arm improvement weakened inference from uncontrolled within-patient changes. Tests the biological product in a shared injection procedure; not a no-procedure control.

~ \\
FOCUS-CCTRN (2012)

N=92 & Matched placebo injection/infusion. Bone-marrow aspiration followed by transendocardial mononuclear-cell or placebo injection.

Primary: Coprimary LV end-systolic volume index, peak oxygen consumption and reversible SPECT defect at six months. & None of the three coprimary endpoints showed superiority.\cite{s86} & Did not confirm broad ventricular, perfusion or functional improvement from earlier small studies. Cell preparation and delivery are specific; cannot generalize to every cell product.

~ \\
TAC-HFT (2014)

N=65 & Matched placebo injection/infusion. Transendocardial mesenchymal-cell or mononuclear-cell injection, each with a matched placebo comparison.

Primary: Treatment-emergent serious adverse events at 30 days. & No treatment-emergent serious event at 30 days; LV volumes and EF did not improve; selected functional/scar findings were exploratory.\cite{s87} & Supported feasibility and informed further trials without establishing definitive efficacy. Small safety study with multiple comparisons and prominent within-group P values; no claim of proven safety equivalence.

Denominator: 65 randomized; 59 injected/analyzed (19 MSC, 19 mononuclear-cell, 21 combined placebo). \\
ixCELL-DCM (2016)

N=126 & Matched placebo injection/infusion. Transendocardial ixmyelocel-T versus vehicle placebo.

Primary: Recurrent death, cardiovascular hospitalization and unplanned visits for acute decompensated HF over 12 months. & Rate ratio 0.63 (95\% CI 0.42--0.97), P=.0344.\cite{s88} & Positive phase-II event signal despite neutral findings in other cell programs. Primary per-protocol exclusions after randomization and limited sample size constrain confirmation.

Denominator: 126 randomized; 114 modified ITT; 109 in the primary per-protocol efficacy analysis. \\
CONCERT-HF (2021)

N=125 & Matched placebo injection/infusion. Transendocardial MSCs, c-kit-positive cardiac cells, their combination or placebo.

Primary: Phase-II safety and efficacy domains over 12 months, including ventricular structure/function, functional capacity, clinical events and quality of life. & Selected cell arms showed MACE or QoL signals; EF, LV volumes, scar, six-minute walk and peak oxygen consumption did not differ.\cite{s89} & Suggested domain-specific hypotheses rather than establishing a general regenerative effect. Four arms and numerous endpoints; avoid presenting a selected favorable result as a single definitive confirmatory primary success.

~ \\
DREAM-HF (2023)

N=565 & Procedural sham. Transendocardial mesenchymal precursor cells versus sham catheterization.

Primary: Recurrent decompensated HF or successfully resuscitated symptomatic ventricular arrhythmia. & HR 1.17 (95\% CI 0.81--1.69), P=.41; hierarchical secondary endpoints were also negative.\cite{s90} & Did not establish the intended recurrent-HF-event benefit despite exploratory vascular/inflammation signals. MI/stroke and inflammation-subgroup findings require confirmation; do not replace the primary hierarchy.

Denominator: 565 in randomized ITT cohort; 537 in the treated analysis population for selected analyses. \\
CUPID phase II (2011)

N=39 & Matched placebo injection/infusion. Intracoronary AAV1-SERCA2a gene transfer at multiple doses versus placebo.

Primary: Prespecified multidomain success criteria across symptoms, function, biomarkers, remodeling and clinical outcomes. & The high-dose group met the prespecified six-month success criteria, with a favorable clinical-event signal.\cite{s91} & Generated the positive prior evidence for CUPID 2 within an already blinded placebo-controlled design. Very small dose-ranging trial with complex composite success rules; later failure is not attributable simply to the introduction of sham.

~ \\
CUPID 2 (2016)

N=250 & Matched placebo injection/infusion. Single intracoronary AAV1-SERCA2a infusion versus placebo.

Primary: Recurrent heart-failure events. & HR 0.93 (95\% CI 0.53--1.65), P=.81.\cite{s92} & Failed to replicate the smaller positive placebo-controlled CUPID signal. Product delivery, dose and biological exposure matter; a neutral primary trial does not isolate the reason for nonreplication.

Denominator: 250 randomized; 243 in modified ITT analysis. \\
\end{longtable}
}

\subsection*{\textbf{Table S1.} Remote ischemic conditioning}
Remote ischemic conditioning: Early cardioprotective evidence included animal experiments and small randomized studies, some already sham controlled. Positive troponin and exploratory clinical signals preceded larger outcome trials; later neutrality cannot be attributed simply to adding sham.

{\footnotesize\setlength{\LTpre}{4pt}
\begin{longtable}[]{@{}
  >{\raggedright\arraybackslash}p{(\linewidth - 6\tabcolsep) * \real{0.1186}}
  >{\raggedright\arraybackslash}p{(\linewidth - 6\tabcolsep) * \real{0.3093}}
  >{\raggedright\arraybackslash}p{(\linewidth - 6\tabcolsep) * \real{0.2268}}
  >{\raggedright\arraybackslash}p{(\linewidth - 6\tabcolsep) * \real{0.3454}}@{}}
\toprule\noalign{}
\begin{minipage}[b]{\linewidth}\raggedright
\textbf{Trial / randomized N}
\end{minipage} & \begin{minipage}[b]{\linewidth}\raggedright
\textbf{Control and primary endpoint}
\end{minipage} & \begin{minipage}[b]{\linewidth}\raggedright
\textbf{Result}
\end{minipage} & \begin{minipage}[b]{\linewidth}\raggedright
\textbf{Knowledge contribution and limitations}
\end{minipage} \\
\midrule\noalign{}
\endhead
\bottomrule\noalign{}
\endlastfoot
Thielmann CABG RIPC (2013)

N=329 & Noninvasive sham procedure. Remote ischemic arm-cuff conditioning versus sham during CABG anaesthesia.

Primary: Postoperative troponin release. & Troponin reduction and an exploratory prognostic signal were reported.\cite{s93} & An example of positive early sham evidence before larger neutral outcome trials. Single-center surrogate-led study; the clinical signal was not definitive outcome confirmation.

~ \\
ERICCA (2015)

N=1612 & Noninvasive sham procedure. Remote arm-cuff conditioning versus simulated cuff procedure before cardiac surgery.

Primary: Cardiovascular death, MI, revascularization or stroke at 12 months. & 26.5\% versus 27.7\%; HR 0.95 (95\% CI 0.79--1.15), P=.58.\cite{s94} & Did not reproduce early cardioprotective optimism on clinical outcomes. Specific surgery/anaesthetic context; does not identify why earlier biomarker findings differed.

~ \\
RIPHeart (2015)

N=1403 & Noninvasive sham procedure. Remote ischemic conditioning versus sham under propofol anaesthesia for cardiac surgery.

Primary: Death, MI, stroke or acute kidney injury. & 14.3\% versus 14.6\%; P=.89.\cite{s95} & A second large masked surgical trial did not establish clinical benefit. Anaesthetic and surgical protocol affect transportability; neutrality is not proof of exact equivalence.

Denominator: 1,403 randomized; 1,385 in the full analysis population. \\
\end{longtable}
}

\subsection*{\textbf{Table S1.} Boundary examples excluded from the included total}
{\footnotesize\setlength{\LTpre}{4pt}
\begin{longtable}[]{@{}
  >{\raggedright\arraybackslash}p{(\linewidth - 4\tabcolsep) * \real{0.1753}}
  >{\raggedright\arraybackslash}p{(\linewidth - 4\tabcolsep) * \real{0.4124}}
  >{\raggedright\arraybackslash}p{(\linewidth - 4\tabcolsep) * \real{0.4124}}@{}}
\toprule\noalign{}
\begin{minipage}[b]{\linewidth}\raggedright
\textbf{Study}
\end{minipage} & \begin{minipage}[b]{\linewidth}\raggedright
\textbf{Actual comparison}
\end{minipage} & \begin{minipage}[b]{\linewidth}\raggedright
\textbf{Reason for separate classification}
\end{minipage} \\
\midrule\noalign{}
\endhead
\bottomrule\noalign{}
\endlastfoot
CHAMPION (2011); N=550 & Clinicians received daily pulmonary-artery pressure data for active management; control clinicians did not use those data. Patients were masked, clinicians were not. & Boundary exclusion: this tests access to information and treatment adjustment, not sham implantation or an inactive therapeutic device. It is not counted in the included sham total.\cite{s96} \\
GUIDE-HF (2021); N=1000 & Pressure-guided clinical management versus usual care with pressure data withheld from control clinicians; patients masked, clinicians aware. & Boundary exclusion: a masked-information care strategy, not a sham therapeutic procedure. Pandemic sensitivity analyses must remain distinct from the overall result.\cite{s97} \\
CONDI-2/ERIC-PPCI (2019); N=5401 & Remote conditioning versus sham in the UK and usual care in other countries. & Not uniformly sham controlled; excluded from the fully masked inventory count.\cite{s98} \\
PRIMA (2016); N=107 & PFO closure versus open medical therapy. & No procedural sham; do not count it with MIST/PREMIUM as a third sham RCT.\cite{s99} \\
PADN-HF-PH (2026); N=264 & PADN plus guideline-directed therapy versus guideline-directed therapy alone. & Not an additional sham trial; clinical interpretation must respect the unblinded treatment strategy.\cite{s100} \\
\end{longtable}
}

CONDI-2/ERIC-PPCI had sham cuff procedures at UK sites and usual care elsewhere; the complete trial was not uniformly sham-controlled. The other rows answer information-access or open-comparator questions.

\subsection*{\textbf{Table S2.} Distinct randomized questions in four atrial fibrillation ablation sham trials}
{\footnotesize\setlength{\LTpre}{4pt}
\begin{longtable}[]{@{}
  >{\raggedright\arraybackslash}p{(\linewidth - 10\tabcolsep) * \real{0.1289}}
  >{\raggedright\arraybackslash}p{(\linewidth - 10\tabcolsep) * \real{0.1649}}
  >{\raggedright\arraybackslash}p{(\linewidth - 10\tabcolsep) * \real{0.1753}}
  >{\raggedright\arraybackslash}p{(\linewidth - 10\tabcolsep) * \real{0.1907}}
  >{\raggedright\arraybackslash}p{(\linewidth - 10\tabcolsep) * \real{0.1649}}
  >{\raggedright\arraybackslash}p{(\linewidth - 10\tabcolsep) * \real{0.1753}}@{}}
\toprule\noalign{}
\begin{minipage}[b]{\linewidth}\raggedright
\textbf{Trial}
\end{minipage} & \begin{minipage}[b]{\linewidth}\raggedright
\textbf{Population}
\end{minipage} & \begin{minipage}[b]{\linewidth}\raggedright
\textbf{Common and control care}
\end{minipage} & \begin{minipage}[b]{\linewidth}\raggedright
\textbf{Monitoring, rescue and estimand}
\end{minipage} & \begin{minipage}[b]{\linewidth}\raggedright
\textbf{Primary result}
\end{minipage} & \begin{minipage}[b]{\linewidth}\raggedright
\textbf{Interpretive limit}
\end{minipage} \\
\midrule\noalign{}
\endhead
\bottomrule\noalign{}
\endlastfoot
ORBITA-AF feasibility (2024) & 20 patients; persistent AF; one UK center; cryoballoon PVI. & Sedation, femoral access, phrenic pacing, cardioversion and usual drug care; no sham transseptal access. & Postprocedure ILR; six-week blanking; persistent recurrence triggered exit/rescue. Late QoL described selected sinus-rhythm survivors. & Masking feasibility demonstrated; efficacy was exploratory and imprecise. & Not a confirmatory efficacy trial; selected late QoL and inconsistent recurrence totals restrict interpretation.\cite{s19} \\
SHAM-PVI (2024) & 126 randomized; approximately 79\% persistent AF; two UK centers; cryoballoon PVI. & Sedation, femoral access, phrenic pacing, cardioversion if AF, and adaptive drug care; no sham transseptal access. & Preprocedure ILR baseline; burden during months 3--6; no sham crossover before endpoint. Treatment-policy comparison. & Adjusted geometric-mean AF-burden ratio 0.25 (95\% CI 0.15--0.42). & QoL was secondary; more controls resumed AADs. Superiority does not establish absence of contextual benefit.\cite{s20,s24} \\
PFA-SHAM (2026) & 60 randomized; 77\% paroxysmal AF; screening AFEQT \ensuremath{\le}50; two Czech centers. & Common EP study, cardioversion when needed, sedation and closure; no sham transseptal access. AAD withdrawal by two months. & Continuous ICM; two-month blanking. Five sham crossovers; pre-rescue questionnaires substituted for six-month assessments. & Recurrence: 2/30 versus 25/30. AFEQT change contrast: 32.6 points (95\% CrI 20.2--44.9). & Selected symptomatic population; rescue-dependent assessment time; model-dependent effect magnitude; small safety sample.\cite{s12,s23} \\
PVI-SHAM-AF (2026) & 262 randomized; 49\% paroxysmal AF; no AFEQT threshold; nine centers; mixed ablation techniques. & Deep sedation, femoral sheaths, cardioversion if AF and discretionary rhythm care; no sham intracardiac catheter. & Seven-day Holter at six months; no baseline burden. Six-month AFEQT with ITT imputation and death assigned worst change. & AFEQT Hodges--Lehmann contrast: 2.6 points (95\% CI \ensuremath{-}2.7 to 8.0; P=.36). & Primary superiority unproven; adjusted sensitivity positive. Rhythm ascertainment and background care differ from smaller trials.\cite{s21,s22} \\
\end{longtable}
}

Primary results are distinct statistical and clinical quantities; they should not be pooled or ranked without aligning the estimand. AAD, antiarrhythmic drug; AFEQT, Atrial Fibrillation Effect on QualiTy-of-life; CI, confidence interval; CrI, credible interval; EP, electrophysiological; ICM/ILR, implantable cardiac monitor/loop recorder; ITT, intention-to-treat; PVI, pulmonary vein isolation; QoL, quality of life.

\subsection*{\textbf{Table S3.} Interpretation audit of AF trial protocols, analysis plans and reports}
{\footnotesize\setlength{\LTpre}{4pt}
\begin{longtable}[]{@{}
  >{\raggedright\arraybackslash}p{(\linewidth - 4\tabcolsep) * \real{0.1753}}
  >{\raggedright\arraybackslash}p{(\linewidth - 4\tabcolsep) * \real{0.4124}}
  >{\raggedright\arraybackslash}p{(\linewidth - 4\tabcolsep) * \real{0.4124}}@{}}
\toprule\noalign{}
\begin{minipage}[b]{\linewidth}\raggedright
\textbf{Trial and issue}
\end{minipage} & \begin{minipage}[b]{\linewidth}\raggedright
\textbf{Documented finding}
\end{minipage} & \begin{minipage}[b]{\linewidth}\raggedright
\textbf{Assessment and implication}
\end{minipage} \\
\midrule\noalign{}
\endhead
\bottomrule\noalign{}
\endlastfoot
PVI-SHAM-AF: primary model and death & Primary rank-based contrast 2.6 points; adjusted longitudinal contrast 5.4. Death assigned \ensuremath{-}100 change in the primary analysis but treated as missing in the mixed model. & Different summaries and death strategies answer partly different questions. Preserve the confirmatory hierarchy while reporting the positive sensitivity analysis.\cite{s21,s22} \\
PVI-SHAM-AF: recurrence composite & Clinical recurrence was inconclusive; Holter outcomes favored ablation. The 27\% versus 48\% clinical-or-Holter composite was post-hoc and counted missing Holters as no detected AF. & Separate prespecified from exploratory results. Absence of detection is not continuously verified AF freedom.\cite{s21} \\
PVI-SHAM-AF: late specifications & Additional outcomes/subgroups submitted April 1, 2026; SAP signed May 15; lock June 1; ethics approval June 2; writing-team unmasking June 8. & Primary endpoint/method remained stable. Document chronology and reconcile subgroup definitions; timing alone does not demonstrate outcome-driven changes.\cite{s21,s22} \\
PFA-SHAM: adaptation and monitoring & Frequentist-to-Bayesian amendment followed enrollment. SAP finalized October 21, 2024, before documented interim analyses; stopping criteria met at 50 evaluable patients, with 60 already randomized. & A transparent adaptation, not an unchanged original design. Delayed endpoints limited recruitment savings; frequentist error properties require evaluation of the complete rule.\cite{s12,s23} \\
PFA-SHAM: rescue and assessment time & Five sham crossovers occurred at days 11, 131, 134, 146 and 157. Pre-rescue QoL replaced six-month values; all patients remained in ITT. & Ethically responsive rescue changes the endpoint to six months or earlier rescue. Complete follow-up and ITT assignment do not ensure a fixed-time treatment-policy estimand.\cite{s12,s23} \\
PFA-SHAM: baseline and model & AFEQT advantage: primary 32.6 points; screening-relative 26.0; baseline-adjusted ANCOVA approximately 19.9. All corresponding credible intervals excluded zero. & Direction was robust; magnitude was not invariant. Baseline imbalance and prerandomization improvement should inform cross-trial comparisons.\cite{s12} \\
PFA-SHAM: Bang denominator & Tabulated indices exclude uncertain guessers; the original three-response index includes them. An explanatory footnote gives values matching the full-denominator calculation. & A reporting/implementation discrepancy needing clarification. Publish raw counts and denominator rules; do not compare the printed indices directly with other AF trials.\cite{s10,s12} \\
SHAM-PVI: effect scale & The 0.25 result is a geometric-mean ratio. Reductions of 60.31 and 35.0 percentage points are descriptive within-group changes; AFEQT was secondary. & Distinguish relative adjusted efficacy, absolute descriptive change and supportive patient-reported outcomes. None measures a placebo fraction.\cite{s20,s24} \\
SHAM-PVI: serial masking & Guesses assessed at discharge, three and six months. Patients became more informative; staff largely remained uncertain. Planned confidence scores and structured reasons are absent from supplied results. & Supports initial masking and information barriers. Symptom-driven recognition is plausible; sustained masking or the cause of later guesses cannot be certified.\cite{s20,s24} \\
ORBITA-AF: feasibility and selection & Twenty-patient pilot prioritized masking. Late QoL was reported among patients maintaining sinus rhythm; recurrence totals and participant-information timing are not fully reconciled. & Feasibility evidence should not become confirmatory efficacy. Postrandomization selection does not preserve ITT; document discrepancies without inferring misconduct.\cite{s19} \\
\end{longtable}
}

This audit separates verified reporting discrepancies, differences in estimands and legitimate design tradeoffs. Missing material in supplied conversions is not evidence that investigators omitted an analysis or procedure. Numerical contrasts are ablation minus sham unless explicitly expressed as ratios. References are numbered throughout this appendix.

\subsection*{\textbf{Table S4.} ORBITA methodological audit}
{\footnotesize\setlength{\LTpre}{4pt}
\begin{longtable}[]{@{}
  >{\raggedright\arraybackslash}p{(\linewidth - 6\tabcolsep) * \real{0.1443}}
  >{\raggedright\arraybackslash}p{(\linewidth - 6\tabcolsep) * \real{0.2887}}
  >{\raggedright\arraybackslash}p{(\linewidth - 6\tabcolsep) * \real{0.2835}}
  >{\raggedright\arraybackslash}p{(\linewidth - 6\tabcolsep) * \real{0.2835}}@{}}
\toprule\noalign{}
\begin{minipage}[b]{\linewidth}\raggedright
\textbf{Issue}
\end{minipage} & \begin{minipage}[b]{\linewidth}\raggedright
\textbf{Documented finding}
\end{minipage} & \begin{minipage}[b]{\linewidth}\raggedright
\textbf{Interpretation}
\end{minipage} & \begin{minipage}[b]{\linewidth}\raggedright
\textbf{Design / reporting implication}
\end{minipage} \\
\midrule\noalign{}
\endhead
\bottomrule\noalign{}
\endlastfoot
Neutral superiority result versus equivalence & ORBITA exercise-time difference 16.6 s (95\% CI \ensuremath{-}8.9 to 42.0; P=0.20), after antianginal optimization. & Failure to demonstrate superiority does not establish no effect or equivalence. ORBITA-2 studied a different medication strategy; cross-trial differences cannot identify a randomized medication--PCI interaction. & Present the estimate and interval alongside the decision threshold; prespecify equivalence margins if equivalence is the question. Randomise background strategy or treatment sequence to compare those strategies directly.\cite{s13,s14} \\
SAP versus Journal-requested analysis and estimand & The SAP specified longitudinal Bayesian transitions. At Journal request the main paper used frequentist ordinal ANCOVA: OR 2.21 (95\% CI 1.41--3.47); Bayesian day-84 OR 1.88 (95\% CrI 1.57--2.27). & Final-state and conditional-transition summaries are distinct; the numerical difference is not a contradiction. The analytical change was explicitly disclosed. & Report both estimands, intercurrent-event strategies and model-specific missing-data handling; give a dated analysis-change rationale and executable specification.\cite{s14} \\
Ordinal outcome preferences and rescue treatment & No symptoms on one medication unit scores 7; \ensuremath{\ge}6 episodes without medication scores 6. Intolerable-angina unblinding, ACS and death override at 77--79; GTN, hospitalization alone, stroke and bleeding are not equivalent primary overrides. & This is a joint symptom/treatment-requirement ordering, not an attack count or complete net-benefit utility. Rescue is incorporated rather than prohibited; prescribed medication is not verified ingestion. & Justify preference ordering with broader patient input; report components and safety separately, medication exposure/adherence, and alternative defensible orderings. State exact rescue and event-persistence coding.\cite{s14} \\
Serial dependence, proportional odds and prior sensitivity & First-order Markov modeling conditions on prior state, baseline, time and treatment. Final code relaxes proportional odds for day and treatment; literal SAP names time only. Near-flat-prior day-84 OR 1.86 (95\% CrI 1.55--2.25) agrees with the main Bayesian estimate. & Patient-days are not treated as independently randomized patients. Prior sensitivity is reassuring within this model; it does not validate all structural assumptions. Treatment non-proportionality is a documented refinement requiring chronology, not proof of biased selection. & Publish model-check outputs, dated refinements, lag/order and alternative-link sensitivity analyses, and distinguish conditional transitions from marginal state occupancy. Interpret posterior probabilities conditional on model and prior.\cite{s14} \\
Missing outcomes, activity and adherence & Daily reporting was 99.7\% complete. Two placebo patients needed final-score LOCF; SAP separately specifies coarsened daily states. COVID-related missing physical tests had concordant imputation analyses. Systematic activity levels were not collected. & Low diary missingness is a strength, but final-score LOCF must not be conflated with longitudinal handling. Diary completion does not prove drug adherence or equal exertion. Exact data preparation remains incompletely reconstructable from extracted output. & Provide an outcome-observation flow and reproducible coarsening/event code; test plausible missingness assumptions; measure activity and medication adherence when central to interpretation.\cite{s14,s15,s18} \\
Masking assessment timing & Discharge Bang indices supported concealment; at 12 weeks patient indices were 0.19 in PCI and 0.24 in placebo, while staff indices remained near zero. & Later guesses may reflect true efficacy or disclosure. They neither prove disclosure bias nor demonstrate identical expectations throughout follow-up. & Prespecify who guesses, when, confidence, information available and whether guesses precede outcome discussions; document accidental disclosure and contextual cues without na\"ively adjusting away symptom-driven guesses.\cite{s14} \\
Protocol development and sample-size targets & Original protocol: 90\% power, 342 randomized; final protocol/SAP: 80\%, 284 randomized, approximately 400 enrolled. Actual 439 enrolled and 301 randomized. Historical washout language includes both 72 hours and one week. & The revised randomization target was exceeded. Mean-difference planning was an approximation for the eventual ordinal model. Unexplained revisions and inconsistent wording are documentation questions, not established conduct errors. & Publish dated rationales for sample-size/model revisions, distinguish enrollment and randomization targets, and reconcile planned with implemented withdrawal/rescue procedures.\cite{s14} \\
Multiplicity, effect modification and repeated publications & Secondary intervals were not multiplicity-adjusted. Physiology (n=279), baseline DSE (n=262) and Rose-symptom (n=267) analyses use subsets of the same randomization. & Interaction ORs compare treatment effects across predictor values; they are not overall treatment effects or independent replications. Stronger benefit in one subgroup does not establish zero benefit in another. & Separate prespecified and exploratory interactions; report predictor contrasts and selection/missingness; validate prediction externally and randomise proposed diagnostic or treatment-selection pathways.\cite{s14,s16,s17,s18} \\
\end{longtable}
}

Documentary questions are not automatically errors or evidence of outcome-driven decisions. Protocol versions, data access and analysis timing are relevant to their assessment.

\subsection*{\textbf{Table S5.} Operational questions for designing, conducting and reporting a cardiovascular procedural-control trial}
{\footnotesize\setlength{\LTpre}{4pt}
\begin{longtable}[]{@{}
  >{\raggedright\arraybackslash}p{(\linewidth - 4\tabcolsep) * \real{0.1237}}
  >{\raggedright\arraybackslash}p{(\linewidth - 4\tabcolsep) * \real{0.3299}}
  >{\raggedright\arraybackslash}p{(\linewidth - 4\tabcolsep) * \real{0.5464}}@{}}
\toprule\noalign{}
\begin{minipage}[b]{\linewidth}\raggedright
\textbf{Stage}
\end{minipage} & \begin{minipage}[b]{\linewidth}\raggedright
\textbf{Question to resolve}
\end{minipage} & \begin{minipage}[b]{\linewidth}\raggedright
\textbf{Concrete output / avoidable failure}
\end{minipage} \\
\midrule\noalign{}
\endhead
\bottomrule\noalign{}
\endlastfoot
Clinical question & Which patient-important decision and causal contrast require masking? & A clinical estimand stating population, complete strategies, outcome/time, rescue/death handling and summary; avoid a control chosen before the question. \\
Control components & Which steps are therapeutic, sensory, diagnostic or potentially harmful? & An arm-by-arm component matrix with the rationale for retaining, omitting or simulating each step; avoid unnecessary invasive duplication. \\
Ethics and consent & Does the additional research exposure have a proportionate scientific justification? & Balanced explanation of actual procedures, uncertainty, proven care, rescue and follow-up; verify comprehension without implying that placebo means harmless. \\
Feasibility & Can patients be recruited and followed without undermining clinical care? & Patient involvement, recruitment pilot and screening/refusal log; avoid claiming broad applicability from a highly selected consenting subset. \\
Active delivery & Can the intended component be delivered reproducibly? & Operator criteria, procedural standards, technical-success definitions and fidelity review; avoid interpreting inadequate target engagement as a definitive test of the mechanism. \\
Randomization & When is eligibility established and what is known before allocation? & Concealed sequence and prespecified stratification; document procedural exclusions before and after assignment; avoid excluding randomized failures from the primary strategy analysis. \\
Masking & Who can know assignment, when and through which records or cues? & Role-specific access, masked follow-up, standardized communication, emergency-unmasking log and prespecified belief questions; avoid an undifferentiated claim of double blinding. \\
Background care & Which cointerventions and rescue rules are part of each strategy? & Medication and rehabilitation policy with clinical exceptions; record actual use by arm; avoid forcing equal realized rescue when treatment changes need. \\
Outcomes & Does each outcome address symptoms, mechanism, function, events or harms? & Prespecified hierarchy, collection times and component reporting; avoid treating physiological success as proof of survival or symptom benefit. \\
Size and monitoring & What precision or decision performance is required? & Clinically justified target effect, attrition assumptions and full interim operating characteristics; avoid observed post-hoc power and unsupported claims of automatic error control. \\
Primary model & Does the model estimate the stated outcome summary? & Signed SAP covering baseline adjustment, repeated-measure dependence, prediction and diagnostics; avoid interpreting conditional odds as marginal benefit without qualification. \\
Missingness / death & Why are measurements unavailable, and what outcomes still exist? & Continued follow-up, compatible imputation and clinically plausible sensitivity analyses; distinguish death from a missed questionnaire and changes of estimand from model sensitivity. \\
Multiplicity & Which result establishes efficacy, and which analyses are exploratory? & Decision rules for primary/co-primary endpoints, secondary outcomes and interactions; avoid promoting a positive subgroup after a neutral primary analysis. \\
Safety & Which hazards arise from common procedures versus the active component? & Randomized and procedural denominators, exposure windows and independent review; avoid interpreting zero events in a small study as safety equivalence. \\
Reporting & Can a reader reconstruct the contrast and every substantive amendment? & Versioned protocol/SAP, amendment chronology, arm-specific control detail, raw masking counts, rescue and missingness; avoid silently substituting later analyses. \\
Synthesis & Are compared studies sufficiently aligned to support a shared estimate? & Population/control/estimand mapping and cohort-overlap checks before pooling; avoid counting substudies as replications or ranking energy sources across unrelated AF trials. \\
\end{longtable}
}

Author-developed synthesis for this review, not a validated checklist or risk-of-bias score. Use alongside current SPIRIT/CONSORT, ASPIRE and TIDieR-Placebo guidance.\cite{s2,s7,s8,s9,s101,s102}

\begin{figure}[!htbp]
\centering
\includegraphics[width=\textwidth,height=0.80\textheight,keepaspectratio]{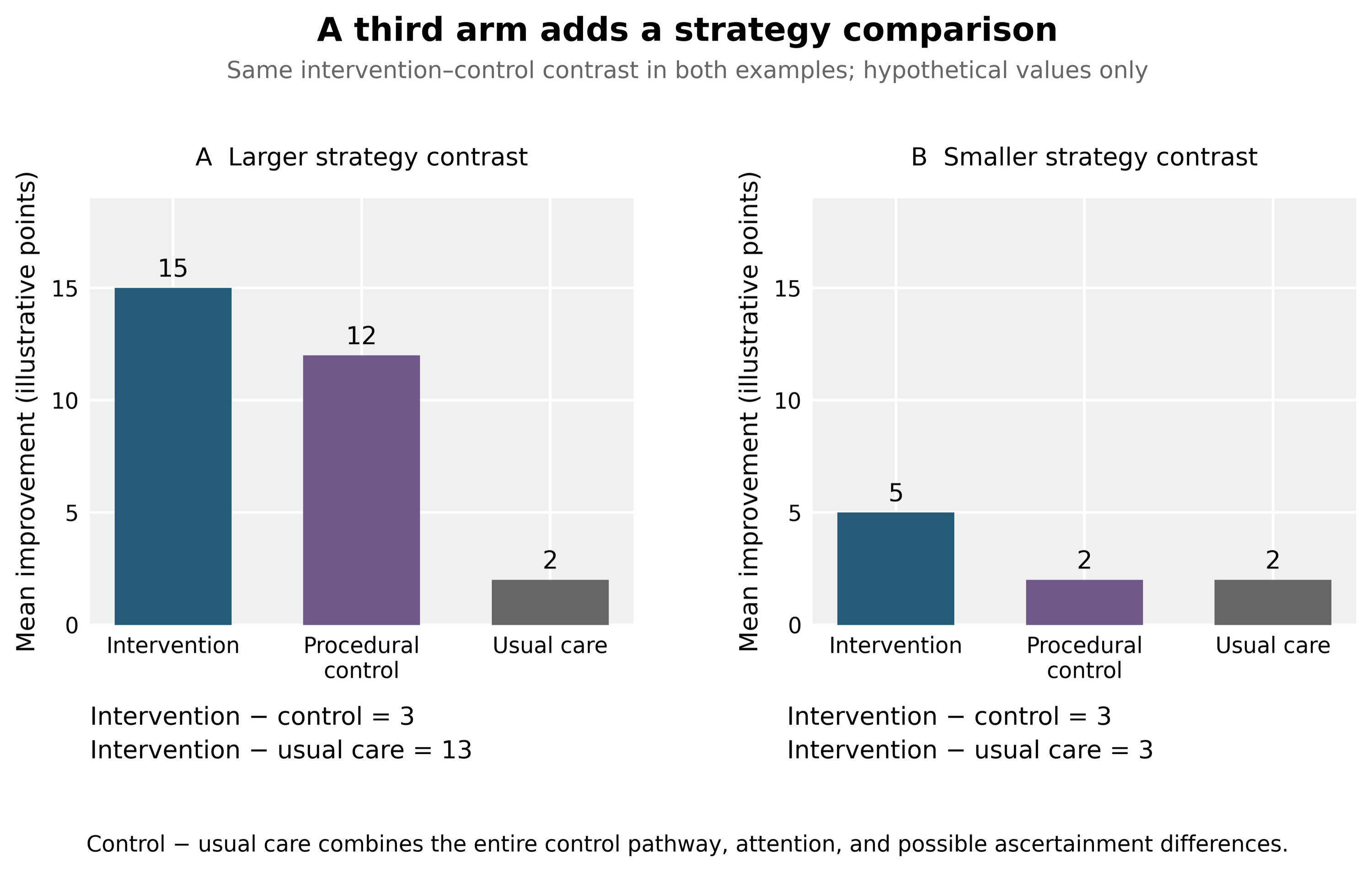}
\caption*{\textbf{Figure S1. A third arm changes the questions that can be answered.} Illustrative mean improvements in active, procedural-control and usual-care groups are 15, 12 and 2 in panel A and 5, 2 and 2 in panel B. The active-minus-procedural-control contrast is 3 in both panels, whereas the active-minus-usual-care contrast is 13 and 3. Thus the same incremental component effect can coexist with different complete-strategy effects. Procedural-control minus usual care still combines procedure, expectations, attention and cointerventions; it does not isolate a pure psychological effect. Values are constructed examples and are not estimates from clinical trials.}
\end{figure}

\begin{figure}[!htbp]
\centering
\includegraphics[width=\textwidth,height=0.80\textheight,keepaspectratio]{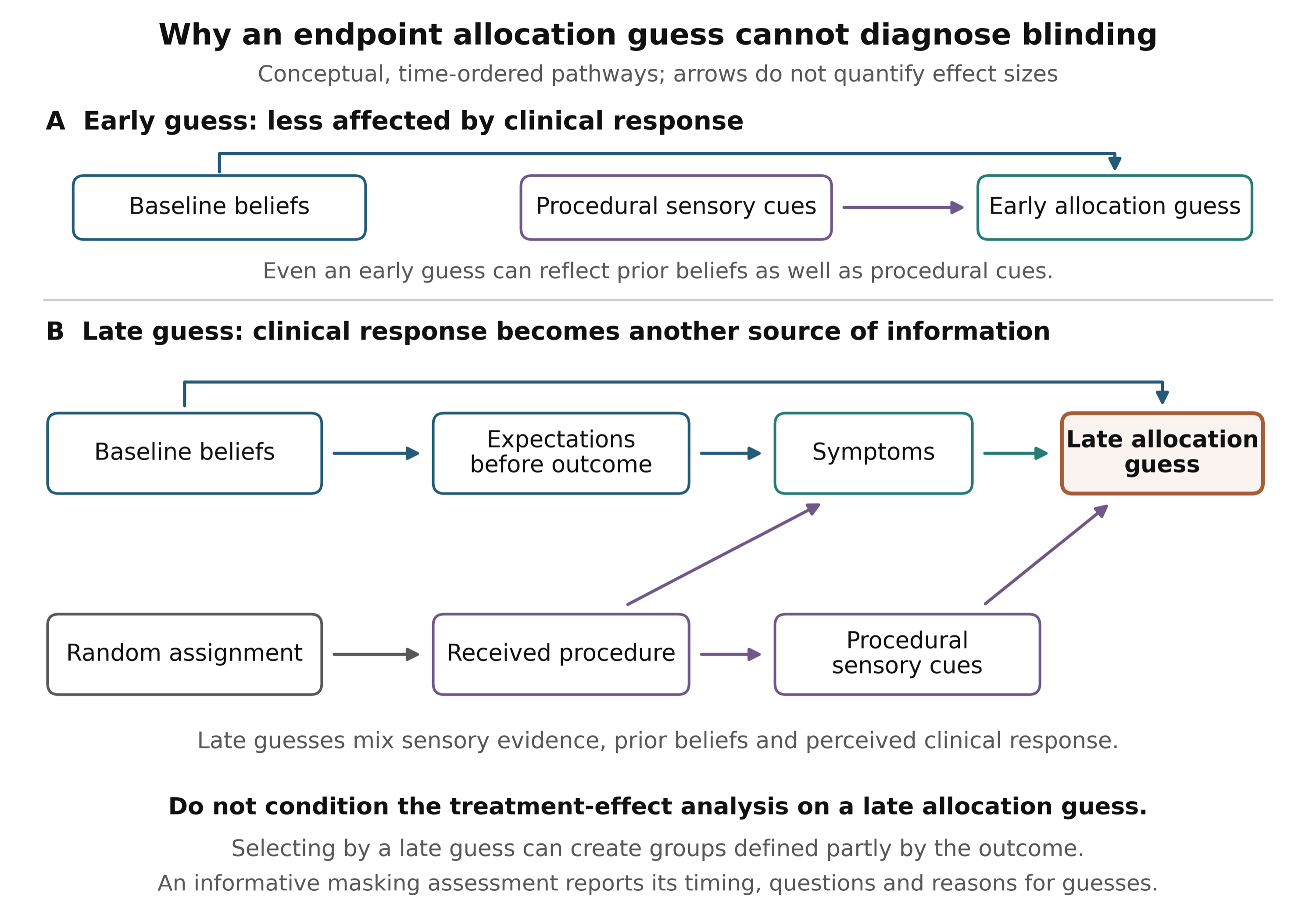}
\caption*{\textbf{Figure S2. Why an endpoint allocation guess cannot independently diagnose blinding.} Early guesses can reveal clues from the procedure before substantial benefit is apparent. At a later assessment, an allocation guess can also be caused by actual symptom response, recognizable adverse effects or information leakage. The causal schematic is explicitly time-ordered: prior expectations can affect later symptoms, and those symptoms can affect a subsequent guess. Conditioning the treatment-effect analysis on that late guess may remove treatment-related information or introduce selection bias. The diagram is conceptual; it does not imply that any observed pathway must be present in a particular trial.\cite{s10,s11}}
\end{figure}

\begin{figure}[!htbp]
\centering
\includegraphics[width=\textwidth,height=0.80\textheight,keepaspectratio]{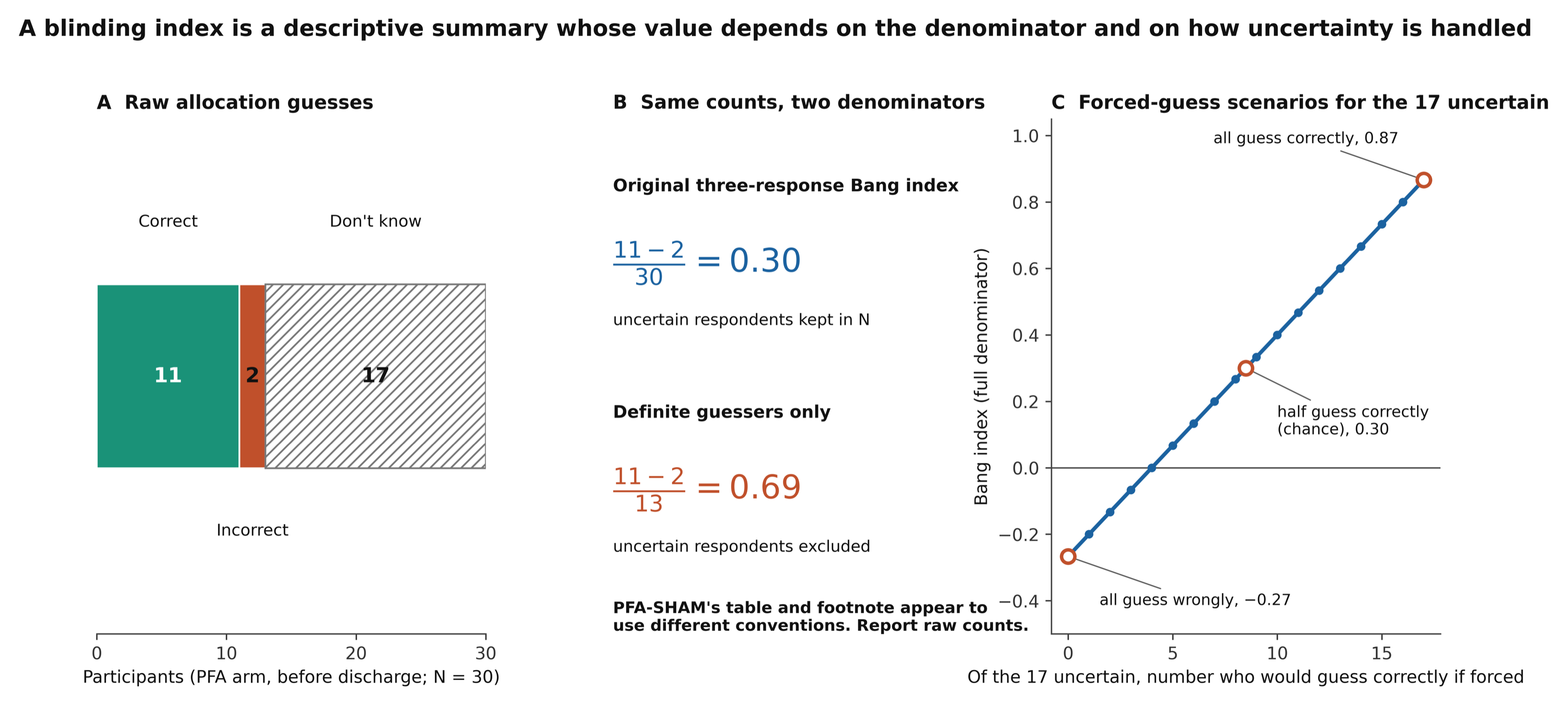}
\caption*{\textbf{Figure S3. A blinding index depends on the denominator and on the handling of uncertainty.} (A) Raw allocation guesses in the PFA arm of PFA-SHAM before discharge (11 correct, 2 incorrect, 17 uncertain). (B) The same counts give an index of 0.30 with the original three-response Bang denominator and 0.69 among definite guessers. (C) Under forced-guess scenarios for the 17 uncertain respondents, the index ranges from \ensuremath{-}0.27 (all incorrect) to 0.87 (all correct), and it is 0.30 if half guess correctly. Reporting raw counts by arm and time lets readers recompute alternative summaries.\cite{s10,s12}}
\end{figure}

\begin{figure}[!htbp]
\centering
\includegraphics[width=\textwidth,height=0.80\textheight,keepaspectratio]{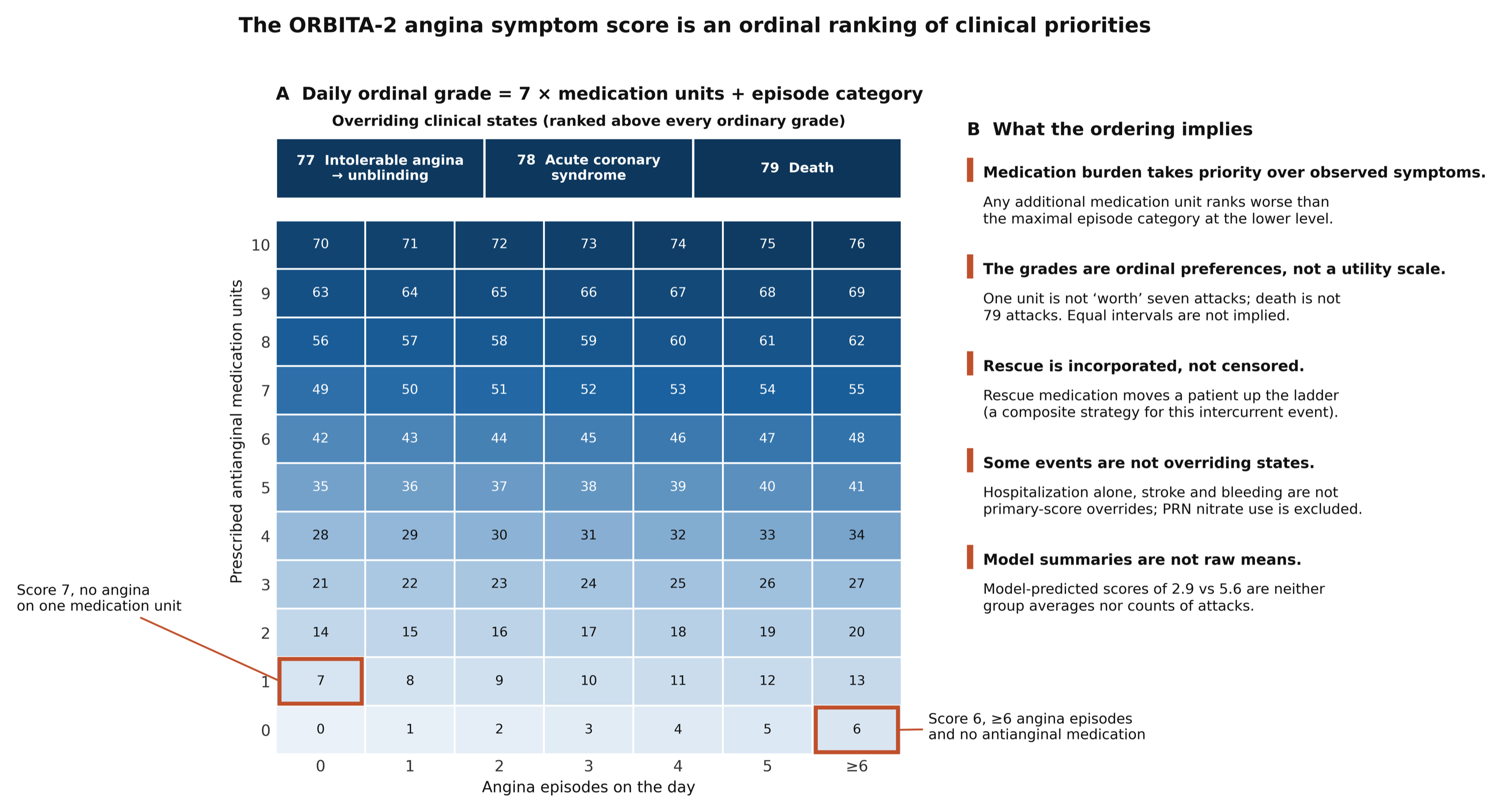}
\caption*{\textbf{Figure S4. Structure of the ORBITA-2 angina symptom score.} Ordinary daily grades (0--76) combine prescribed antianginal medication units (rows) with the number of angina episodes on that day (columns), so that the grade equals 7 \ensuremath{\times} medication units + episode category. Grades 77--79 are overriding clinical states. Outlined cells show that a day without angina on one medication unit (grade 7) ranks worse than a day with six or more episodes and no medication (grade 6). The ordering encodes clinical priorities, and the spacing between grades has no cardinal meaning. The structure follows the ORBITA-2 statistical analysis plan.\cite{s14}}
\end{figure}

\begin{figure}[!htbp]
\centering
\includegraphics[width=\textwidth,height=0.80\textheight,keepaspectratio]{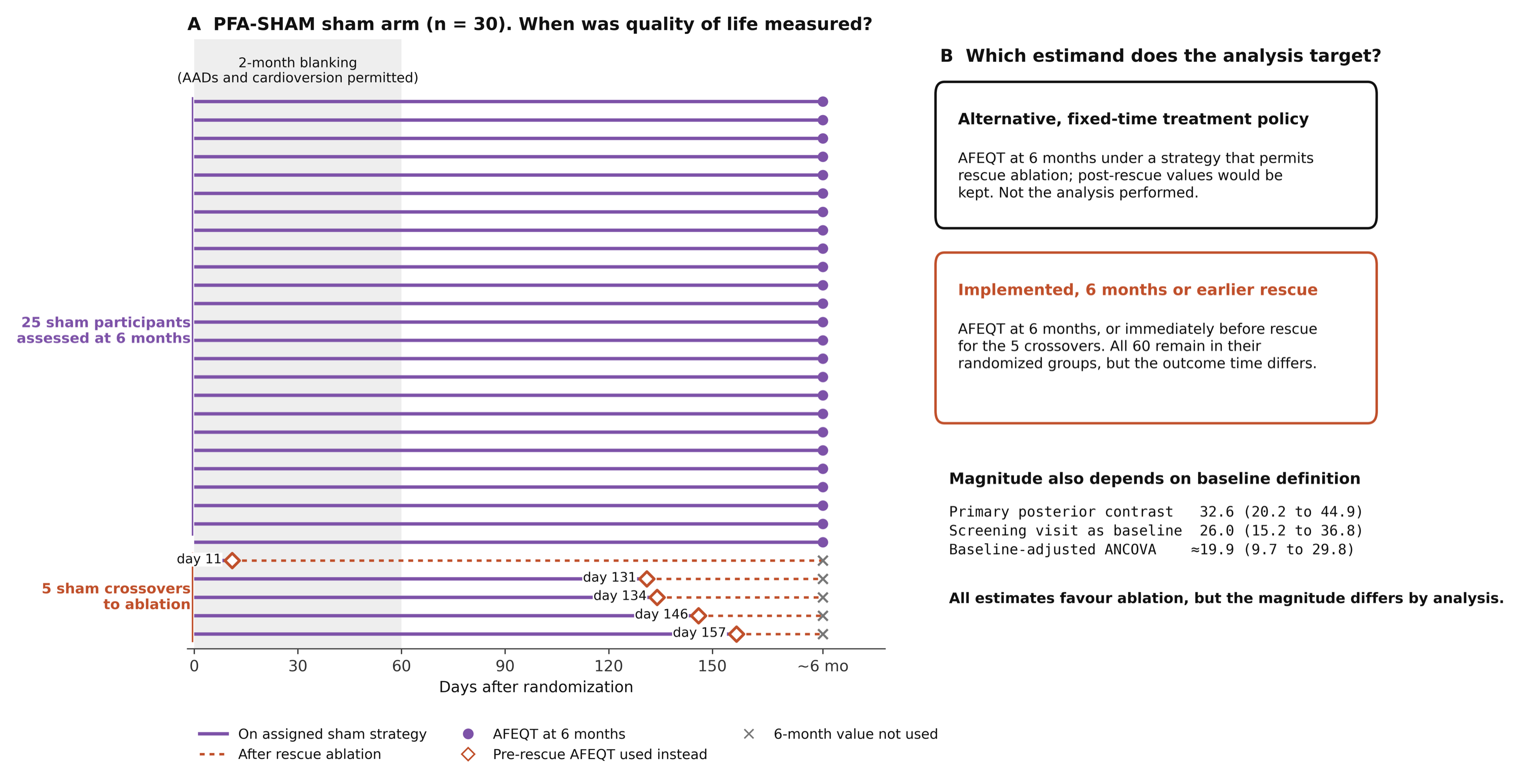}
\caption*{\textbf{Figure S5. Rescue changed the time at which quality of life was measured in PFA-SHAM.} (A) Sham-arm participants (n = 30). Five crossed over to ablation, on days 11, 131, 134, 146 and 157, and their pre-rescue AFEQT replaced the six-month value. The other 25 participants were assessed at the nominal six-month visit, and their individual lines are schematic. The shaded interval is the two-month blanking period. (B) The implemented analysis targets AFEQT at six months or at earlier rescue, not a fixed-time treatment-policy estimand. The estimated magnitude also depends on the baseline definition, although all estimates favor ablation.\cite{s12,s23}}
\end{figure}

\begin{figure}[!htbp]
\centering
\includegraphics[width=\textwidth,height=0.80\textheight,keepaspectratio]{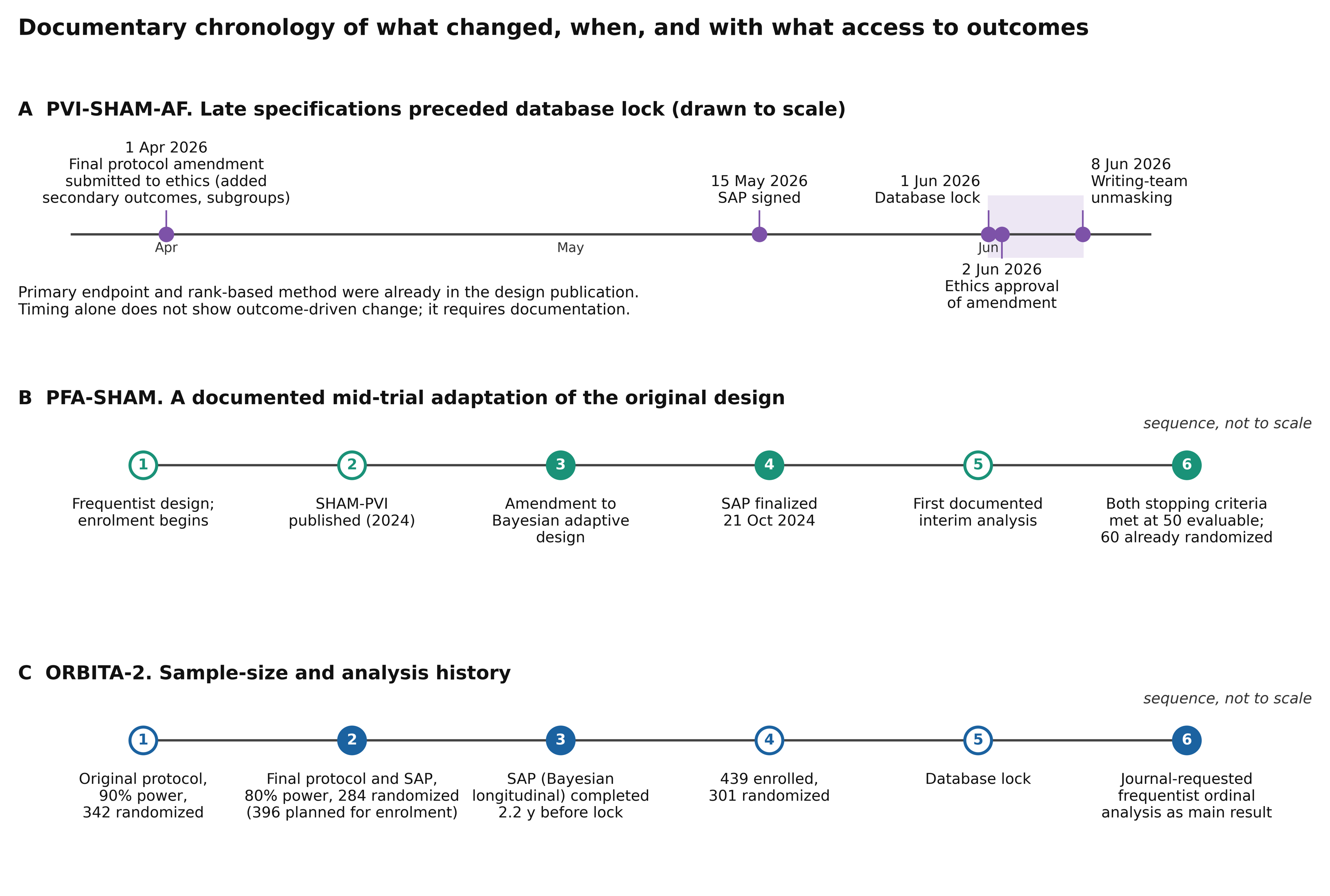}
\caption*{\textbf{Figure S6. Documentary chronology of analysis and design changes.} (A) PVI-SHAM-AF events drawn to scale. The final protocol amendment, which added secondary outcomes and subgroups, was submitted to the ethics committee on April 1, 2026 and approved on June 2, 2026. The primary endpoint and rank-based method were already specified in the design publication. (B) PFA-SHAM changed from a frequentist to a Bayesian adaptive design, and both stopping criteria were met at 50 evaluable participants after 60 had been randomized. (C) ORBITA-2 sample-size and analysis history. The main report gives a planned enrollment of 396 patients. Filled markers indicate steps discussed in the text. Panels B and C show order only, not elapsed time. Timing alone does not establish outcome-driven change, but these chronologies show what readers need in order to assess it.\cite{s12,s14,s21,s22,s23}}
\end{figure}

\begin{figure}[!htbp]
\centering
\includegraphics[width=\textwidth,height=0.80\textheight,keepaspectratio]{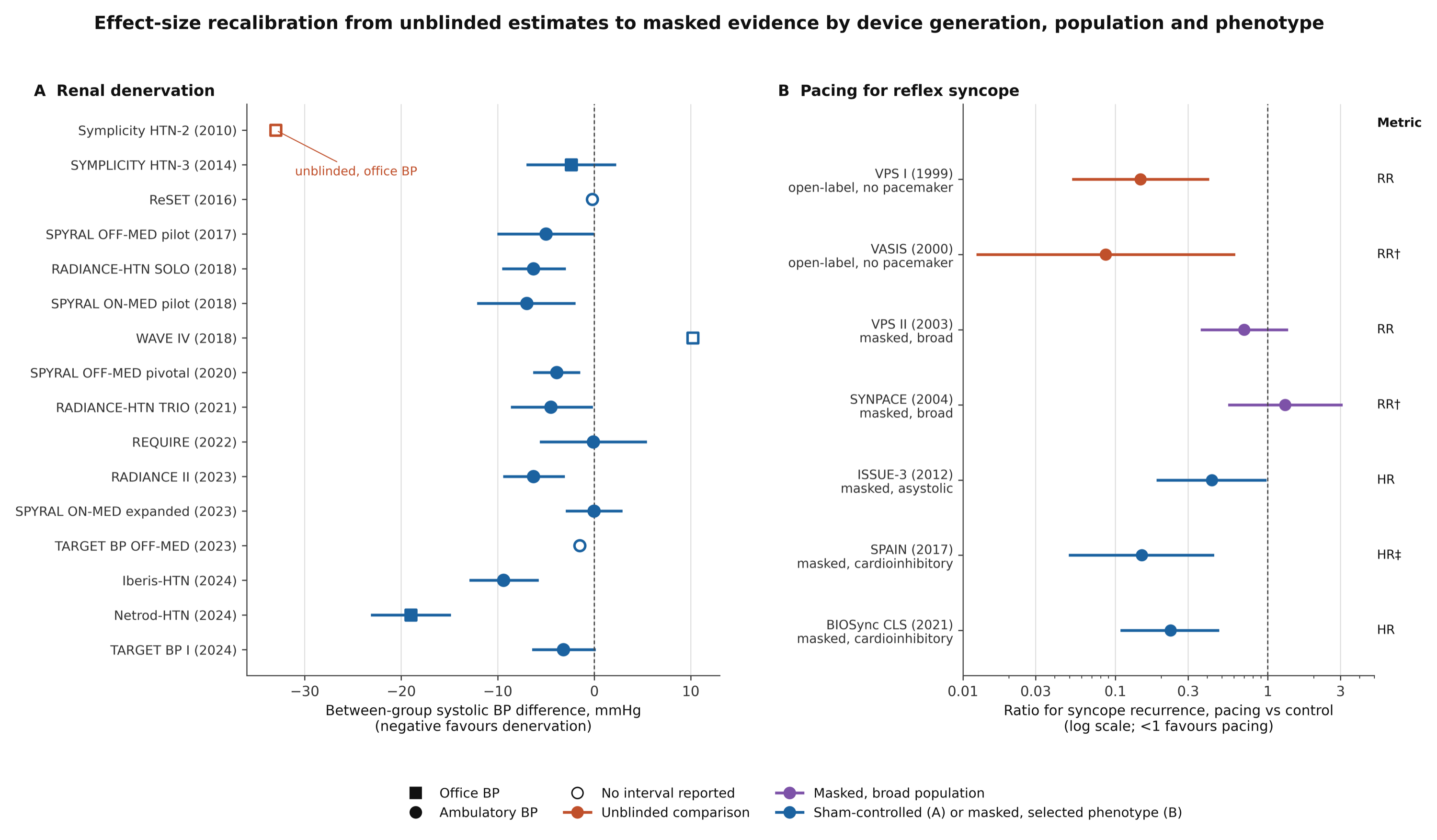}
\caption*{\textbf{Figure S7. Effect-size recalibration from unblinded to masked evidence.} (A) Renal denervation. Between-group differences in systolic blood pressure (negative values favor denervation) are shown for the unblinded Symplicity HTN-2 trial and the sham-controlled trials in Supplementary Table S1. Squares show office and circles ambulatory blood pressure, and open markers indicate that no interval was reported. (B) Pacing for reflex syncope. The ratio for syncope recurrence with pacing versus control is plotted on a log scale. VPS I and VASIS (main-text references 65 and 66) compared pacing with no pacemaker without masking. Later trials were masked, with a pacemaker implanted in all participants, and the selected-phenotype trials enrolled patients with asystolic or cardioinhibitory syncope. RR\dag{}, crude risk ratio computed from reported counts (Katz interval). HR\ddag{}, SPAIN hazard ratio inverted from the published sham-versus-pacing estimate. The estimates use different metrics and designs and are not pooled. Sources are listed with Supplementary Table S1.}
\end{figure}

\begin{figure}[!htbp]
\centering
\includegraphics[width=\textwidth,height=0.80\textheight,keepaspectratio]{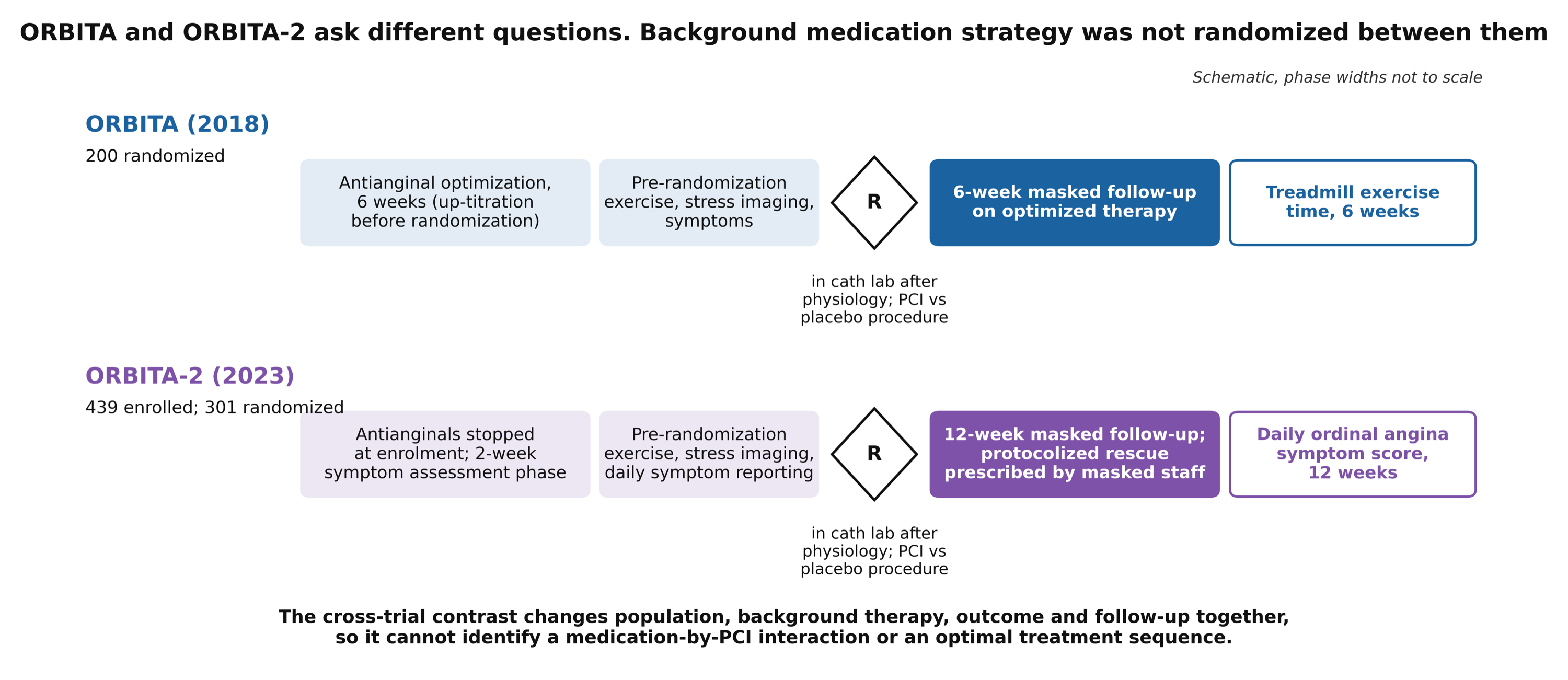}
\caption*{\textbf{Figure S8. ORBITA and ORBITA-2 answer different questions.} Schematic trial pathways, with phase widths not to scale. ORBITA randomized patients after six weeks of antianginal optimization and assessed treadmill exercise time at six weeks. ORBITA-2 stopped antianginal medication at enrollment, followed patients through a 2-week symptom assessment phase before randomization, allowed protocolized rescue prescribed by masked staff and assessed a daily ordinal symptom score over 12 weeks. Background strategy was not randomized between the trials, so their comparison cannot identify a medication-by-PCI interaction or an optimal treatment sequence.\cite{s13,s14}}
\end{figure}

\clearpage
\renewcommand{\refname}{Supplementary references}

\end{document}